\documentclass[12pt]{iopart}

\expandafter\let\csname equation*\endcsname\relax
\expandafter\let\csname endequation*\endcsname\relax 

\usepackage[english]{babel}
\usepackage{amsmath}
\usepackage{amssymb}
\usepackage{graphicx}
\usepackage{color}
\begin{document}

\title{Record statistics and persistence for a random walk with a drift}

\author{Satya N. Majumdar$^1$, Gr\'egory Schehr$^1$ and Gregor Wergen$^2$}

\address{$\;^1$Laboratoire de Physique Th\'eorique et Mod\`eles Statistiques, UMR 8626, Universit\'e Paris Sud 11 and CNRS, B\^at. 100, Orsay F-91405, France}
\address{$\;^2$Institut f\"ur Theoretische Physik, Universit\"at zu
K\"oln, 50937 K\"oln, Germany}
\ead{satya.majumdar@u-psud.fr,gregory.schehr@u-psud.fr,gw@thp.uni-koeln.de}
\begin{abstract}

We study the statistics of records of a one-dimensional random walk of $n$ steps, starting 
from the origin, and in presence of a constant bias $c$. At each time-step the walker makes a 
random jump of length $\eta$ drawn from a continuous distribution $f(\eta)$ which is symmetric 
around a constant drift $c$. We focus in particular on the case were $f(\eta)$ is a symmetric 
stable law with a L\'evy index $0 < \mu \leq 2$. The record statistics depends crucially on 
the persistence probability which, as we show here, exhibits different behaviors depending on 
the sign of $c$ and the value of the parameter $\mu$. Hence, in the limit of a large number of 
steps $n$, the record statistics is sensitive to these parameters ($c$ and $\mu$) of the jump 
distribution. We compute the asymptotic mean record number $\langle R_n \rangle$ after $n$ 
steps as well as its full distribution $P\left(R,n\right)$. We also compute the statistics of 
the ages of the longest and the shortest lasting record. Our exact computations show the 
existence of five distinct regions in the $(c, 0 < \mu \leq 2)$ strip where these quantities 
display qualitatively different behaviors. We also present numerical simulation results
that verify our analytical predictions.

\end{abstract}

\maketitle

\section{Introduction}

The statistical properties of record-breaking events in stochastic processes have been a 
popular subject of research in recent years. The theory of records has found many interesting 
applications. Record events are very important in sports \cite{Gembris2002,Gembris2007} and 
climatology \cite{Redner2006,Meehl2009,WK,AK}, but have also been found relevant in biology 
\cite{KJ}, in the theory of spin-glasses \cite{Oliveira2005,Sibani2006} and in models of 
growing networks \cite{GL1}. Also in finance, record-breaking events, e.g., 
when the price of a stock breaks its previous records, can lead to
increased financial activities~\cite{WBK,WMS}.
In all of these 
fields researchers have recently made progress in understanding and modeling the statistics of 
records by comparing the records in observational data with various kinds of stochastic 
processes. In this context it has become increasingly important to improve our understanding 
of the record statistics of elementary stochastic processes. In this
paper we focus on one such elementary stochastic process namely a random
walk in presence of a constant bias. We show that even for such a simple process, its
record statistics is considerably nontrivial and rich. 

In general, one is interested in the record events of a discrete-time series 
of random variables (RV's) $x_0,x_1,...,x_n$.
An (upper) record is an entry $x_k$, which exceeds 
all previous entries: $x_k > \textrm{max}\left(x_0,x_1,...,x_{k-1}\right)$. Until the end of 
the last century record statistics was fully understood only in the case
when the entries of the time series are 
independent and identically distributed (i.i.d.) RV's (see for 
instance \cite{FS54,Arnold,Nevzorov}). For i.i.d. RV's from a continuous distribution $p(x)$ 
the 
probability $r_n$ of a record in the $n$-th time step is given by~\cite{FS54}
\begin{equation}
 r_n := \textrm{Prob} \left[x_n > \textrm{max}\left(x_0,x_1,...,x_{n-1}\right)\right] 
= \frac{1}{n+1} \;,
\end{equation}
which is universal, i.e., independent of the parent distribution $p(x)$. This
universality follows simply from the isotropy in ordering, i.e., 
any one of the $(n+1)$ entries is equally probable to be a record.
Let $R_n$ denote the total number of records up to step $n$. The mean record number
is then simply 
$\langle R_n\rangle = \sum_{m=0}^n r_m$, which grows asymptotically
as $\sim \ln n$ for large $n$.   

Due to the numerous applications of the theory of records it became interesting to consider 
more general models. There has been a lot of interest in the record statistics of RV's which 
are uncorrelated but not identical anymore. For instance Ballerini~et~al. considered 
uncorrelated RV's with a linear drift \cite{Ballerini1985}. More recently Franke~et~al. 
studied the same problem as well and found numerous new results~\cite{FWK1,WFK,FWK2} by also 
considering the correlations between individual record events. This model was then 
successfully applied to the statistics of temperature records in the context of global warming 
\cite{WK}. In 2006 Krug studied the statistics of records of uncorrelated RV's with a 
time-increasing 
standard deviation, a model with important biological implications \cite{Krug1}.

Another important issue is the study of record statistics for {\em correlated}
random variables. For {\em weak} correlations, with a finite correlation time,
one would expect that the record statistics for a large sequence to be
asymptotically similar to the uncorrelated case. This is no longer true
when there are {\em strong} correlations between the entries.  
Perhaps, one of the simplest and most natural time series with
strong correlations between its entries corresponds to
the positions of a one dimensional random walk~\cite{Weiss}.
Despite the 
striking importance and abundance of random walk in various areas of research, 
the record 
statistics of a 
single, discrete-time random walk with a symmetric jump distribution was not computed and 
understood until only a few years ago. In 2008, Majumdar and Ziff \cite{MZrecord} 
computed exactly the record statistics of a one dimensional symmetric random walk model and 
showed that 
the 
record rate of such a process is completely universal for any continuous
and symmetric jump distribution, 
thanks to the so called Sparre Andersen theorem~\cite{SA}.
They considered a time series of RV's $x_m$ given by:
\begin{equation}
 x_m = x_{m-1} + \eta_m,
 \label{markov.symm}
\end{equation}
where $\eta_m$'s are i.i.d. RV's drawn from a 
symmetric and continuous jump distribution $f\left(\eta\right)$ (it includes
even L\'evy flights where $f(\eta)\sim 1/|\eta|^{\mu+1}$ with $0< \mu < 2$). 
Then, the record 
rate $r_n$ for such a process is given by the universal formula~\cite{MZrecord}
\begin{equation}
 r_n = \binom{2n}{n}2^{-2n} \xrightarrow{n\rightarrow\infty} \frac{1}{\sqrt{\pi n}} \;,
\end{equation}
independently of the jump distribution $f(\eta)$. 
They also computed exactly the mean record number $\langle R_n\rangle$ and even its full 
distribution~\cite{MZrecord}. In addition, there exists nice connection
between the record statistics and the extreme value statistics for the
one dimensional symmetric jump processes and many universal results
can be subsequently derived using the Sparre Andersen theorem (see 
\cite{SM10} for a review).

Following Ref. \cite{MZrecord}, there has been considerable interests in generalising them to 
more general set of strongly correlated stochastic processes. For instance, Sabhapandit 
discussed symmetric
random walks with a random, possibly heavy tailed, waiting time between the individual jumps 
(the so called Continuous Time Random Walk model)~\cite{SS}. Recently the
present authors considered the 
record 
statistics of an ensemble of $N$ independent and symmetric random walks \cite{WMS}. There, in 
contrast to the case of a single random walker, the record statistics of $N$ L\'evy flights 
with a heavy-tailed jump distribution was found to be different from the one of $N$ Gaussian 
random walkers 
with a jump distribution that has a finite second moment.

Another important generalization is to consider a single one dimensional
random walker but with asymmetric jump distribution, for instance,
in presence of a constant bias $c$. First steps towards this generalization
were taken by Le Doussal and Wiese in 2009~\cite{PLDW2009} who
derived the exact record statistics for a biased random walker with
a Cauchy jump distribution (a special case of L\'evy flights with
L\'evy index $\mu=1$).
More recently in 2011, Wergen et al. showed that a biased random walk is useful to model 
record-breaking events in daily stock prices \cite{WBK}. They were able to obtain
results in some special limits of a biased random walker with a Gaussian
jump distribution. Apart from these two special cases, namely the Cauchy and
the Gaussian jump distribution, there are no other analytical results available, to our 
knowledge, for other jump distributions for a biased random walker.
Recently, the record statistics for a biased random walker was also
studied numerically in order to quantify the contamination spread in
a porous medium via the particle tracking simulations~\cite{EKB2011}.
  
In this article we present a complete analysis of the record statistics
for a biased random walker with arbitrary jump distributions. As we will 
see, the record statistics depends crucially on the persistence probability $Q(n)$ [see Eq.~(\ref{qminus}) below], the probability that the biased walker stays to the left of
its initial starting position up to $n$ steps. While persistence probability
for various stochastic processes have been extensively studied in the
recent past~\cite{pers_review}, it seems
that for this simple biased jump process, it has not been systematically
studied in the literature to the best of our knowledge. Here we provide exact results 
for the
persistence probability $Q(n)$ for a biased random walk arbitrary jump distributions [see
Eq. (\ref{qn_asymp})], 
which subsequently leads to the exact record statistics for the same process.   

The rest of the paper is organized as follows. Since the paper is long with many detailed 
results, we provide in section \ref{results.section} a short review on the record 
statistics for random walks both with
and without bias, followed by a summary of 
the main results of this paper. Readers not interested in the details
of the calculations can skip the rest of the paper.
In 
section \ref{Gen_Sparre_Andersen.section}, we will show how to use the renewal property of the 
random walk and a generalized version of the Sparre Andersen theorem~\cite{SA} to compute the 
persistence of random walks in presence of both positive and negative drift. The results for 
the persistence are interesting on their own and will be discussed in detail in section~\ref{persistence.section}, but they will also allow us to compute the record statistics. In 
particular we will show that, in the presence of drift, the complete universality found for 
the record statistics in the unbiased case \cite{MZrecord} breaks down and there are five 
different types of asymptotic behaviors which emerge depending on the two parameters of the model, 
namely the drift $c$ and the index $0<\mu\le 2$ characterizing the tail of the jump 
distribution. This record statistics will be discussed in detail in section 
\ref{Asymm_distr.section}. Later, in section \ref{ages.section}, we will also discuss the 
extreme value statistics of the ages of the longest (section \ref{longest_records.section}) 
and the shortest lasting records (section \ref{shortest_records.section}) in each of the 
regimes. We will show that the asymptotic behavior of these quantities is also systematically 
different in the five regimes. Finally in section \ref{conclusion.section}, we will conclude 
with some open problems. 

\section{Record statistics for random walks: A short review and a summary of new results}
\label{results.section}

In this section, we provide a short review on the record statistics 
of a one dimensional random walk model, with and without external drift.
This will also serve to set up our notations for the rest of the paper. At the end
of this section, we summarize the main new results obtained in this work.

Let us first start with the driftless case.
Consider a sequence of random variables $\{x_0=0,x_1,x_2,\ldots, x_n\}$ 
where $x_m$ represents the position of a discrete-time {\em unbiased} random walker
at step $m$. The walker starts at the origin and its position evolves
via the Markov rule $x_m= x_{m-1}+ \eta_m \;,$
where $\eta_m$ represents the stochastic jump at the $m$-th step. The
jump variables $\eta_m$'s are i.i.d.
random variables, each drawn from the common probability distribution function (pdf)
$f(\eta)$, normalized to unity. The pdf $f(\eta)$ is continuous and symmetric with 
zero mean.
Let ${\hat f}(k)= \int_{-\infty}^{\infty} f(\eta)\, e^{ik\eta}\, d\eta$
denote the Fourier transform of the jump distribution. 
We will henceforth focus on jump distributions $f(\eta)$ whose Fourier transform has
the following small $k$ behavior
\begin{equation}
{\hat f}(k)= 1- (l_\mu\,|k|)^{\mu}+\ldots
\label{smallk.1}
\end{equation}
where $0< \mu\le 2$ and $l_\mu$ represents a typical length scale associated
with the jump. 
The exponent $0<\mu\le 2$ dictates
the large $|\eta|$ tail of $f(\eta)$. For jump densities with a finite
second moment $\sigma^2= \int_{-\infty}^{\infty} \eta^2\, f(\eta)\,d\eta$,
such as Gaussian, exponential, uniform etc,
one evidently has $\mu=2$ and $l_2=\sigma/\sqrt{2}$. In contrast, $0<\mu<2$ 
corresponds to jump densities with fat tails 
$f(\eta)\sim |\eta|^{-1-\mu}$ as $|\eta|\to \infty$.
A typical example is ${\hat f}(k)=\exp[-|k|^\mu]$ where $\mu=2$ corresponds to the Gaussian jump distribution, while $0<\mu<2$ corresponds
to L\'evy flights (for reviews on these jump processes see \cite{BG90,MK00}).

A quantity that will play a crucial role later is 
$P_n(x)$  which denotes the 
probability density of the position of the symmetric random walk at step $n$.
Using the Markov rule in Eq. (\ref{markov.symm}), it is easy to see that
$P_n(x)$ satisfies the recursion relation
\begin{equation}
P_n(x)= \int_{-\infty}^{\infty} P_{n-1}(x')\, f(x-x')\, dx' \;,
\label{ffp.1}
\end{equation}
starting from $P_0(x)=\delta(x)$. This recurrence relation can be trivially solved 
by taking Fourier transform
and using the convolution structure. Inverting the Fourier transform, one gets
\begin{equation}
P_n(x) = \int_{-\infty}^{\infty} \frac{dk}{2\pi}\, \left[{\hat f}(k)\right]^n \,
e^{-i\,k\, x}\,.
\label{pdf.x}
\end{equation}
In the limit of large $n$, the small $k$ behavior of ${\hat f}(k)$ dominates
the integral on the right hand side (rhs) of Eq. (\ref{pdf.x}). Substituting 
the small $k$ behavior from Eq. (\ref{smallk.1}), one easily finds that 
for $0<\mu <2$, typically $x\sim l_\mu n^{1/\mu}$ and $P_n(x)$ approaches the 
scaling form~\cite{BG90}
\begin{equation}
P_n(x) \to \frac{1}{l_\mu\, n^{1/\mu}}\, {\cal 
L}_\mu\left(\frac{x}{l_\mu\, n^{1/\mu}}\right) \;,\quad 
{\rm where}\quad {\cal L}_\mu(y) = \int_{-\infty}^{\infty} \frac{dk}{2\pi}\, 
e^{-|k|^\mu}\, e^{-i\,k\,y}\, .
\label{scaling.1}
\end{equation}  
For $0<\mu<2$, the scaling function ${\cal L}_{\mu}(y)$ decays as a power law for large 
$|y|$~\cite{BG90}
\begin{equation}
{\cal L}_\mu(y) \xrightarrow[y\to \infty]{} \frac{A_\mu}{|y|^{\mu+1}} \;, \quad {\rm 
where}\; A_\mu= \frac{1}{\pi}\,\sin(\mu\pi/2)\,\Gamma(1+\mu).
\label{lmu_asymp}
\end{equation}
In particular, for $\mu=1$, the scaling function ${\cal L}_1(y)$ is precisely
the Cauchy density itself
\begin{equation}
{\cal L}_1(y)= \frac{1}{\pi}\,\frac{1}{1+y^2}\, .
\label{cauchy.0}
\end{equation}
In contrast, for $\mu=2$, the central limit theorem holds, $x\sim {\sigma}\,n^{1/2}$\;,
and $P_n(x)$ approaches a Gaussian scaling form
\begin{equation}
P_n(x) \to \frac{1}{\sigma\,n^{1/2}}{\cal 
L}_2\left(\frac{x}{\sigma\, n^{1/2}}\right) \;,\quad
{\rm where}\quad {\cal L}_2(y)=\frac{1}{\sqrt{2\pi}}\, \exp(-y^2/2)\, .
\label{clt.1}
\end{equation}

From the sequence of symmetric random variables representing the
position of a discrete-time {\em unbiased} random walker, we next construct
a new sequence of random variables $\{y_0=0,y_1,y_2,\ldots, y_n\}$ 
where 
\begin{equation}
y_m= x_m+ c\, m\quad {\rm implying}\quad y_m=y_{m-1}+ c +\eta_m  \;,
\label{bias.1}
\end{equation}
where $\eta_m$'s are symmetric i.i.d. jump variables each drawn from the
pdf $f(\eta)$.
Clearly, $y_m$ then represents the position of a discrete-time random walker at step 
$m$ in presence of a constant bias $c$.

In this paper, we are interested in the record statistics of this biased sequence
$\{y_0=0,y_1,y_2,\ldots, y_n\}$. A record happens at step $m$
if $y_m> {\rm max}(y_0=0,y_1,y_2,\ldots, y_{m-1})$, i.e., if the
position of the biased walker $y_m$ at step $m$ is bigger than all previous 
positions, with the convention that the initial position $y_0=0$ is counted as a record. Let 
$R_n$ denote the number of records up to step $n$.
Clearly, $R_n$ is a random variable and we denote its distribution by
\begin{equation}
P(R,n)= {\rm Proba.}\,[R_n=R]\,.
\label{def_dist}
\end{equation} 
We would like to compute
the asymptotic properties of this record number distribution $P(R,n)$
for large $n$, for arbitrary drift $c$ and for arbitrary symmetric 
and continuous jump
density $f(\eta)$ whose Fourier transform ${\hat f}(k)$ has the
small $k$ behavior as in Eq. (\ref{smallk.1}) with the index $0<\mu\le 2$. 

In absence of a drift, i.e., for $c=0$, the distribution $P(R,n)$ was
computed exactly in Ref.~\cite{MZrecord} for all $R$ and $n$, using
a renewal property of the record process. 
Amazingly, the distribution was found to be completely universal, i.e.,
independent of the jump distribution $f(\eta)$ (as long as it is symmetric 
and continuous) for all $R$ and $n$~\cite{MZrecord}. In particular, for large $n$,
it was shown that $P(R,n)$ has a scaling form~\cite{MZrecord} 
\begin{equation}
P(R,n)\approx \frac{1}{\sqrt{n}}\,g_0\left(\frac{R}{\sqrt{n}}\right) \;,
\label{symm_dist.1}
\end{equation}
where the universal scaling function 
\begin{equation}
g_0(x)= \frac{1}{\sqrt{\pi}}\,\exp (-x^2/4)\;, \; {\rm for}\; x\ge 0
\label{half_gauss.1}
\end{equation}
is a half-Gaussian. Consequently, the mean and the variance of the 
number of records grows asymptotically as~\cite{MZrecord}
\begin{equation}
\langle R_n\rangle \approx \frac{2}{\sqrt{\pi}}\, n^{1/2},\quad \langle 
R_n^2\rangle-\langle R_n\rangle^2\approx 2\left(1-\frac{2}{\pi}\right)\, n\, .
\label{symm_mv.1}
\end{equation}

The renewal property of the record process derived originally for the
unbiased random walker in Ref.~\cite{MZrecord} was then generalized
to the case with a nonzero drift $c$ in Ref.~\cite{PLDW2009}. 
In particular, the authors of Ref.~\cite{PLDW2009} studied
in detail the special
case of the Cauchy jump distribution $f_{\rm Cauchy}(\eta)= 1/[\pi (1+\eta^2)]$
[which belongs to the $\mu=1$ family of jump densities in Eq. (\ref{smallk.1})]
and 
found that the mean number of records $\langle R_n\rangle$
grows algebraically with $n$ for large $n$ with an exponent
that depends continuously on $c$~\cite{PLDW2009}
\begin{equation}
\langle R_n\rangle \approx \frac{1}{\Gamma(1+\theta(c))}\, n^{\theta(c)},\quad {\rm 
where}\quad \theta(c)= \frac{1}{2}+\frac{1}{\pi}\,\arctan(c)\,.
\label{cauchy_mean.1}
\end{equation}
In addition, the asymptotic distribution $P(R,n)$ for large $n$ was 
found~\cite{PLDW2009} to have a scaling
distribution, $P(R,n)\sim n^{-\theta(c)}\, g_c\left(R\, 
n^{-\theta(c)}\right)$
with a nontrivial scaling function $g_c(x)$ which reduces, for $c=0$, to the
half-Gaussian in Eq. (\ref{half_gauss.1}).

For jump densities with a finite second moment $\sigma^2$ and in presence
of a nonzero positive drift $c>0$, the mean number of records $\langle R_n\rangle$
was analysed in Ref.~\cite{WBK} and was found to grow linearly with $n$ for large 
$n$, $\langle R_n\rangle \approx a_2(c)\, n$ where the prefactor $a_2(c)$ was
computed approximately for the Gaussian jump distribution. However, an exact 
expression of the prefactor for arbitrary jump densities with a finite
$\sigma^2$ is missing. In 
addition, these
results were then applied~\cite{WBK} to analyse the record statistics of stock 
prices from the Standard and Poors 500. The distribution of the record 
number $P(R,n)$ for large $n$ has not been studied for jump densities
with a finite second moment.

\begin{figure}
\includegraphics[width=0.8\textwidth]{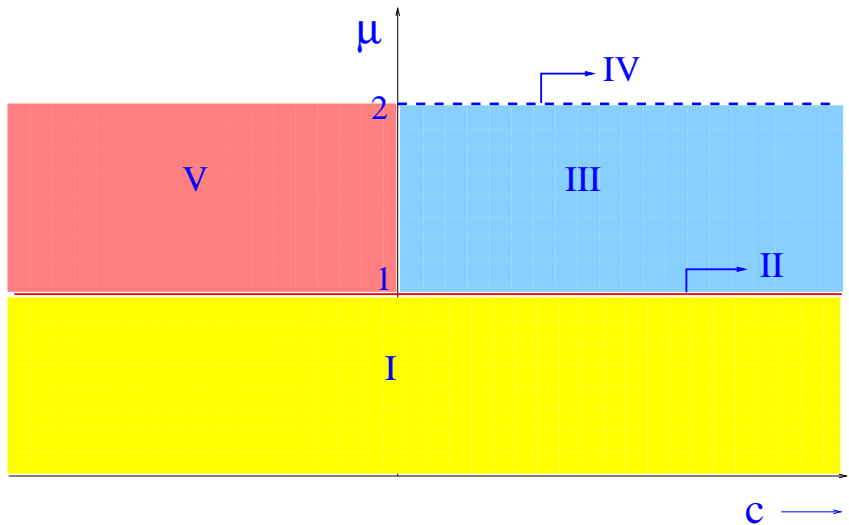}
\caption{ Phase diagram in the $(c,0<\mu\le 2)$ strip
depicting $5$ regimes: (I) $0<\mu<1$ and $c$ arbitrary (II) the line
$\mu=1$ and $c$ arbitrary (III) $1<\mu<2$ and $c>0$ (IV) the semi-infinite line
$\mu=2$ and $c>0$ and (V) $1<\mu\le 2$ and $c<0$. The persistence $Q(n)$, the record number distribution $P(R,n)$ and the mean 
ages of the longest and the shortest lasting record exhibit different asymptotic behaviors
in these $5$ regimes (see text).} 
\label{phd.fig}
\end{figure}

In this paper, we present detailed exact results for the asymptotic record number 
distribution $P(R,n)$ for large $n$, for arbitrary drift $c$ (both positive and 
negative) and for arbitrary symmetric and continuous jump densities $f(\eta)$
with Fourier transform ${\hat f}(k)$ having a small $k$ behavior
as in Eq. (\ref{smallk.1}) parametrized by the exponent $0<\mu\le 2$.
We find a variety of rather rich behaviors for $P(R,n)$
depending on the value of $c$ and the exponent $\mu$.
On the strip $(c,0<\mu\le 2)$ (see Fig. \ref{phd.fig}), we find five distinct 
regimes: (I) when $0<\mu<1$ with $c$ arbitrary
(II) when $\mu=1$ and $c$ arbitrary (III) when $1<\mu<2$ and $c>0$
(IV) when $\mu=2$ and $c>0$ and (V)
when $1<\mu\le 2$ and $c<0$. In these five regimes
the record statistics behave differently, resulting in
different asymptotic forms for the record number distribution $P(R,n)$.
The line $\mu=1$ (regime II above) is a critical line on which the record 
statistics exhibits
marginal behavior. These five regimes are summarized in the phase diagram 
in the 
$(c,0<\mu\le 2)$ strip in Fig. \ref{phd.fig}.

As we will see later, a quantity that plays a crucial role in the study 
of record statistics
is the persistence $Q(n)$ which denotes the probability that the process $y_m$ in
Eq. (\ref{bias.1}) stays below its initial value $y_0$ up to step $n$, i.e.,
\begin{equation}
Q(n) = {\rm Proba.}\,\left[y_i<y_0, {\rm for\, all}\,\, i=1,2,\ldots,n\right]\,.
\label{qminus}  
\end{equation}
Due to the translational invariance of the process, $Q(n)$ does
not depend on $y_0$. The persistence probability 
has been studied quite extensively in recent years in a variety of theoretical
and experimental systems~\cite{pers_review}. We will see that
even for the simple stochastic process $y_m$ representing the position
of a discrete-time random walker in presence of a drift, the persistence
$Q(n)$ has a rather rich asymptotic behavior depending on the parameters
$\mu$ and $c$. Hence, even though here our main interest is in the
record statistics, we include the results for the persistence $Q(n)$ as 
a byproduct.

We also analyse the statistics of 
waiting times between individual record events. In particular we are 
interested in the expected ages of the longest and the shortest lasting records. 
The age of the longest lasting record is 
defined as:
\begin{equation}
 l_{\textrm{max},n} = \textrm{max}\left(l_1,l_2,...,l_R\right),
\end{equation}
where $l_i$ is the 
length of the time interval between the $i$-th and the $\left(i+1\right)$-th record. 
Similarly one defines the age of the shortest lasting record as
\begin{equation}
 l_{\textrm{min},n} = \textrm{min}\left(l_1,l_2,...,l_R\right).
\end{equation}
In \cite{MZrecord}, the mean values of $l_{\textrm{max},n}$ and $l_{\textrm{min},n}$
were computed exactly for the symmetric random walk with arbitrary jump distribution.
It was found that~\cite{MZrecord} for large $n$
\begin{equation}
 \langle l_{\textrm{max,n}} \rangle \sim C_0 \, n \;,
 \label{unbiased_lmax}
\end{equation}
where $C_0\approx0.626508...$ is a universal constant independent of the jump 
distribution. Interestingly, the same constant $C_0$ also appears in other 
related problems \cite{pitman_yor,godreche_excursion}. In contrast, 
the shortest record exhibits different behavior for large $n$~\cite{MZrecord}
\begin{equation}
\langle l_{\textrm{min,n}} \rangle \sim \sqrt{n/\pi} \;.
\end{equation}
In this paper we generalize 
these results to the case of a biased random walk and as in the case of record number 
distribution, we find five different asymptotic behaviors depending on $c$ and $\mu$. 

\vskip 0.3cm

\noindent {\bf Summary of the new results:}
Let us then summarize the main new results in this paper for the asymptotic behavior
of the persistence $Q(n)$, the record number distribution $P(R,n)$ and 
the extremal ages of records in the $5$ regimes in the $(c,\mu)$ strip
mentioned above.  

\vskip 0.3cm

\noindent {\bf Regime I ($0<\mu<1$ and $c$ arbitrary):} In this regime,
we find that the persistence $Q(n)$ decays algebraically for large $n$
\begin{equation}
Q(n) \approx \frac{B_I}{\sqrt{n}} \;,
\label{pers.I}
\end{equation}
where the prefactor $B_I$ depends on the details of the jump distribution
$f(\eta)$ and the drift $c$ and can be computed explicitly [see Eq. (\ref{BI})].
The mean record number up to $n$ 
steps 
grows asymptotically for large $n$ as 
\begin{equation}
\langle R_n \rangle \approx A_{\rm I}\, \sqrt{n}\, .
\label{mean.I}
\end{equation}
While the growth exponent $1/2$ is universal, i.e. independent of $c$ and the
precise form of the jump distribution $f(\eta)$, the prefactor $A_{\rm I}$ depends on
$c$ and on the details of the density $f(\eta)$. In addition, the two prefactors 
$A_{\rm I}$ and $B_{\rm I}$ are related simply via $B_I= 2/(\pi A_{\rm I})$. 
We find the following exact 
expression for the prefactor $A_{\rm I}$   
\begin{equation}
A_{\rm I}= \frac{2}{\sqrt{\pi}}\, \exp\left[\frac{1}{\pi}\,\int_0^{\infty} 
\frac{dk}{k}\,
\arctan\left(\frac{{\hat f}(k)\sin(kc)}{1-{\hat f}(k)\cos(kc)}\right)\right] \;.
\label{ac.1}
\end{equation}
In the scaling limit when $n\to \infty$ and $R\to \infty$, but with
the ratio $R/\sqrt{n}$ fixed, we find that the distribution $P(R,n)$ approaches
the scaling form 
\begin{equation}
P(R,n) \approx \frac{2}{A_{\rm I}\sqrt{\pi\, n}}\, 
g_0\left(\frac{2\,R}{A_{\rm I}\sqrt{\pi\,n}}\right)\;,\quad 
{\rm where}\quad g_0(x)= \frac{1}{\sqrt{\pi}}\, \exp(-x^2/4)\,.
\label{dist.I}
\end{equation}
Averaging over $R$ evidently reproduces the result in Eq. (\ref{mean.I}). 
Thus, the record number, rescaled by the nonuniversal scale factor $R\to R/A_{\rm I}$,
approaches asymptotically the same universal half-Gaussian scaling distribution
as in the driftless case $c=0$ in Eq.~(\ref{half_gauss.1}). 

The statistics of the longest lasting record is completely unaffected by the drift $c$ in this regime. For the mean value $\langle l_{\textrm{max},n}\rangle$ we find that
\begin{equation}
  \langle l_{\textrm{max},n}\rangle \sim C_{\rm I}\,n, 
\end{equation}
where the same constant $C_I=C_0\approx0.626508...$ was also found in the unbiased case [see Eq. (\ref{unbiased_lmax})]. The age of the shortest lasting record is given by
\begin{equation}
  \langle l_{\textrm{min},n}\rangle \sim D_{\rm I} \, \sqrt{n}, 
\end{equation}
with a prefactor $D_{\rm I} = B_{\rm I}$. Therefore, in contrast to $\langle l_{\textrm{max},n}\rangle$, $\langle l_{\textrm{min},n}\rangle$ slightly differs from the unbiased case and has a prefactor that depends non-trivially on $c$.

\vskip 0.3cm

\noindent {\bf Regime II (the line $\mu=1$ and $c$ arbitrary):}  
On this line, we find that the persistence $Q(n)$ decays algebraically
for large $n$ but with an exponent that depends continuously on $c$
\begin{equation}
Q(n) \approx \frac{B_{\rm II}}{n^{\theta(c)}}\,, 
\label{pers.II}
\end{equation}
where the exponent $0\le \theta(c)\le 1$ is given in Eq. (\ref{cauchy_mean.1}).
In this sense the behavior is marginal. 
The prefactor $B_{\rm II}$ can be computed exactly [see Eq. (\ref{cl.thetac})].
The mean record number also
grows marginally for large $n$
\begin{equation}
\langle R_n\rangle \approx \frac{A_{\rm II}}{\Gamma[1+\theta(c)]}\, n^{\theta(c)} \;,
\label{mean.II}
\end{equation}
where the prefactor $A_{\rm II}= 1/\left[\Gamma[1-\theta(c)]\, B_{\rm II}\right]$. 
The record number distribution exhibits an asymptotic scaling form
\begin{equation}
P(R,n) \approx \frac{1}{A_{\rm II}\, n^{\theta(c)}}\, 
g_c\left(\frac{R}{A_{\rm II}\,n^{\theta(c)}}\right) \;,
\label{dist.II}
\end{equation}
where one can obtain a formal exact expression (\ref{prnscaling.cl}) and explicit tails of the
scaling function $g_c(x)$ which also exhibits marginal behavior, i.e., depends 
continuously on $c$.

Like in regime I we find that the mean age of the longest lasting record grows linearly in $n$, but this time with a non-trivial $c$ dependent prefactor. We find that
\begin{equation}
  \langle l_{\textrm{max},n}\rangle \sim C_{\rm II} \, n \;,
\end{equation}
where $C_{\rm II} $ is given in Eq. (\ref{lmax_regime2}). The mean age of the shortest lasting record is more strongly affected by the drift. Here we find that $\langle l_{\textrm{min},n}\rangle$ grows algebraically with $n$ with an exponent which depends continuously on $c$:
\begin{equation}
  \langle l_{\textrm{min},n}\rangle \sim D_{\rm II} \,n^{1-\theta\left(c\right)},
\end{equation}
with $D_{\rm II}=B_{\rm II}$ as in Eq. (\ref{pers.II}) and $\theta\left(c\right)$ as defined in Eq. (\ref{cauchy_mean.1}).

\vskip 0.3cm

\noindent {\bf Regime III ($1<\mu<2$ and $c>0$):}
In this regime, the persistence $Q(n)$ decays for large $n$ as
\begin{equation}
Q(n) \approx \frac{B_{\rm III}}{n^{\mu}} \;,
\label{pers.III}
\end{equation}
where the prefactor $B_{\rm III}$ depends on the details of the jump distribution
and can be computed [see Eq. (\ref{BIII})].  
The mean number of records grows linearly with increasing $n$
\begin{equation}
\langle R_n\rangle \approx a_\mu(c)\, n \;,
\label{avgrec.III}
\end{equation}
where the prefactor $a_\mu(c)$ can be computed explicitly [see Eq. (\ref{amuc})]. 
The 
record number 
distribution $P(R,n)$, for large $n$,
behaves as
\begin{equation}
P(R,n) \approx  \frac{1}{a_\mu(c) n^{1/\mu}}\, V_\mu\left( \frac{R-a_\mu(c) 
n}{a_\mu(c)\, n^{1/\mu}}\right) \;,
\label{dist.III}
\end{equation}
where the scaling function $V_\mu(u)$ can be computed exactly and it 
has a non-Gaussian form 
with highly 
asymmetric tails
\begin{eqnarray}
V_\mu(u) & \approx & A_\mu\, |u|^{-\mu-1}\quad {\rm as}\quad u\to -\infty 
\label{vmuleft} 
\\
& \approx & c_1\, u^{(2-\mu)/{2(\mu-1)}}\, \exp\left[-c_2 \, 
u^{\mu/(\mu-1)}\right]\quad {\rm as}\quad u\to \infty \label{vmuright} \;,
\end{eqnarray}
where $A_\mu$ is the same constant as in Eq. (\ref{lmu_asymp}) and the constants 
$c_1$
and $c_2$ are given explicitly by
\begin{eqnarray}
c_1 &=& \left[2\pi (\mu-1) (\mu B_\mu)^{1/(\mu-1)}\right]^{-1/2} \label{c1} \:,\\
c_2 &=& (1-1/\mu)\, (\mu B_\mu)^{-1/(\mu-1)} \label{c2} \;,
\end{eqnarray}
where  
\begin{equation}
B_\mu= -\frac{1}{2\cos(\mu\pi/2)}>0\quad {\rm for} \quad 1<\mu<2\,.
\label{bmu}
\end{equation} 
Thus, in this regime, while the mean record number grows linearly with $n$, the
fluctuations around the mean are anomalous $\sim n^{1/\mu}$ and described 
by a non-Gaussian distribution.

Also the extremal ages of records have an interesting behavior in this regime. In particular we find that the average age of the longest lasting record grows like
\begin{equation}
  \langle l_{\textrm{max},n}\rangle \sim C_{\rm III} \, n^{\frac{1}{\mu}},
\end{equation}
where the constant $C_{\rm III}$ can be computed explicitly [see Eq. (\ref{c3})]. On the other hand and in contrast to the results of regime I and II, the mean age of the shortest lasting record converges to a finite value:
\begin{equation}
  \langle l_{\textrm{min},n}\rangle \sim D_{\rm III} = 1 - a_\mu(c) \;,
\end{equation}
which thus depends continuously on $c$. The strongly different $n$ dependence of $\langle l_{\textrm{max},n}\rangle$ and $\langle l_{\textrm{min},n}\rangle$ in the regime I and in the regime III is a consequence of the fact that while in regime I the asymptotic behavior is dominated by the fluctuations, in regime III the effect of the drift is stronger in the large $n$ limit. 

\vskip 0.3cm

\noindent {\bf Regime IV ( the semi-infinite line $\mu=2$ and $c>0$):} 
On this semi-infinite line the variance $\sigma^2$ of the jump pdf is finite.
This leads to an exponential tail of the persistence $Q(n)$ for large $n$. 
More precisely we show that
\begin{equation}
Q(n) \approx \frac{B_{\rm IV}}{n^{3/2}}\, \exp[- (c^2/2\sigma^2)\, n] \;, \label{pers.IV}
\end{equation}
where the nonuniversal prefactor $B_{\rm IV}$ can be computed exactly [see Eq. 
(\ref{BIV})]. 
We also show that the mean and the variance of the record number both grow
linearly for large~$n$
\begin{equation}
\langle R_n\rangle \approx a_2(c)\, n \quad {\rm and}\quad \langle R_n^2\rangle 
-\langle 
R_n\rangle^2\approx b_2(c)\, n \;,
\label{mean_var_IV}
\end{equation}
where the amplitudes $a_2(c)$ and $b_2(c)$ are nonuniversal and depend on the 
details
of the jump distribution $f(\eta)$. We provide exact expressions for these amplitudes
respectively in Eqs. (\ref{a2c}) and (\ref{variance.reg4}) as well as in  \ref{appendix_a2}. The distribution of 
the record 
number $P(R,n)$ 
approaches a Gaussian form asymptotically for large $n$
\begin{equation}
P(R,n)\approx \frac{1}{\sqrt{2\,\pi\,b_2(c)\, n}}\,\exp\left[-\frac{1}{2 
b_2(c)n}(R-a_2(c)\,n)^2\right].
\label{dist_IV}
\end{equation}
Thus, in this regime, the mean record number grows linearly with $n$ with
normal Gaussian fluctuations $\sim n^{1/2}$ around the mean.

It is interesting to see that the asymptotic behavior of $\langle l_{{\rm max},n}\rangle$
in regime IV is qualitatively different from regime III. Here we find that $\langle l_{\textrm{max},n}\rangle$ grows only logarithmically with $n$ for $n\rightarrow\infty$:
\begin{equation}
  \langle l_{\textrm{max},n}\rangle \sim C_{\rm IV} \ln n \;,
\end{equation}
with an $n$ independent constant $C_{\rm IV}=\frac{2\sigma^2}{c^2}$. Like in regime III, the 
average age of the shortest lasting record approaches a (different) constant value depending on $c$:
\begin{equation}
  \langle l_{\textrm{min},n}\rangle \sim D_{\rm IV} = 1 - a_2(c)\;,
\end{equation}
 which depends continuously on $c$. 
 
\vskip 0.3cm

\noindent {\bf Regime V ($1<\mu\le 2$ and $c<0$):} In this case, the walker 
predominantly moves towards the negative axis due to the drift.
Consequently, the events where the walker crosses the origin
from the negative to the positive side become extremely rare. 
As a result, with a finite probability the walker stays 
forever on the negative side. Thus, the persistence $Q(n)$ approaches
a constant for large $n$
\begin{equation}
Q(n) \to \alpha_\mu (c)\, .
\label{pers.V}
\end{equation}
Similarly,   
the
occurrence of the records (with positive record values) are also rare.
Subsequently, we find that the mean record number also approaches a
constant for large $n$
\begin{equation}
\langle R_n\rangle \to \frac{1}{\alpha_\mu(c)} \;,
\label{mean_IV}
\end{equation}
where the constant $\alpha_\mu(c)$ is given by
\begin{eqnarray}
\alpha_\mu(c)&= & a_\mu(|c|)\quad {\rm for}\quad 1<\mu<2 \:, \label{meanconst1} \\
& =& a_2(|c|)\quad {\rm for}\quad \mu=2 \;, \label{meanconst2}
\end{eqnarray} 
where $a_\mu(c)$ and $a_2(c)$ are precisely the amplitude of the linear growth of 
the mean record number respectively in regime III and IV [respectively 
in Eqs. (\ref{avgrec.III}) and (\ref{mean_var_IV})]. An explicit 
expression for $\alpha_\mu(c)$ is given in Eq. (\ref{reg5.qn}).
The record 
number distribution 
$P(R,n)$
also approaches a steady state, i.e., $n$-independent distribution as
$n\to \infty$. This distribution has a purely geometric form
\begin{equation}
P(R,n\to \infty)= \alpha_\mu(c)\,[1-\alpha_\mu(c)]^{R-1}\,. 
\label{dist_V}
\end{equation} 
Physically this result is easy to understand because for $c<0$ and $\mu>1$, the 
walker typically moves away from the origin on the negative side with
very rare and occasional excursions to the positive side caused by rare
large jumps. As a result, the occurrence of a record is like a Poisson process
which eventually leads to a geometric distribution as in Eq. (\ref{dist_V}).

In this regime the statistics of the longest and the shortest lasting records are particularly simple. Since the record number is finite, the longest lasting record will grow linearly in $n$:
\begin{equation}
  \langle l_{\textrm{max},n}\rangle \sim  C_{\rm V}\, n \;, C_{\rm V} = 1 \;. 
\end{equation}
For the shortest lasting record we find a similar behavior:
\begin{equation}
  \langle l_{\textrm{min},n}\rangle \sim \alpha_{\mu}(c) \, n,
\end{equation}
with the same $c$ dependent constant $\alpha_{\mu}\left(c\right)$ as in Eq. (\ref{pers.V}). Here, the main contributions to these quantities 
come from trajectories that never cross the origin and stay negative for all $n$.

The five regimes in the $(c,0<\mu\le 2)$ strip
are depicted in Fig.~\ref{phd.fig}. As mentioned above, 
the line $\mu=1$  is a special `critical' line with marginal exponents that
depend continuously on the drift $c$. 
It is not difficult to understand physically why
$\mu=1$ plays a special role. Indeed, writing $y_n=x_n+c\,n$ where
$x_n$ represents a symmetric random walk, we see that the two terms
$x_n$ and $c\,n$ compete with each other for large $n$. Since $x_n\sim n^{1/\mu}$
for $0<\mu\le 2$ [see Eq. (\ref{scaling.1})], it is clear that for $0<\mu<1$, the 
term $x_n$ dominates over the drift and the presence of a nonzero drift only
leads to subleading asymptotic effect. In contrast, for $\mu>1$, the drift term
starts to play an important role in governing the asymptotic record statistics. 
In the region $1<\mu<2$ and $c>0$ (regime III), while the mean record number 
increases linearly with $n$ 
due to the
dominance of the drift term, the typical fluctuation around the mean
is still dominated by the $x_n\sim n^{1/\mu}$ term [see Eq. (\ref{dist.III})].
However when $\mu=2$ and $c>0$ (regime IV), the drift term completely dominates
over the $x_n$ term leading to Gaussian fluctuations around the mean.  
This
competition between $x_n$ and $c\,n$ thus leads to (i) a `phase transition' in the 
asymptotic 
behavior of record statistics of $y_n$
at the critical value $\mu=1$ and (ii) an anomalous region with non-Gaussian
fluctuations around the mean in the region $1<\mu<2$ and $c>0$.

\section{Record Number distribution via renewal property and the generalized Sparre 
Andersen theorem}
\label{Gen_Sparre_Andersen.section}

The idea of using the renewal property of random walks to compute the distribution
of record number was first used in Ref.~\cite{MZrecord} for symmetric random walks 
and was subsequently 
generalized to biased random walks~\cite{PLDW2009}.
We briefly summarize below the main idea. 

Consider the random sequence
$\{y_0,y_1,y_2,\ldots,\}$ representing the successive positions of
a discrete-time biased random walker evolving via Eq. (\ref{bias.1}), starting from 
an arbitrary initial position $y_0$. Consider first the persistence
$Q(n)$ defined in Eq. (\ref{qminus}). 
Let us also define 
\begin{equation}
F(n)= {\rm Proba.}\,\left[y_1<y_0,\, y_2<y_0,\,\ldots,\, y_{n-1}<y_0,\, y_n>y_0\right] 
\label{fminus}
\end{equation}
which denotes the probability that the walker crosses its initial position
$y_0$ from {\em below} for the first time at step $n$. Clearly
\begin{equation}
F(n)= Q(n-1)-Q(n) \,.
\label{qfrelation.1}
\end{equation}
It is also useful to define the generating functions
\begin{equation}
{\tilde Q}(z) = \sum_{n=0}^{\infty} Q(n)\, z^n \;, \quad {\tilde F}(z)= 
\sum_{n=1}^{\infty} F(n)\, z^n\,.
\label{def_gen.1}
\end{equation}
Using the relation in Eq. (\ref{qfrelation.1}) it follows that
\begin{equation}
{\tilde F}(z)= 1- (1-z){\tilde Q}(z)\,.
\label{qfrelation.2}
\end{equation}

\begin{figure}
\includegraphics[width=0.8\textwidth]{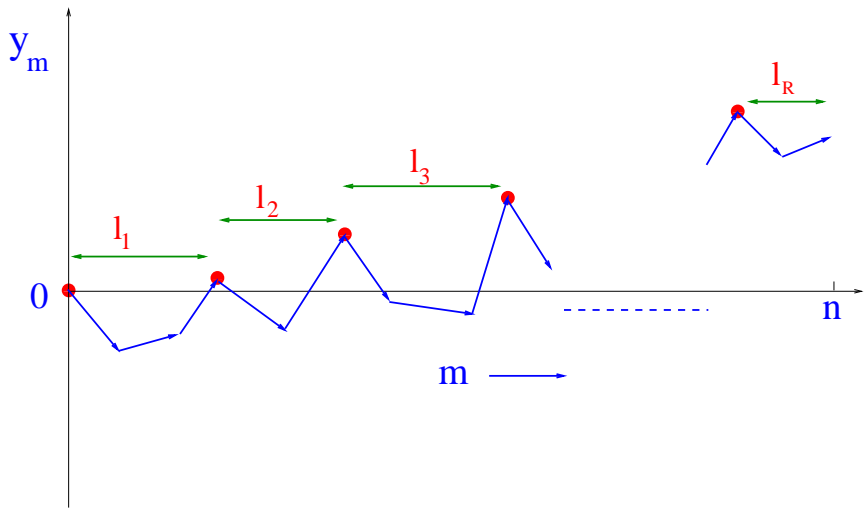}
\caption{ A typical realization of the biased random walk sequence 
$\{y_0=0,y_1,y_2,\ldots, y_n\}$ of $n$ steps with $R$ records. Each record
is represented by a filled circle. The set $\{l_1,l_2,\ldots,l_{R-1}\}$ represents
the time intervals between the successive records and $l_R$ is the age of 
the last record which is still a record till step $n$.} 
\label{renewal.fig}
\end{figure}

Consider now any realization of the sequence $\{y_0=0,y_1,y_2,\ldots,y_n\}$ up to 
$n$ steps and let $R_n$ be the number of records in this realization. Let
$\vec l= \{l_1,l_2,\ldots, l_R\}$ denote the time intervals between successive 
records in this sequence (see Fig. \ref{renewal.fig}). Clearly $l_i$ denotes
the age of the $i$-th record, i.e., the time up to which the $i$-th record survives.
The last record, i.e. the $R$-th record, stays a record till step $n$.
Let $P(\vec l, R|n)$ denote the joint distribution of the ages and the number
of records up to step $n$. Using the two probabilities $Q(n)$ and $F(n)$
defined earlier and the fact that the 
successive intervals
between records are statistically independent due to the Markov nature of the
process, it follows immediately that 
\begin{equation}
P(\vec l, R|n)= F(l_1)F(l_2)\ldots F(l_R)\,Q(l_R)\, \delta_{\sum_{i=1}^R l_i,N} \;,
\label{renewal.1}
\end{equation}
where the Kronecker delta enforces the global constraint that the sum of the
time intervals equals $n$. The fact that the last record, i.e. the $R$-th record,
is still surviving as a record at step $n$ indicates that the distribution
$Q(l_R)$ of $l_R$ is different from the preceding ones. It is easy to check
that $P(\vec l, R|n)$ is normalized to unity when summed over $\vec l$ and $R$.
The record number distribution $P(R,n)= \sum_{\vec l}P(\vec l,R|n)$ is just the 
marginal of the joint 
distribution when one sums over the interval lengths. Due to the presence of
the delta function, this sum is most easily carried out by considering
the generating function with respect to $n$. Multiplying Eq. (\ref{renewal.1})
by $z^n$ and summing over $\vec l$ and $n$, one arrives at the fundamental
relation
\begin{equation}
\sum_{n=0}^{\infty} P(R,n)\, z^n = \left[{\tilde F}(z)\right]^{R-1}\, {\tilde 
Q}(z)=\left[1-(1-z) {\tilde Q}(z)\right]^{R-1}\, {\tilde Q}(z) \;,
\label{renewal.genf}
\end{equation}
where we used the relation in Eq. (\ref{qfrelation.2}). Note that, by definition,
$R\le (n+1)$, i.e. $P(R,n)=0$ if $n<R-1$. Hence, the sum in Eq.~(\ref{renewal.genf}) actually runs from $n=R-1$ to~$\infty$. 

Thus the basic object is the generating function ${\tilde Q}(z)$. Once
this is determined, one can, at least in principle, compute other
quantities such as the statistics of records or their ages using the fundamental
renewal equation (\ref{renewal.genf}). Fortunately, there exists a beautiful
combinatorial identity first derived by Sparre Andersen~\cite{SA} that allows
one to compute ${\tilde Q}(z)$
\begin{equation}
{\tilde Q}(z) = \sum_{n=0}^{\infty} Q(n)\, z^n= 
\exp\left[\sum_{n=1}^{\infty}\frac{z^n}{n}\, p(n)\right] \;,
\label{SA.1}
\end{equation}
where $p(n)= {\rm Proba.}\,[y_n<0]$. Using the relation $y_n=x_n+cn$ 
where
$x_n$ represents the symmetric random walk at step $n$ in Eq. (\ref{markov.symm})
one gets, $p(n)= {\rm Proba.}\,[x_n<-cn]$. 
Then, using the pdf $P_n(x)$ of the symmetric walk $x_n$ at step $n$ in Eq. 
(\ref{pdf.x}), one gets
\begin{equation}
p(n)= {\rm Proba.}\;[x_n<-cn]= \int_{-\infty}^{-cn}P_n(x)\,dx= 
\int_{cn}^{\infty} P_n(x)\, dx \;,
\label{SA.symm}
\end{equation}
where, in obtaining the last equality we used the symmetry $P_n(x)=P_n(-x)$.
Substituting this expression of $p(n)$ in Eq. (\ref{SA.1}) gives
\begin{equation}
{\tilde Q}(z) = \sum_{n=0}^{\infty} Q(n)\, 
z^n=\exp\left[\sum_{n=1}^{\infty}\frac{z^n}{n}\,\int_{cn}^{\infty} P_n(x)\, 
dx\right].
\label{SA.2}
\end{equation}

Eq. (\ref{SA.2}), with $P_n(x)$ given by Eq. (\ref{pdf.x}), determines ${\tilde 
Q}(z)$ in terms of the Fourier transform ${\hat f}(k)$ of the jump distribution 
$f(\eta)$. Subsequently Eq. (\ref{renewal.genf}) then determines, in principle, the
record number distribution $P(R,n)$.
In the driftless case $c=0$, great simplification occurs, since by
symmetry $\int_0^{\infty} P_n(x)dx=1/2$. This gives, from Eq. (\ref{SA.2}),
${\tilde Q}(z)= 1/\sqrt{1-z}$. This is completely universal as all
the dependence on the jump distribution $f(\eta)$ drops out. Subsequently, 
Eq. (\ref{renewal.genf}) provides, for $c=0$, the universal result for the record 
number distribution~\cite{MZrecord}
\begin{equation}
\sum_{n=0}^{\infty} P(R,n)\, z^n= \frac{\left(1-\sqrt{1-z}\right)^{R-1}}{\sqrt{1-z}} \;,
\label{rnd.symm}
\end{equation}
which, when inverted, yields~\cite{MZrecord} for large $n$ the scaling behavior
in Eq. (\ref{symm_dist.1}) with the scaling function given by the
half-Gaussian form in Eq. (\ref{half_gauss.1}).
 
However, in presence of a nonzero bias $c$,
extraction of the precise large $n$ behavior of $P(R,n)$
from the set of equations (\ref{renewal.genf}), (\ref{SA.2}) and (\ref{pdf.x})
is more complicated. For the special case of the Cauchy distribution, this
was performed in Ref.~\cite{PLDW2009} which showed nontrivial behavior.
The rest of this paper is devoted precisely to this technical task
of extracting the large $n$ behavior of $P(R,n)$ for a general
jump distribution $f(\eta)$
and we will see that a variety of rather rich asymptotic behavior 
emerges 
depending on the value of the drift $c$ and the exponent $\mu$
characterizing the small $k$ behavior of ${\hat f}(k)$ in Eq. (\ref{smallk.1}).

Before finishing this section, let us remark that from Eq. (\ref{renewal.genf})
one can also compute the generating functions of the moments of the number
of records. For example, multiplying Eq. (\ref{renewal.genf}) by $R$, summing 
over $R$ and using the identity $\sum_{n=0}^{\infty} n x^{n-1}= 1/(1-x)^2$ we get
for the first moment
\begin{equation}
\sum_{n=0}^{\infty} \langle R_n\rangle\, z^n = \frac{1}{(1-z)^2 {\tilde Q}(z)}.
\label{mom1.genf}
\end{equation}
Similarly, multiplying Eq. (\ref{renewal.genf}) by $R^2$ and summing over $R$ one 
gets for the second moment
\begin{equation}
\sum_{n=0}^{\infty} \langle R_n^2\rangle\, z^n= \frac{2-(1-z){\tilde 
Q}(z)}{(1-z)^3\, {\tilde Q}^2(z)}.
\label{mom2.genf}
\end{equation}
We will use these two results later in Section IVB.

\section{Asymptotic behavior of persistence $Q(n)$ for large $n$}
\label{persistence.section}

\begin{figure} \includegraphics[angle=-90,width=1\textwidth]{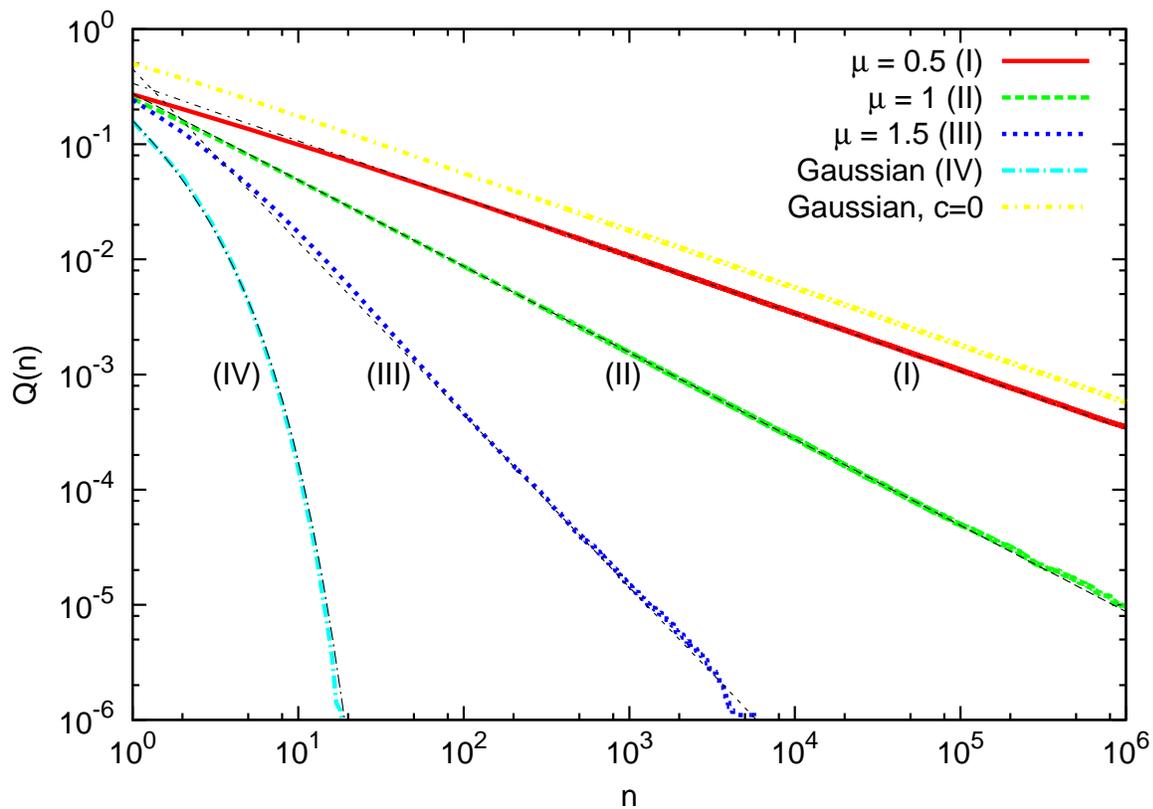} 
\caption{Numerical simulations of the persistence $Q\left(n\right)$, i.e. the
probability that a random walker with a bias $c$ stays {\em below} its initial position
up to step $n$. We considered $4$ different L\'evy-stable jump distributions
characterized respectively by the L\'evy index  
$\mu=0.5$, $1$, $1.5$ and $\mu=2$ (in the last case it is just Gaussian jump distribution). 
In all cases, we
had a constant positive bias $c=1$ and the data were obtained by averaging over $10^7$ samples. 
For comparison, we also present the result for the unbiased case ($c=0$) with
a Gaussian jump distribution (the top curve). 
The thin dashed lines give our analytical 
predictions 
from Eq. (\ref{qn_asymp}) with fitted prefactors $B_{\rm I}$, $B_{\rm II}$,
$B_{\rm III}$ and 
$B_{\rm IV}$. For the 
$\mu=1$ case we used $\theta\left(c=1\right)\approx 0.7498...$.} 
\label{persistence_I_to_IV.fig} 
\end{figure}

The persistence $Q(n)$, i.e. the probability that the process $y_n$
stays below its initial value $y_0$ up to step $n$  and its 
generating function ${\tilde Q}(z)$
is the key ingredient to determine the
record number distribution $P(R,n)$ via Eq. (\ref{renewal.genf}).
Apart from its key role as an input for the record statistics, the
persistence $Q(n)$ for this process is, by itself, an interesting quantity 
to study. 
We will see in this section that
even for the simple stochastic process $y_n$, representing the position
of a discrete-time random walker in presence of a drift, the persistence
$Q(n)$ has a rather rich asymptotic behavior depending on the parameters
$\mu$ and $c$. Before getting into the details of the derivation, it is useful to 
summarize these asymptotic results. We find that for large $n$, the
persistence $Q(n)$ has the following asymptotic tails depending
on $\mu>0$ and $c$
\begin{eqnarray}
Q(n) &\sim & B_{\rm I}\, n^{-1/2}\quad {\rm for}\quad 0<\mu<1\,\, {\rm and}\,\, c\,\,{\rm 
arbitrary}\quad ({\rm regime}\,\, {\rm I}) \;,
\nonumber \\
&\sim & B_{\rm II}\, n^{-\theta(c)} \quad {\rm for}\quad \mu=1 \,\, {\rm and}\,\,c \,\,{\rm
arbitrary}  
\quad ({\rm regime}\,\, {\rm II}) \;,
\nonumber \\
&\sim & B_{\rm III}\, n^{-\mu} \quad {\rm for}\quad 1<\mu<2\,\, {\rm and}\,\, c>0 
\quad ({\rm regime}\,\, {\rm III}) \;,
\nonumber \\
&\sim & B_{\rm IV}\, n^{-3/2}\, \exp[-(c^2/{2\sigma^2})\, n] \quad  {\rm for}\quad \mu=2 \,\, 
{\rm and}\,\, c>0 
\quad ({\rm regime}\,\, {\rm IV}) \;,
\nonumber \\
&\sim & \alpha_\mu(c) \quad {\rm for}\quad 1<\mu\le 2 \,\, {\rm and}\,\, c<0 
\quad ({\rm regime}\,\, {\rm V}) \;,
\nonumber \\
\label{qn_asymp}
\end{eqnarray}
where the prefactors $B_{\rm I}$, $B_{\rm II}$, $B_{\rm III}$, $B_{\rm IV}$ 
can be explicitly computed.
In regime ${\rm V}$, $\alpha_\mu(c)$ is a constant independent of $n$ that 
can also be computed explicitly
[see Eq. (\ref{reg5.qn}) and \ref{appendix_a2} for $\alpha_2(\mu)$].  
The exponent $\theta(c)$
depends continuously on $c$ and is given in Eq. (\ref{cauchy_mean.1})
[see also Eq. (\ref{cl.11})]. In Fig. \ref{persistence_I_to_IV.fig} these results are 
confirmed numerically for the regimes I-IV. 

\begin{figure}
\includegraphics[width=0.8\textwidth]{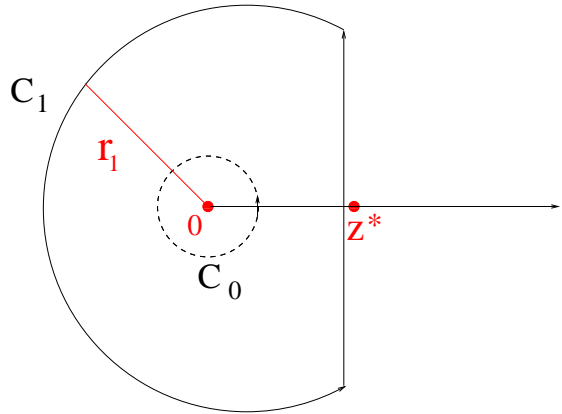}
\caption{The contour $C_0$ in the complex $z$ plane can be deformed to the 
contour $C_1$. In the large $n$ limit, the dominant contribution to the
Cauchy integral in Eq. (\ref{cauchy.1}) comes from the vertical part
of $C_1$.}
\label{contour.fig}
\end{figure}

To derive these asymptotic behaviors of $Q(n)$ for large $n$, we start
with the key result in Eq. (\ref{SA.2}). Using Cauchy's inversion formula
in the complex $z$ plane one can write
\begin{equation}
Q(n) = \int_{C_0} \frac{dz}{2\pi i}\, \frac{1}{z^{n+1}}\, {\tilde Q}(z) \quad 
{\rm with}\quad {\tilde 
Q}(z)=\exp\left[\sum_{n=1}^{\infty}\frac{z^n}{n}\,\int_{cn}^{\infty} 
P_n(x)\, dx\right] \;,
\label{cauchy.1}
\end{equation}
where the contour $C_0$ encircles the origin $0$ and is free of any singularity
of ${\tilde Q}(z)$ (see Fig.~\ref{contour.fig}). Let $z^*$ denote the singularity of ${\tilde Q}(z)$ on the real axis closest to the
origin. Then, one can deform the contour $C_0$ to $C_1$ (see Fig.~\ref{contour.fig}) 
such that the vertical part of $C_1$ is located just left of
$z^*$ and the circular part has radius $r_1$. By taking the $r_1\to \infty$ limit,
it follows from Eq.~(\ref{cauchy.1}) that for large $n$, the contribution
from the circular part vanishes exponentially. Thus for large $n$, the
leading contribution comes from the vertical part of $C_1$, i.e the
imaginary axis located just left of $z^*$. Next we make a change of variable
$z=e^{-sn}$ and define
\begin{eqnarray}
&&{\tilde q}(s)={\tilde Q}(z=e^{-s})=\sum_{n=0}^{\infty} Q(n)\, e^{-sn}= 
\exp\left[W_{c, \mu}(s)\right]\;, \label{SA.3bis} \\
&&\quad {\rm where}\quad 
W_{c, \mu}(s)= \sum_{n=1}^{\infty}\frac{e^{-sn}}{n}\,\int_{cn}^{\infty} P_n(x)\,dx\, .
\label{SA.3}
\end{eqnarray}
Using this expression in the integrand in Eq. (\ref{cauchy.1})
and retaining only the contribution from the vertical part of the contour
$C_1$ for large $n$, we get
\begin{equation}
Q(n) \approx \int_{s^*-i\infty}^{s^*+i\infty} \frac{ds}{2\pi i}\, e^{s\,n}\, 
\exp\left[W_{c, \mu}(s)\right] \;,
\label{qn.brom}
\end{equation}
where $W_{c, \mu}(s)$ is given in Eq. (\ref{SA.3}) and
$s^*=-\ln(z^*)$ is the singularity of ${\tilde q}(s)=\exp[W_{c, \mu}(s)]$
on the real axis closest to $s=0$. Identifying the integral
on the rhs of Eq. (\ref{qn.brom}) as a standard Bromwich integral
in the complex $s$ plane, we see that for large $n$, $Q(n)$ is
essentially given by the inverse Laplace transform of the
function ${\tilde q}(s)=\exp[W_{c, \mu}(s)]$. 
To make further progress, we need to first identify the 
position of the singularity $s^*$ of $W_{c, \mu}(s)$ and
then analyse the dominant contribution 
in the Bromwich integral coming from the neighborhood
of $s^*$ for large $n$. We see below that the singular behavior
of $W_{c, \mu}(s)$ as a function of $s$ depends on the parameters
$c$ and $\mu>0$ and there are essentially $5$ regimes in the
$(c, 0<\mu\le 2)$ strip as shown in Fig.~\ref{phd.fig}.
Below we discuss these regimes separately.

\subsection{Regime I: $0<\mu<1$ and $c$ arbitrary}

To analyse the leading singularity of $W_{c, \mu}(s)$ as a function of 
$s$ in
this regime, it is first convenient to use the normalization condition
$\int_{-\infty}^{\infty} P_n(x)dx=1$ and the symmetry $P_n(x)=P_n(-x)$
to rewrite
\begin{equation}
\int_{cn}^{\infty} P_n(x)\,dx = \frac{1}{2}-\int_0^{cn} P_n(x)\,dx\,.
\label{reg1.1}
\end{equation}
Substituting this in Eq. (\ref{SA.3}) gives
\begin{equation}
W_{c, \mu}(s) = -\frac{1}{2}\, \ln\left(1-e^{-s}\right) - 
\sum_{n=1}^{\infty}\frac{e^{-sn}}{n}\,\int_{0}^{cn} P_n(x)\,dx \,.
\label{reg1.2}
\end{equation}
Now, as $s\to 0$, the sum in Eq. (\ref{reg1.2}) converges to a constant
for $0<\mu<1$
\begin{equation}
S_0=\sum_{n=1}^{\infty}\frac{1}{n}\,\int_{0}^{cn} P_n(x)\,dx \, .
\label{sumS}
\end{equation}
To see this, let us see how the
integral $\int_0^{cn} P_n(x)\,dx$ behaves for large $n$. For large $n$,
we can use the scaling form for $P_n(x)$ in Eq. (\ref{scaling.1}). One finds
that $\int_0^{cn} P_n(x) dx\to \int_0^{c n^{(1-1/\mu)}} {\cal L}_\mu(y)\, dy$ 
as $n\to \infty$. For $0<\mu<1$, clearly this integral decreases leading
to the convergence of the series in Eq. (\ref{sumS}). Thus, the leading 
singularity of $W_{c, \mu}(s)$ occurs near $s=s^*=0$ where it behaves as
\begin{equation}
W_{c, \mu}(s)\approx -\frac{1}{2}\, \ln(s) - S_0 \;.
\label{reg1.singular1}
\end{equation}
Substituting this result in Eq. (\ref{SA.3}) gives
\begin{equation}
{\tilde q}(s)= \sum_{n=0}^{\infty} Q(n)\, e^{-sn}\xrightarrow[s\to 0]{} 
\frac{e^{-S_0}}{\sqrt{s}}\,.
\label{qs.1}
\end{equation}
We now substitute this singular behavior of the integrand in Eq. (\ref{qn.brom})
after setting $s^*=0$ and perform the standard Bromwich integral
(one can use the fact that the inverse Laplace transform
$LT^{-1}_{s\to n}[s^{-1/2}]= 1/\sqrt{\pi n}$\,)
\begin{equation}
Q(n) \xrightarrow[n\to \infty]{} \frac{B_{\rm I}}{\sqrt{ n}} \;,
\label{reg1.qn}
\end{equation}  
where the prefactor $B_{\rm I}$ is given by
\begin{equation}
B_{\rm I}= \frac{e^{-S_0}}{\sqrt{\pi}}= \frac{1}{\sqrt{\pi}}\, 
\exp\left[-\sum_{n=1}^{\infty} \frac{1}{n}\, \int_0^{cn} P_n(x)\, dx\right]\,.
\label{BI}
\end{equation}

\subsection{Regime II: $\mu=1$ and $c$ arbitrary}

The case $\mu=1$ is rather special and marginal as we demonstrate now.
Consider the sum $W_{c,1}(s)$ in Eq. (\ref{SA.3}).
In this case, it follows from Eq. (\ref{scaling.1}) that
$P_n(x)\to (1/n) {\cal L}_1(x/n)$ as $n\to \infty$, where
${\cal L}_1(y)=1/[\pi (1+y^2)]$ for all $y$ and hence is integrable.
Thus the integral $\int_{cn}^{\infty} P_n(x) dx$ converges to
a constant for large $n$ 
\begin{equation}
\int_{cn}^{\infty} P_n(x) dx \xrightarrow[n\to \infty]{} 
\int_{c}^{\infty} {\cal L}_1(y)\, dy=1-\theta(c),
\label{cl.1}
\end{equation}
where
\begin{equation}
\theta(c)=\int_{-\infty}^c {\cal L}_1(y) dy= \frac{1}{2}+\frac{1}{\pi}\,\arctan(c)\,. 
\label{cl.11}
\end{equation}
Hence, the $n$-th term of the sum in $W_{c,1}(s)$ behaves, for large $n$,
as $T_n\to (1-\theta(c))\, e^{-sn}/n$. Consequently, the sum $W_{c,1}(s)=\sum_{n \geq 1} T_n$
has a singularity at $s=s^*=0$. The leading asymptotic behavior near this 
singularity reads
\begin{equation}
W_{c,1}(s)\xrightarrow[s\to 0]{} -(1-\theta(c))\, \ln (s) - \gamma_0 \;,
\label{cl.singular}
\end{equation}
where $\gamma_0$ is a constant that depends on the details of $P_n(x)$, in 
particular on the difference between $P_n(x)$ and its large $n$ asymptotic form
$(1/n) {\cal L}_1(x/n)$ for finite $n$
\begin{equation}
\gamma_0= \sum_{n=1}^{\infty}\left[1-\theta(c)-\int_{cn}^{\infty} P_n(x)\, 
dx\right]\, .
\label{gamma0}
\end{equation}
Using this result on the right hand side (rhs) of Eq. (\ref{SA.3}) gives
\begin{equation}
{\tilde q}(s)= \xrightarrow[s\to 0]{}
\frac{e^{-\gamma_0}}{s^{1-\theta(c)}}\,.
\label{qs.cl}
\end{equation}
Substituting this result in the Bromwich contour in Eq. (\ref{qn.brom})
(after setting $s^*=0$) and performing the Bromwich integral gives
\begin{equation}
Q(n)\xrightarrow[n\to \infty]{} \frac{B_{\rm II}}{n^{\theta(c)}} \;,
\label{cl.qn}
\end{equation}
where 
\begin{equation}
B_{\rm II}=\frac{e^{-\gamma_0}}{\Gamma[1-\theta(c)]}\, 
\quad {\rm and} \quad \theta(c)= \frac{1}{2}+\frac{1}{\pi}\,\arctan(c) \;,
\label{cl.thetac}
\end{equation}
and $\gamma_0$ in given in Eq. (\ref{gamma0}).

Thus, for $\mu=1$, the persistence $Q(n)$ decays algebraically for large $n$
but with an exponent $\theta(c)$ that depends continuously on the drift $c$.
In this sense the line $\mu=1$ is {\em marginally}
critical. The exponent $\theta(c)$ in Eq. (\ref{cl.thetac}) increases 
continuously with $c$ from
$\theta(c\to -\infty)=0$ to $\theta(c\to \infty)=1$. 

Let us end this subsection with the following remark on
the special case of
pure Cauchy jump distribution, $f_{\rm Cauchy}(\eta)= 1/[\pi (1+\eta^2)]$.
As mentioned before, the record statistics for this case was studied in detail in 
Ref. \cite{PLDW2009}. For the Cauchy case, it is well known that
$P_n(x)=(1/n) f_{\rm Cauchy} (x/n)= (1/n) {\cal L}_1(x/n)$ for {\em all} $n$. 
As a result, it follows from Eq. (\ref{gamma0}) that the constant $\gamma_0=0$ in 
this case. However, in the general
$\mu=1$ case (not necessarily the Cauchy case), 
$\gamma_0$ is generically nonzero. 
Thus, while the persistence exponent $\theta(c)=1/2+\frac{1}{\pi}\,\arctan(c)$
is universal for all jump densities belonging to the $\mu=1$ case, the amplitude
$B_{\rm II}$ is {\em nonuniversal} and depends on the details of the jump
density.

\subsection{Regime III: $1<\mu<2$ and $c>0$}

To analyse the singular behavior of the sum $W_{c, \mu}(s)$ in Eq. (\ref{SA.3}) in 
this regime, we consider the $n$-th term of the sum $T_n= 
(e^{-sn}/n)\int_{cn}^{\infty} P_n(x) dx$ and substitute,
for large $n$, the scaling behavior
of $P_n(x)$ in Eq. (\ref{scaling.1}). This gives
$T_n\approx (e^{-sn}/n) \int_{c n^{(1-1/\mu)}}^{\infty} {\cal L}_\mu(y) dy$.
For $1<\mu<2$, the lower limit of the integral in $T_n$ becomes large
as $n\to \infty$ and we can use the tail behavior in Eq. (\ref{lmu_asymp})
to estimate, $T_n \approx (A_\mu/\mu c^{\mu}) e^{-sn}/n^\mu$ for large $n$.
Hence the sum, $W_{c, \mu}(s)= \sum _{n=1}^\infty T_n $ clearly converges to a 
constant $W_{c, \mu}(0)$ as $s\to 0$. For small $s$, one can replace the sum by an
integral and estimate exactly the first singular correction to this constant. This 
gives
\begin{equation}
W_{c, \mu}(s) \xrightarrow[s\to 0]{} W_{c, \mu}(0) - B_{\mu}\, s^{\mu-1} \;,
\label{reg3.singular}
\end{equation}
where the constant $B_\mu= A_{\mu} \Gamma(2-\mu)/[\mu(\mu-1) c^\mu]$.
Using the exact expression of $A_{\mu}$ from Eq. (\ref{lmu_asymp}) and simplify,
one finds $B_\mu= -1/[2\cos(\mu\pi/2)]>0$ as in Eq. (\ref{bmu}).
Note also that from the definition in (\ref{SA.3})
\begin{equation}
{\tilde q}(0)= \exp[W_{c, \mu}(0)]=\exp\left[\sum_{n=1}^{\infty} 
\frac{1}{n}\int_{cn}^{\infty} P_n(x)\, dx\right]\, .
\label{qs0.3}
\end{equation} 
Substituting the small $s$ behavior from Eq. (\ref{reg3.singular}) in Eq. 
(\ref{SA.3}) gives
\begin{equation}
{\tilde q}(s)\xrightarrow[s\to 0]{}
{\tilde q}(0)\left[1 - B_{\mu}\, s^{\mu-1}+\dots\right]\, .
\label{qs.3}
\end{equation}
Substituting this singular behavior of ${\tilde q}(s)=\exp[W_{c, \mu}(s)]$
in the Bromwich integral in Eq.~(\ref{qn.brom}) (upon setting $s^*=0$) and performing
the integral by standard method provides the following large $n$ power law tail
for $Q(n)$
\begin{equation}
Q(n)\xrightarrow[n\to \infty]{} \frac{B_{\rm III}}{n^{\mu}} \;,
\label{reg3.qn}
\end{equation}
where the prefactor $B_{\rm III}$ is given by
\begin{equation}
B_{\rm III}= \frac{(\mu-1)B_\mu}{\Gamma(2-\mu) c^\mu}\, {\tilde 
q}(0)=-\frac{(\mu-1)}{2\cos(\mu\pi/2)\, \Gamma(2-\mu) c^\mu}\, 
\exp\left[\sum_{n=1}^{\infty}
\frac{1}{n}\int_{cn}^{\infty} P_n(x)\, dx\right]\, .
\label{BIII}
\end{equation}
  
\subsection{Regime IV: $\mu=2$ and $c>0$}

In this regime, the leading singularity $s^*$ of $W_{c, \mu}(s)$ occurs not at 
$s=0$, but
at $s=s^*=-s_1$ where $s_1=c^2/2\sigma^2$. To see this, let us again consider the 
$n$-th term
of the sum $W_{c, \mu}(s)$, i.e. $T_n= (e^{-sn}/n)\int_{cn}^{\infty} P_n(x) dx$.
For large $n$, $P_n(x)$ now has the Gaussian scaling form in Eq. (\ref{clt.1})
due to the central limit theorem. Substituting this Gaussian form and carrying out
the integration one gets, 
\begin{equation}
T_n\to \frac{e^{-sn}}{2n}\,{\rm 
erfc}\left(\frac{c}{\sigma\sqrt{2}}\,\sqrt{n}\right), \quad {\rm where}\quad 
{\rm erfc}(y)= \frac{2}{\sqrt{\pi}}\, \int_y^{\infty} e^{-x^2}\,dx\, .
\label{reg4.tn1}
\end{equation}
Using the asymptotic behavior ${\rm erfc}(y) \approx e^{-y^2}/{y\sqrt{\pi}}$ for 
large $y$, one finds that
\begin{equation}
T_n\xrightarrow[n\to \infty]{} 
\frac{\sigma}{c\sqrt{2\pi}}\,\frac{e^{-(s+s_1)n}}{n^{3/2}}, \quad {\rm where}\quad
s_1= \frac{c^2}{2\sigma^2}\, .
\label{reg4.tn2}
\end{equation} 
Consequently, the sum $W_{c, \mu}(s)=\sum_{n\geq 1} T_n$ actually, while perfectly analytic 
near $s=0$, has a singularity near $s=s^*=-s_1$. Close to this singular 
value, by taking the limit $s+c^2/2\sigma^2\to 0$ whereby replacing the
sum by an integral over $n$, one finds the following leading singular behavior of
$W_{c, \mu}(s)$ near $s=-s_1$
\begin{equation}
W_{c, \mu}(s)\xrightarrow[s\to -s_1]{} W_{c, \mu}(-s_1) 
-\sqrt{2}\,\frac{\sigma}{c}\,\sqrt{s+s_1} \;,
\label{reg4.singular}
\end{equation}
where $W_{c, \mu}(-s_1)$ is just a constant.  
Substituting this leading singular behavior on the rhs of Eq. (\ref{SA.3}) gives
\begin{equation}
{\tilde q}(s) \xrightarrow[s\to 0]{}
e^{W_{c, \mu}(-s_1)}\left[1- 
\sqrt{2}\,\frac{\sigma}{c}\,\sqrt{s+s_1}+\dots\right]\,.
\label{qs.4}
\end{equation}
We set $s^*=-s_1$ in the Bromwich contour in Eq. (\ref{qn.brom}), 
substitute the singular behavior of ${\tilde q}(s)$
in Eq. (\ref{qs.4}) and perform the Bromwich integral to get
\begin{equation}
Q(n)\xrightarrow[n\to \infty]{}
\frac{B_{\rm IV}}{n^{3/2}}\, e^{-s_1 n}\, \quad {\rm where}\quad 
s_1=\frac{c^2}{2\sigma^2}
\label{reg4.qn}
\end{equation}
and the prefactor
\begin{equation}
B_{\rm IV}= \frac{\sigma e^{W_{c, \mu}(-s_1)}}{c\sqrt{2\pi}}= 
\frac{\sigma}{c\sqrt{2\pi}}\, \exp\left[\frac{e^{s_1 n}}{n}\int_{cn}^{\infty} 
P_n(x)\, dx\right]\,.
\label{BIV}
\end{equation}
Thus, contrary to regimes I, II and III, here the persistence $Q(n)$ has a leading 
exponential tail (modulated by a power law $n^{-3/2}$).

\subsection{Regime V: $1<\mu\le 2$ and $c<0$}

In this regime $c=-|c|<0$ and $\mu>1$. It is convenient,
using the normalization condition $\int_{-\infty}^{\infty} P_n(x)dx=1$,
to first reexpress the sum
$W_{c, \mu}(s)$ in Eq. (\ref{SA.3}) as
\begin{equation}
W_{c, \mu}(s)= \sum_{n=1}^{\infty}\frac{e^{-sn}}{n}\, \int_{-|c|n}^{\infty} 
P_n(x)dx= \sum_{n=1}^{\infty}\frac{e^{-sn}}{n}\,\left[1-\int_{|c|n}^{\infty} P_n(x) \, dx \right]\,.
\label{reg5.1}
\end{equation}
Performing the sum, and using the definition of $W_{c, \mu}(s)$ in Eq. 
(\ref{SA.3}) one gets
\begin{equation}
W_{c, \mu}(s) = -\ln\left(1-e^{-s}\right)-W_{|c|,\mu}(s)\,.
\label{reg5.2}
\end{equation}
For $\mu>1$, $W_{|c|,\mu}(0)$ is a constant as was demonstrated in the
previous two subsections. Hence, one gets from Eq. (\ref{reg5.2}), the
leading singular behavior for small $s$ 
\begin{equation}
W_{c, \mu}(s)\xrightarrow[s\to 0]{} -\ln (s) - W_{|c|,\mu}(0)
\label{reg5.3}
\end{equation}
which yields, via Eq. (\ref{SA.3})
\begin{equation}
{\tilde q}(s)\xrightarrow[s\to 0]{} 
\frac{\exp[-W_{|c|,\mu}(0)]}{s}\,.
\label{qs.5}
\end{equation}
Thus, in this regime, the leading singularity of ${\tilde q}(s)$ occurs
at $s=s^*=0$. Setting $s^*=0$ and the result (\ref{qs.5}) in the Bromwich
integral in Eq. (\ref{qn.brom}) gives
\begin{equation}
Q(n)\xrightarrow[n\to\infty]{} \alpha_\mu(c)= 
\exp[-W_{|c|,\mu}(0)]=\exp\left[-\sum_{n=1}^{\infty} 
\frac{1}{n}\int_{|c|n}^{\infty} 
P_n(x)\,dx\right]\,.
\label{reg5.qn}
\end{equation}
The fact that the persistence $Q(n)$ approaches to a constant for large $n$ in 
this regime can be understood physically because for $c<0$ and $\mu>1$, a finite 
fraction of trajectories escape to $-\infty$ as $n\to \infty$.

\section{Asymptotic Record Number distribution $P(R,n)$ for large $n$}
\label{Asymm_distr.section}

In this section, we analyse the asymptotic large $n$ properties of 
the mean record number $\langle R_n\rangle $ and its full distribution
$P(R,n)$
for arbitrary $c$ by analysing
the set of equations (\ref{pdf.x}), (\ref{renewal.genf}), (\ref{SA.2}) and (\ref{mom1.genf}) 
with arbitrary jump distribution $f(\eta)$. 
Consider first the mean record number. As in Section IV, 
we invert Eq. (\ref{mom1.genf})
by using the Cauchy inversion formula,
deform the contour (as in Fig. \ref{contour.fig}), keep
only the vertical part of the contour $C_1$ for large $n$ and
finally make the substitution $z=e^{-s}$ to obtain the
following Bromwich formula
\begin{equation}
\langle R_n \rangle \approx \int_{s^*-i\infty}^{s^*+i\infty} \frac{ds}{2\pi i}\, 
e^{s\,n}\, \frac{1}{(1-e^{-s})^2 {\tilde q}(s)} \;,
\label{mean.cauchy}
\end{equation}
where ${\tilde q}(s)$ is given in Eqs. (\ref{SA.3bis}) and (\ref{SA.3}) and its small $s$ properties   
have already been analysed in section IV in different regimes in the 
$(c,0<\mu\le 2)$  
strip.
As in section IV, $s^*$ denotes the singularity of ${\tilde q}(s)$ on the
real line in the complex plane that is closest to the origin at $s=0$.

Similarly, the record number 
distribution is obtained by inverting Eq. (\ref{renewal.genf}) in the same way
\begin{equation}
P(R,n) \approx \int_{s^*-i\infty}^{s^*+i\infty} \frac{ds}{2\pi i}\, e^{s\,n}\, 
{\tilde q}(s)\,\left[1-(1-e^{-s}){\tilde q}(s)\right]^{R-1}\, .
\label{SA.dist}
\end{equation}

In this section, we use the already derived results for ${\tilde q}(s)$ in Section 
IV and analyse the asymptotic behavior of $\langle R_n\rangle$ and $P(R,n)$ 
respectively in 
Eqs. 
(\ref{mean.cauchy}) and (\ref{SA.dist}) in different regimes of the $(c, 0<\mu \leq 2)$ strip and on the critical line $\mu=1$.    

\subsection{Regime I: $0<\mu<1$ and $c$ arbitrary}

\begin{figure}
\includegraphics[angle=-90,width=0.5\textwidth]{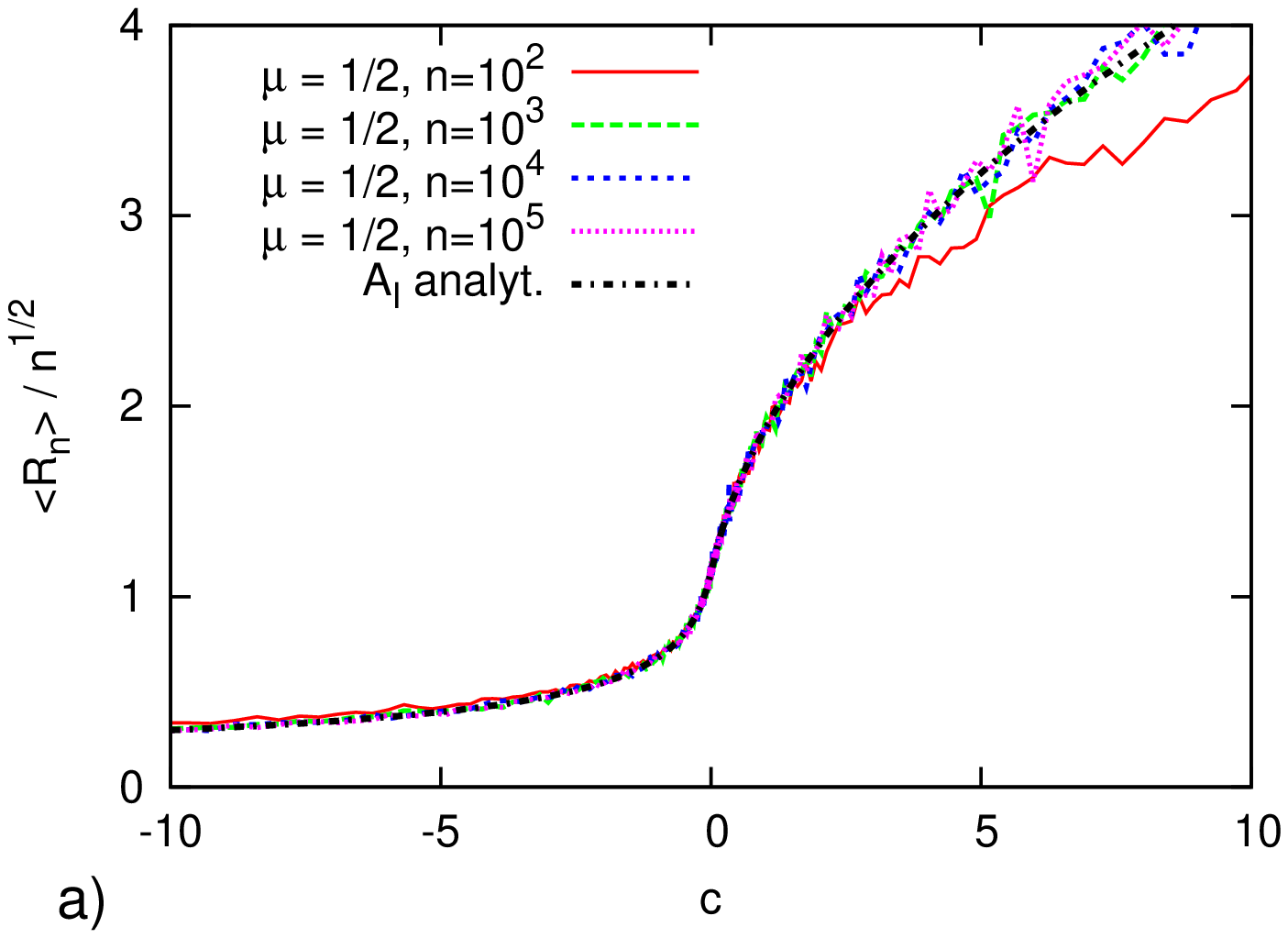}\includegraphics[angle=-90,width=0.5\textwidth]{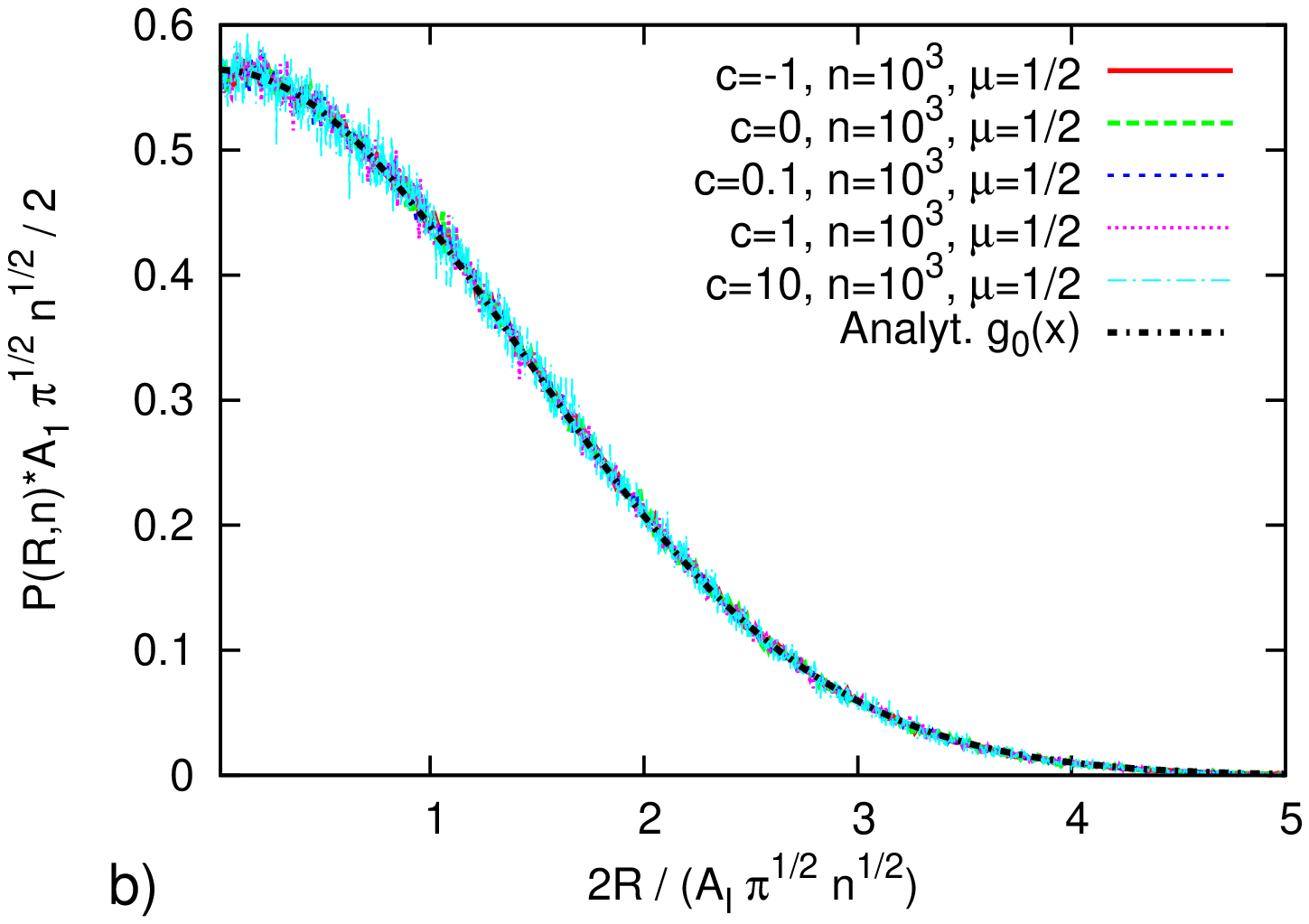}
\caption{\textbf{a):} Rescaled mean record number $\langle R_n\rangle / \sqrt{n}$ for a L\'evy-stable distribution with L\'evy index $\mu=1/2$ and different series length $n=10^2,10^3,10^4$ and $10^5$. For each $n$ the average was performed over $10^3$ samples. For $n\gg1$ the results collapse and agree with the predicted analytical behavior for $A_{\rm I}\left(c\right)$ in Eq. (\ref{meanrec_2.reg1}). \textbf{b): } Rescaled distribution $A_{\rm I} \sqrt{\pi n}\,P(R,n)/2$ as a function of $2R/A_{\rm I}\sqrt{\pi n}$ of the record number $R_n$ after $n$ steps for a random walk with a L\'evy-stable jump distribution of L\'evy index $\mu=1/2$, $n=10^3$ and different values of the drift $c=-1,0,0.1,1$ and $10$. We also plotted the asymptotic analytical result $g_0(x)$ given in Eq. (\ref{dist.I}). All curves collapse nicely. In regime I, the record number has a half-gaussian distribution. }
\label{I_Ac_and_PRn.fig}
\end{figure}

Let us first consider the asymptotic behavior of the mean number of records $\langle 
R_n\rangle$ for large $n$ in this regime. Consider the Bromwich integral
in Eq. (\ref{mean.cauchy}). For large $n$, this integral can be shown to
be dominated by the small $s$ region of the integrand. Taking $s\to 0$
limit in the integrand, 
substituting
the result (\ref{qs.1}) on the rhs of Eq. 
(\ref{mean.cauchy}), and performing the Bromwich integral we get 
the leading asymptotic behavior for large $n$
\begin{equation}
\langle R_n \rangle \approx A_{\rm I}\, \sqrt{n} \;,\quad {\rm where}\quad A_{\rm I}= 
\frac{2}{\sqrt{\pi}}\, e^{S_0}= \frac{2}{\sqrt{\pi}}\,\exp\left[\sum_{n=1}^\infty 
\frac{1}{n} \int_0^{cn} P_n(x)\, dx\right].
\label{meanrec_2.reg1}
\end{equation}
Comparing this to the amplitude of persistence in Eq. (\ref{BI}) we see
that the two prefactors are related simply via
$B_{\rm I}= 2/(\pi A_{\rm I})$.
The prefactor $A_{\rm I}$ can further be expressed explicitly in terms of the Fourier 
transform of
the jump distribution ${\hat f}(k)$ as in Eq. (\ref{ac.1}). 
This is shown in Appendix A where we also compute the asymptotic behavior
of $A_{\rm I}$ for large $|c|$ [see Eq. (\ref{AI.largec.1})]. In Fig. \ref{I_Ac_and_PRn.fig} a) we 
compare 
this result for $\langle R_n\rangle$ to numerical simulations. The numerical 
results for $n\gg1$, $\langle R_n\rangle / \sqrt{n}$ agree nicely with our analytical 
values for~$A_{\rm I}\left(c\right)$.

Next we turn to $P(R,n)$ in
the limit of large $n$. To extract the scaling behavior of $P(R,n)$
from Eq. (\ref{SA.dist}), we substitute on the rhs the small $s$ behavior of 
${\tilde q}(s)$ from Eq. (\ref{qs.1}) and use
the notation $e^{-S_0}= (2/\sqrt{\pi})\,A_{\rm I}$. The appropriate scaling limit
is clearly $R\to \infty$, $s\to 0$ but keeping the product $\sqrt{s}\, R$
fixed. Taking this limit in Eq. (\ref{SA.dist}) gives,
\begin{equation}
P(R,n)\approx \int_{-i\infty}^{+i\infty} \frac{ds}{2\pi i}\, e^{s\,n}\,
\frac{2}{A_{\rm I}\sqrt{\pi\, s}}\,\exp\left[-\frac{2}{A_{\rm I}\sqrt{\pi}}\, 
\sqrt{s}\,R\right]\, .
\label{dist_1.reg1}
\end{equation}
One can simply evaluate the Bromwich integral by using the identity,
$LT_{s\to n}^{-1}[e^{-b R\sqrt{s}}/\sqrt{s}]= e^{-b^2 R^2/4n}/\sqrt{\pi 
n}$. This leads to the asymptotic result announced in Eq. (\ref{dist.I}) in the   
scaling limit $n\to \infty$, $R\to \infty$ with the ratio $R/\sqrt{n}$ fixed. 
In Fig. \ref{I_Ac_and_PRn.fig} b) we computed numerically the rescaled 
distribution $A_{\rm I} \sqrt{\pi n}\,P(R,n)/2$ as a function of $2R/A_{\rm I}\sqrt{\pi n}$
and compared it with $g_0(x)$ Eq. (\ref{dist.I}). The figure confirms that in regime I, the record number has a half-Gaussian 
distribution with a width that depends non-trivially on the drift $c$ and 
the L\'evy-index $\mu$. 

In summary, for $0<\mu<1$, the drift is not strong enough to change the
$\sqrt{n}$ growth of the mean record number. The presence of drift just modifies
the prefactor of the $\sqrt{n}$ growth. Similarly, the distribution of the record
number in Eq. (\ref{dist.I}) in presence of a drift, when appropriately scaled, 
remains unchanged from the universal half-Gaussian form in the driftless case.

\subsection{ Regime II: $\mu=1$ and $c$ arbitrary}

\begin{figure}
\includegraphics[angle=-90,width=0.8\textwidth]{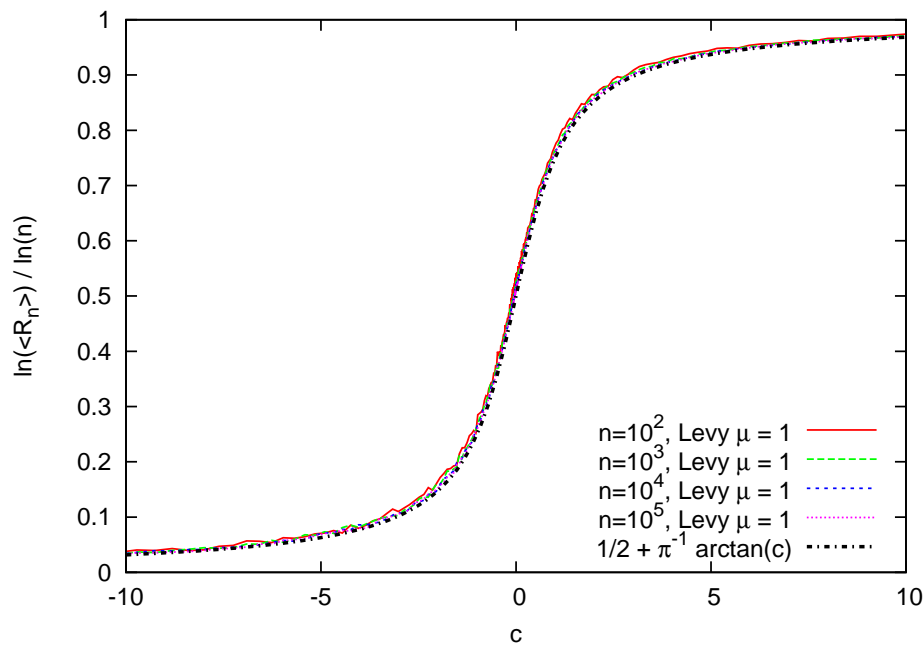}
\caption{$\ln \langle R_n \rangle / \ln n$ as a function of the drift $c$ for the Cauchy distribution with L\'evy index $\mu=1$ and for different values of $n=10^2,10^3,10^4$ and $10^5$. For each $n$ and $c$, the average was performed over $10^3$ samples. The results from the numerical simulations collapse and agree with the predicted analytical behavior of $\ln \langle R_n \rangle /\ln n = \theta\left(c\right)$ and $\theta\left(c\right) = \frac{1}{2} + \frac{1}{\pi} \textrm{arctan}\left(c\right)$ as in Eq. (\ref{cl.thetac}).}
\label{II_Ac.fig}
\end{figure}

As mentioned in the introduction, on the critical line $\mu=1$, the
record statistics was investigated in detail in Ref.~\cite{PLDW2009}
for the special case of Cauchy jump distribution $f_{\rm Cauchy}(\eta)=1/[\pi 
(1+\eta^2)]$. For a general jump distribution with $\mu=1$ (not necessarily of the
Cauchy form), the record statistics has a very similar mathematical
structure that can be derived from the general framework
developed in this paper. 

Let us first consider the growth of the mean record number $\langle R_n\rangle$
in Eq. (\ref{mean.cauchy}). Substituting the small $s$ behavior of ${\tilde q}(s)$
from Eq. (\ref{qs.cl}) and performing the Bromwich integral upon setting $s^*=0$
we get for large $n$
\begin{equation}
\langle R_n\rangle \approx \frac{A_{\rm II}}{\Gamma(1+\theta(c))}\, 
n^{\theta(c)}\quad {\rm where}\quad A_{\rm II}= e^{\gamma_0}\, . 
\label{meanrec1.cl}
\end{equation}
Note that $\gamma_0$ is a distribution dependent constant while the exponent 
$\theta(c)= 
\int_{-\infty}^c {\cal L}_1(y) dy=1/2+ \frac{1}{\pi}\,\arctan(c)$ is universal. In Fig. (\ref{II_Ac.fig}) this exponent is plotted and compared with numerical simulations of random walks with a Cauchy jump distribution ($\mu=1$).

Turning now to the distribution $P(R,n)$ in Eq. (\ref{SA.dist}), as before,
we substitute the small $s$ expansion of ${\tilde q}(s)$ from Eq. (\ref{qs.cl}).
It turns out that the appropriate scaling limit for $P(R,n)$ is
$n\to \infty$, $R\to \infty$ but keeping the ratio $R/n^{\theta(c)}$ fixed.
To see this, we first set
$s^*=0$, set $R$ large but fixed, and keep the leading terms for 
small $s$ to get
\begin{equation}
P(R,n)\approx e^{-\gamma_0}\, \int_{-i\infty}^{+i\infty} \frac{ds}{2\pi i}\, 
e^{s\,n}\, \frac{1}{s^{1-\theta(c)}}\, \exp\left[-e^{-\gamma_0}\, s^{\theta(c)}\, 
R\right]\, .
\label{prn1.cl}
\end{equation}
Rescaling $s\,n\to s$ and keeping the scaled variable $R/n^{\theta(c)}$ fixed
gives the asymptotic scaling distribution 
\begin{equation}
P(R,n)\approx \frac{1}{A_{\rm II}\,n^{\theta(c)}}\, 
g_c\left(\frac{R}{A_{\rm II}\,n^{\theta(c)}}\right) \;,
\label{prn2.cl}
\end{equation}
where the scaling function $g_c(u)$, which depends continuously on $c$, is given 
by the formal Bromwich integral
\begin{equation}
g_c(u) = \int_{-i\infty}^{+i\infty} \frac{ds}{2\pi i}\,s^{\theta(c)-1}\, e^{s- u 
s^{\theta(c)}}\, \quad {\rm with} \quad u\ge 0 \;,
\label{prnscaling.cl}
\end{equation}
where we recall that $0\le \theta(c) \le 1$.

One can easily extract the tail behavior of the scaling function $g_c(u)$
by analysing the integral in Eq. (\ref{prnscaling.cl}). For instance,
when $u\to 0$, $g_c(u)$ approaches a constant
\begin{equation}
g_c(0)= \int_{-i\infty}^{+i\infty} \frac{ds}{2\pi i}\, s^{\theta(c)-1}\, e^{s}=
\frac{1}{\pi}\,\Gamma[\theta(c)]\, \sin[\pi \theta(c)] = \frac{1}{\Gamma[1-\theta(c)]}\;.
\label{gc0}
\end{equation}
The integral in Eq. (\ref{gc0}) can be performed by wrapping the contour around
the branch cut on the negative real $s$ axis. 

In the opposite limit, when $u\to \infty$, the integral in Eq. 
(\ref{prnscaling.cl}) can be performed using the standard steepest descent method.
Skipping details and using the  shorthand notation $\theta=\theta(c)$ we get
\begin{equation}
g_c(u\to \infty) \approx \left[2\pi 
(1-\theta)\,\theta^{(1-2\theta)/(1-\theta)}\right]^{-1/2}\, 
u^{-(1-2\theta)/{2(1-\theta)}}\, \exp\left[-(1-\theta)\, 
\theta^{\theta/(1-\theta)}\, u^{1/(1-\theta)}\right]\,.
\label{gcrightail}
\end{equation}
\begin{figure}
\includegraphics[angle=-90,width=\textwidth]{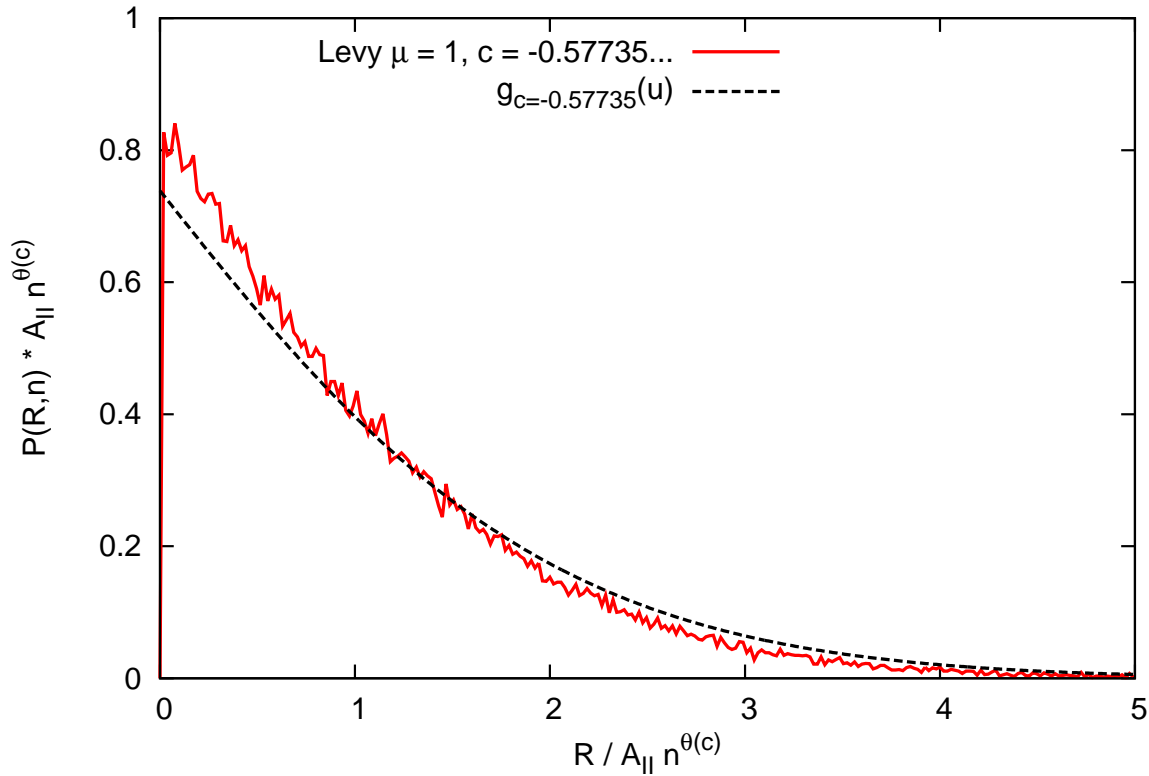}
\caption{Rescaled plot of $A_{\rm II} \, n^{\theta(c)} \, P(R,n)$ as a function of $R/A_{\rm II} n^{\theta(c)}$ for $\mu=1$ and $c= - 1/\sqrt{3}=-0.57735É$, and $\theta(c)=1/3$ (regime II). These data have been obtained for $n=10^5$ and averaged over $10^5$ samples. The dotted line corresponds to our exact result in Eq. (\ref{exact.airy}).}\label{fig:comp.airy}
\end{figure}
Thus the distribution has a non-Gaussian tail. The function $g_c(u)$ can be expressed in terms of the one-sided L\'evy distribution, which was discussed for instance in Ref. \cite{schehr_pld_rsrg}. In some particular cases, the Bromwich integral in Eq. (\ref{prnscaling.cl}) can be evaluated explicitly. For rational values of $\theta(c)$, $g_c(u)$ can be expressed as a finite sum of hypergeometric functions. A very special case corresponds to $c= -1/\sqrt{3}$ where one has $\theta = 1/3$, such that 
\begin{eqnarray}\label{exact.airy}
g_{c=-1/\sqrt{3}}(u) = 3^{2/3} {\rm Ai}\left(\frac{u}{3^{1/3}} \right) \;, \; u \geq 0 \;.
\end{eqnarray}
where ${\rm Ai}(x)$ is the Airy function. Its asymptotic behaviors are then given by
\begin{eqnarray}
&&g_{c = -1/\sqrt{3}}(u) \sim 1/\Gamma(2/3) \;, \; u \to 0 \\
&&g_{c = -1/\sqrt{3}}(u) \sim \frac{3^{3/4}}{2 \sqrt{\pi}} u^{-1/4} \exp\left(-\frac{2}{3\sqrt{3}} u^{3/2} \right) \;,
\end{eqnarray}
which agree with the general analysis presented above (\ref{gc0}, \ref{gcrightail}). In Fig. \ref{fig:comp.airy} we show a plot of the rescaled probability $A_{\rm II} \, n^{\theta(c)} \, P(R,n)$ as a function of $R/A_{\rm II} n^{\theta(c)}$ computed numerically for $c=-1/\sqrt{3}$, which agrees reasonably well with our exact analytical result in Eq.~(\ref{exact.airy}).  

\subsection{Regime III: $1<\mu<2$ and $c>0$}

\begin{figure}
\includegraphics[angle=-90,width=1\textwidth]{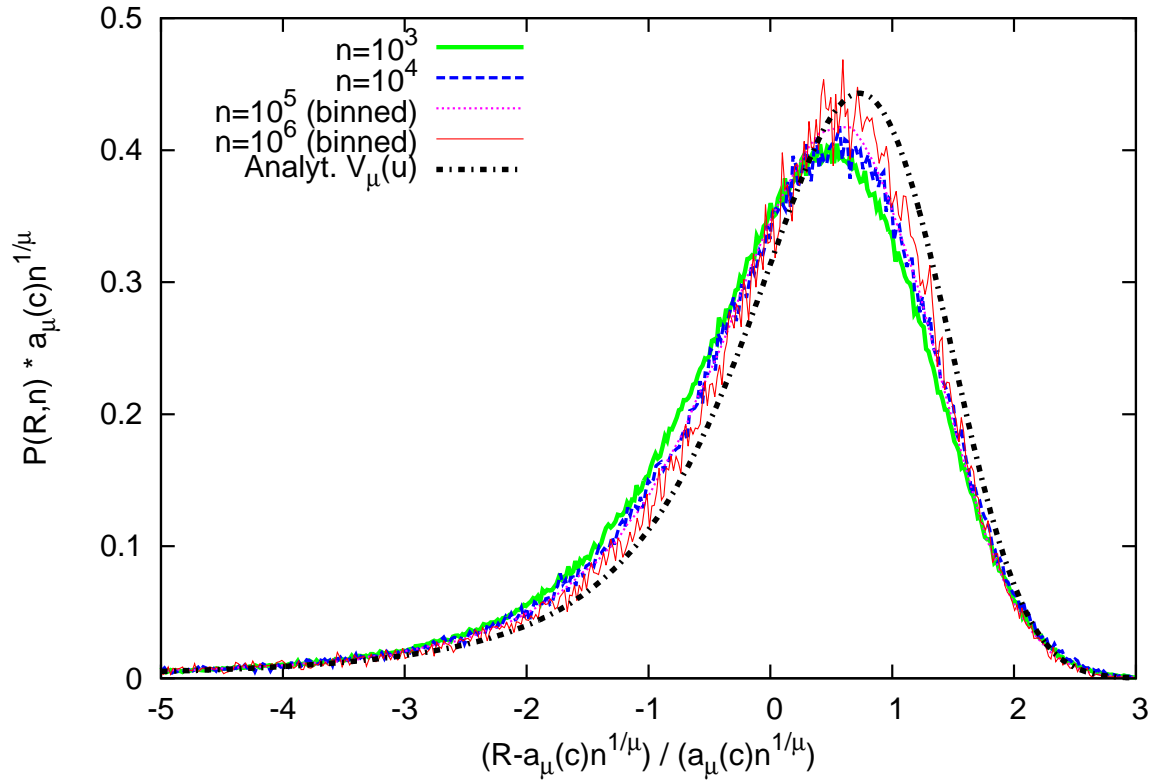}
\caption{Rescaled distribution $a_\mu(c)\, n^{1/\mu} P(R,n)$ of the record number $R_n$ after $n$ steps for a random walk with a L\'evy-stable jump distribution of L\'evy index $\mu=1.5$. The data are plotted as a function of the shifted and scaled variable $u= ({R-a_\mu(c)\,n})/({a_\mu(c)\,n^{1/\mu}})$. The different curves correspond to different values of $n=10^3,10^4,10^5$ and $10^6$ and for a drift $c=1$. They were obtained by averaging over $10^6$ samples. For $n=10^5$ and $n=10^6$ the numerical results were binned for technical reasons. We also plotted our analytical results for the scaling function $V_{\mu}\left(u\right)$ given by Eq. (\ref{vmu_plot}). While for smaller values of $n$, there is still a significant difference between the simulations and our analytical result, it converges to the behavior in Eq.~(\ref{vmu_plot}) when $n$ increases.}
\label{III_rescaled_distr_levy1p5.fig}
\end{figure}

We first compute the asymptotic growth of the mean number of records in
this regime. Substituting the leading singular behavior of ${\tilde 
q}(s)$ from Eq. (\ref{qs.3}) on the rhs of Eq.
(\ref{mean.cauchy}) and performing the Bromwich integral gives
\begin{equation}
\langle R_n\rangle \approx a_\mu(c)\, n \quad {\rm where}\quad 
a_\mu(c)=\frac{1}{{\tilde q}(0)}= \exp\left[-\sum_{n=1}^{\infty}
\frac{1}{n}\int_{cn}^{\infty} P_n(x)\, dx\right]\,.
\label{amuc}
\end{equation}
Note that we used above the expression of ${\tilde q}(0)$ in Eq. (\ref{qs0.3}). We have checked numerically this linear growth and in 
Fig. \ref{IV_A_c.fig} the bottom curve shows a plot of $\langle R_n\rangle / n$ as a function of $c$, although we have not tried to evaluate $a_{\mu}(c)$ numerically.

We next consider the distribution $P(R,n)$ in Eq. (\ref{SA.dist}).
We substitute the small $s$ behavior of ${\tilde q}(s)$ from
Eq. (\ref{qs.3}) on the rhs of Eq. (\ref{SA.dist}), set $s^*=0$, $R$ large and
keep only leading small $s$ terms to get
\begin{equation}
P(R,n) \approx {\tilde q}(0)\,\int_{-i\infty}^{+i\infty} \frac{ds}{2\pi i}
\, \exp\left[-s\,({\tilde q}(0)R-n)+ B_\mu {\tilde q}(0) R s^{\mu}\right]\,. 
\label{lptinv1.reg3}
\end{equation}
Next we set
\begin{equation}
R= a_\mu(c)\, n + a_\mu(c)\, n^{1/\mu}\, u \;,
\label{rscaling1.reg3}
\end{equation}
where $a_\mu(c)=1/{\tilde q}(0)$ 
and take the limit $R\to \infty$, $n\to \infty$ but keeping the scaled
variable $u$ above fixed. We substitute Eq. (\ref{rscaling1.reg3}) on the rhs
of Eq. (\ref{lptinv1.reg3}). Keeping only the two leading terms for large $n$
and fixed $u$ gives 
\begin{equation}
P(R,n) \approx {\tilde q}(0)\,\int_{-i\infty}^{+i\infty} \frac{ds}{2\pi i}
\, \exp\left[- s n^{1/\mu} u + B_\mu\, n\, s^{\mu}\right]\,.
\label{lptinv2.reg3}
\end{equation}
Note that for fixed $u$, both terms inside the exponential are of the same 
order.
In fact, the scaling in Eq. (\ref{rscaling1.reg3}) is chosen so as to make the
two leading terms precisely of the same order for large $n$. Rescaling $s$
by $n^{1/\mu}$, i.e., $s\,n^{1/\mu}\to s$ and using $a_\mu(c)=1/{\tilde 
q}(0)$ reduces Eq. (\ref{lptinv2.reg3})
to a nicer scaling form announced in Eq. (\ref{dist.III})
\begin{equation}
P(R,n)\approx \frac{1}{a_\mu(c) n^{1/\mu}}\, V_\mu(u), \quad {\rm where} \quad 
u= \frac{R-a_\mu(c)\,n}{a_\mu(c)\,n^{1/\mu}} \;,
\label{rscaling2.reg3}
\end{equation}
and the scaling function $V_{\mu}(u)$ is formally given by the Bromwich integral
\begin{equation}
V_\mu(u)= \int_{-i\infty}^{i\infty} \frac{ds}{2\pi i}\, e^{-u\, s+ B_\mu\, 
s^{\mu}} \;,
\label{rscaling3.reg3}
\end{equation}
where the constant $B_\mu>0$ is given in Eq. (\ref{bmu}). 

\begin{figure}
\includegraphics[angle=-90,width=0.8\textwidth]{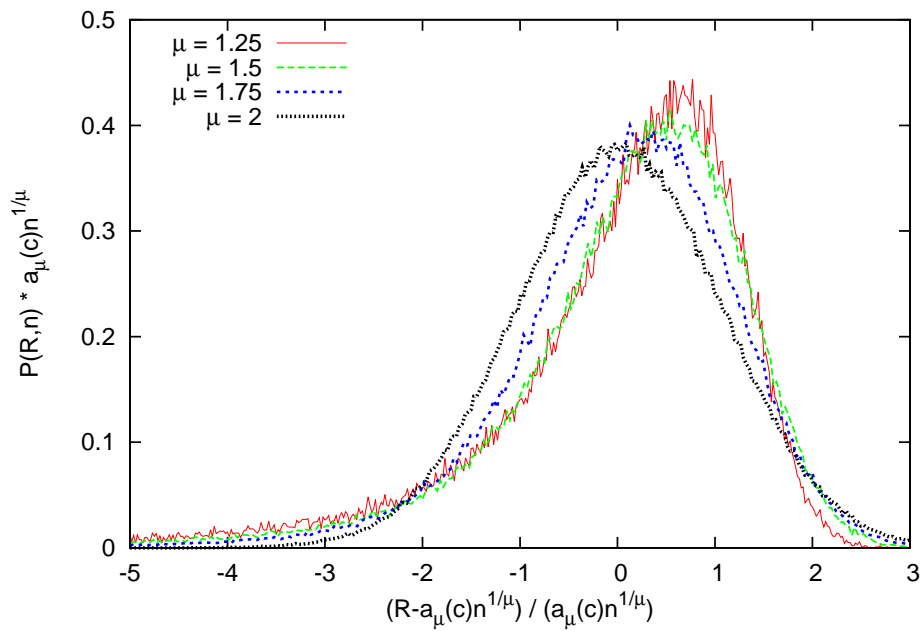}
\caption{Rescaled distribution $a_\mu(c)\, n^{1/\mu} P(R,n)$ of the record number $R_n$ after $n=10^4$ steps for a random walk with a L\'evy-stable jump distribution with different L\'evy indices $\mu=1.25,1.5,1.75$ and $\mu=2$. The data are plotted as a function of the shifted and scaled variable $u= ({R-a_\mu(c)\,n})/({a_\mu(c)\,n^{1/\mu}})$. For all these data, the value of the drift is $c=1$ and they have been obtained by averaging over $10^6$ samples. The figure shows that for $\mu\rightarrow2$ this rescaled distribution approaches the Gaussian form given in Eq. (\ref{rscaling4.reg4}). }
\label{III_rescaled_distr_var_mu.fig}
\end{figure}

Interestingly, the same scaling function $V_\mu(u)$ also appeared
in Ref.~\cite{MEZ2006} in the context of the partition function
of the zero range process on a ring. The asymptotic tails
of the function $V_\mu(u)$ were analysed in great detail in \cite{MEZ2006}
(see Eqs. (78)-(83) and Fig. 5 in Ref.~\cite{MEZ2006} and note that in 
\cite{MEZ2006}, the index 
$\mu$ was denoted by $\gamma-1$).
We do not repeat the computations here, but just quote the results. It was found 
that $V_{\mu}(u)$ has highly asymmetric tails. For $u\to -\infty$, it decays
as a power law, $V_{\mu}(u) \to K_\mu |u|^{-\mu-1}$ where the prefactor
$K_\mu= B_\mu \Gamma(1+\mu)\sin[\pi(\mu+1)]/\pi$. Using our expression
$B_\mu= -1/(2\cos(\mu\pi/2))$ from Eq. (\ref{bmu}), it is easy to show
that $K_\mu=A_\mu$ where the constant $A_\mu$ is defined in Eq. (\ref{lmu_asymp}).
This leads to Eq.~(\ref{vmuleft}). In contrast, when $u\to \infty$, $V_\mu(u)$
has a faster than Gaussian tail as described precisely in Eq. (\ref{vmuright}).
To plot this scaling function, a convenient real space representation 
can be used from Ref.~\cite{MEZ2006}. Replacing $\gamma-1$ by $\mu$
in Eq. (84) of Ref.~\cite{MEZ2006} and using $B_\mu=-1/{2\cos(\mu\pi/2)}$,
we obtain
\begin{equation}
V_{\mu}(u)= \frac{1}{\pi}\,\int_0^{\infty} dy\, e^{-y^{\mu}/2}\, \cos\left[
\frac{1}{2}\tan(\mu\pi/2)\, y^{\mu}+ y\, u\right]\,.
\label{vmu_plot}
\end{equation}
We compared this result for a L\'evy index of $\mu=1.5$ to numerical simulations 
in Fig.~\ref{III_rescaled_distr_levy1p5.fig}. 
Even though the convergence of the numerically obtained 
distributions is slow, it is clear that the asymptotic distribution 
$V_{\mu}\left(u\right)$ is approached for $n\rightarrow\infty$. 
In Fig.~\ref{III_rescaled_distr_var_mu.fig} we plotted numerical simulations of the 
rescaled record number distribution for different values of $\mu$. One finds both numerically and 
by taking the limit in Eq. (\ref{rscaling4.reg4}) that, for $\mu\rightarrow 2$, this rescaled distribution approaches a Gaussian form (see regime IV).

To summarize, in this regime the mean record number increases linearly with
increasing $n$, but the typical fluctuations around the mean are anomalously
large of ${\cal O}(n^{1/\mu})$ (superdiffusive) as described in Eq. 
(\ref{rscaling1.reg3}). In addition, the probability distribution 
of these typical fluctuations around the mean are described by a
highly non-Gaussian form described precisely in Eq. (\ref{rscaling2.reg3}).

\subsection{Regime IV: $\mu=2$ and $c>0$}

\begin{figure}
\includegraphics[angle=-90,width=0.8\textwidth]{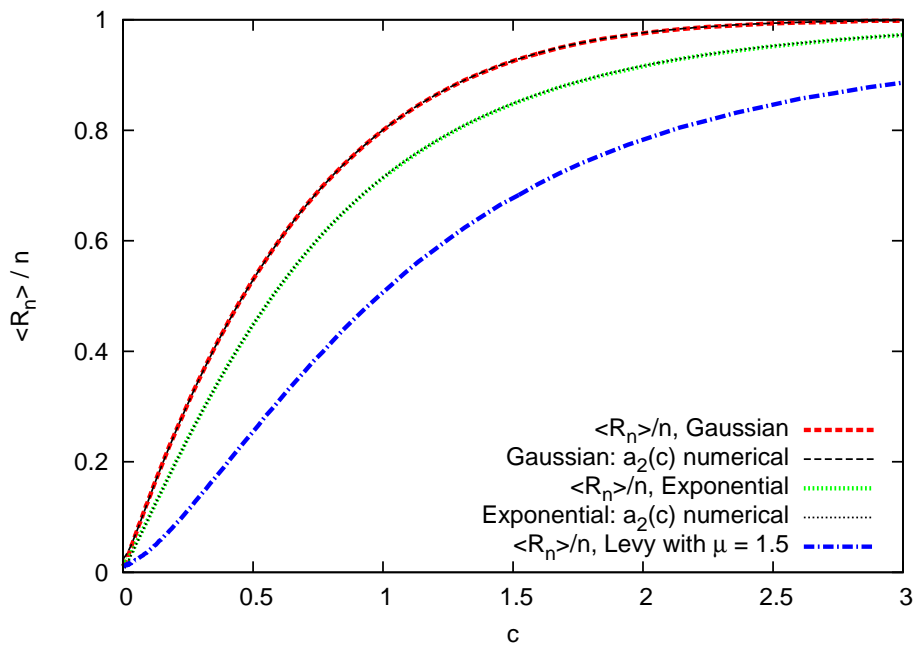}
\caption{Numerical simulations of $\langle R_n\rangle / n$ for random walks with a Gaussian (with variance $\sigma = 1$), an exponential [with parameter $b=1$, see its definition below Eq.~(\ref{a2cgaussian})], both regime IV, and a L\'evy-stable jump distribution with $\mu=1.5$, in regime III, with positive drift $c>0$. For each distribution we show data for 
$n=10^4$ which were obtained by averaging over $10^4$ samples. For the Gaussian and the exponential distribution we also plotted a numerical evaluation of our exact formula for $a_2\left(c\right)$ using Eq. (\ref{a2cgaussian}) for the Gaussian case and Eq. (\ref{a2.exp}) for the exponential case. Those curves agree perfectly with the numerical simulations. }
\label{IV_A_c.fig}
\end{figure}

In this regime, as explained in section IV.C, ${\tilde q}(s)=\exp[W_{c, \mu}(s)]$ 
in Eqs. (\ref{SA.3bis}) and (\ref{SA.3}) is 
analytic at $s=0$. This can be seen by expanding the sum $W_{c, \mu}(s)$
in Eq. (\ref{SA.3}) in a Taylor series in $s$
\begin{equation} 
W_{c, \mu}(s)=\sum_{m=0}^{\infty} d_m \, s^m, \quad {\rm where}\quad
d_m= \frac{(-1)^m}{m!}\,\sum_{n=1}^{\infty} n^{m-1}\,\int_{cn}^{\infty} 
P_n(x)\,dx\,.
\label{series.reg4}
\end{equation}
The coefficient $d_m$, for each $m$, is finite as the sum over $n$ is convergent
since the integral $\int_{cn}^{\infty} P_n(x)\,dx$ decreases with $n$ faster
than exponentially for large $n$ (see section IV.C), as long as $\mu = 2$ and $c>0$. 
Consequently, for small $s$, ${\tilde q}(s)$ also has a Taylor series
expansion
\begin{equation}
{\tilde q}(s) = {\tilde q}(0) + {\tilde q}'(0)\, s+ \frac{1}{2}{\tilde q}''(0) 
s^2+ \dots
\label{qs.taylor}
\end{equation}

Let us start with the asymptotic behavior of the mean record number $\langle 
R_n\rangle$ in Eq.~(\ref{mean.cauchy}). Once again, the dominant contribution to 
the integral in Eq. (\ref{mean.cauchy}) for large $n$ comes from the small
$s$ region. Taking the $s\to 0$ limit in the integrand and using the small $s$
expansion in Eq. (\ref{qs.taylor}), keeping only the leading terms 
and performing the Bromwich integral term by term
one gets for large $n$
\begin{equation}
\langle R_n \rangle \approx \int_{s^*-i\infty}^{s^*+i\infty} \frac{ds}{2\pi i}\,
e^{s\,n}\, \frac{1}{{\tilde q}(0) s^2}\,\left[1+ (1-\frac{{\tilde q}'(0)}{{\tilde 
q}(0)})\, s+ O(s^2)\right]\approx a_2(c) n + \kappa_2(c) + {\cal O}(1/n)
\label{meanrec1.reg4}
\end{equation}
where 
\begin{equation}
a_2(c)= \frac{1}{{\tilde q}(0)}= \exp\left[-\sum_{n=1}^{\infty} 
\frac{1}{n}\,\int_{cn}^{\infty} P_n(x)\, dx\right]
\label{a2c}
\end{equation}
and $\kappa_2(c)= \left[1- {\tilde q}'(0)/{\tilde
q}(0)\right]/{{\tilde q}(0)}$. 

For example, for a Gaussian jump distribution
$f(\eta) = (2 \pi \sigma^2)^{-1/2} e^{- \eta^2/2\sigma^2}$,
we have $P_n(x)= (2 \pi n \sigma^2)^{-1/2}\, e^{-x^2/{2\sigma^2 n}}$
and hence $a_2(c)$ in Eq. (\ref{a2c}) is given by the explicit formula
\begin{equation}
a_2(c) =
\exp\left[-\sum_{n=1}^\infty \frac{1}{2n}  
{\rm erfc}\left(\frac{c \, \sqrt{n}}{\sigma\, \sqrt{2}} \right)\right]\,. 
\label{a2cgaussian}
\end{equation} 
For instance, for $c=1$ and $\sigma=1$, one gets $a_2(c=1)= 0.800543\dots$.
Another example is the exponential jump distribution $f(\eta) = (2\,b)^{-1} \exp(-|x|/b)$.
In this case, one can also compute (see the Appendix B) the constant $a_2(c)=\lambda$ where
$\lambda$ is given by the solution of the transcendental equation
$\exp(-\lambda \,c/b)=1-\lambda^2$. For example, for $c=1$, $b=1$, one gets
$\lambda=0.714556\ldots$. For these two examples, we have confirmed
the leading asymptotic result for the mean record number in Eq. (\ref{meanrec1.reg4})
with the exactly computed prefactors $a_2(c)$ (as discussed above) in our numerical
simulations (see Fig. \ref{IV_A_c.fig}). 

In a similar way, one can also analyse Eq. (\ref{mom2.genf}) for the large 
$n$ behavior of the second moment
$\langle R_n^2\rangle$. Skipping details, we get the following leading large $n$ 
behavior
\begin{equation}
\langle R_n^2 \rangle \approx a_2^2(c)\, n^2 + \rho_2(c)\, n + {\cal O}(1) \;, \quad 
{\rm where} \quad \rho_2(c)= \frac{1}{{\tilde q}^2(0)}\left[3- {\tilde 
q}(0)-4\frac{ {\tilde q}'(0)}{{\tilde q}(0)}\right]\,.
\label{secondmom.reg4}
\end{equation}
Consequently, the variance of the record number grows for large $n$ as 
\begin{equation}
\langle R_n^2 \rangle-{\langle R_n\rangle}^2 \approx b_2(c)\, n \quad {\rm 
where}\quad b_2(c)= \frac{1}{{\tilde q}^2(0)}\left[1- {\tilde 
q}(0)-2\frac{ {\tilde q}'(0)}{{\tilde q}(0)}\right]\, .
\label{variance.reg4}
\end{equation} 
Thus, in this regime, the mean record number grows linearly with $n$ for large $n$
while the size of typical fluctuations around this mean grows as $\sim \sqrt{n}$.

\begin{figure}
\includegraphics[angle=-90,width=0.8\textwidth]{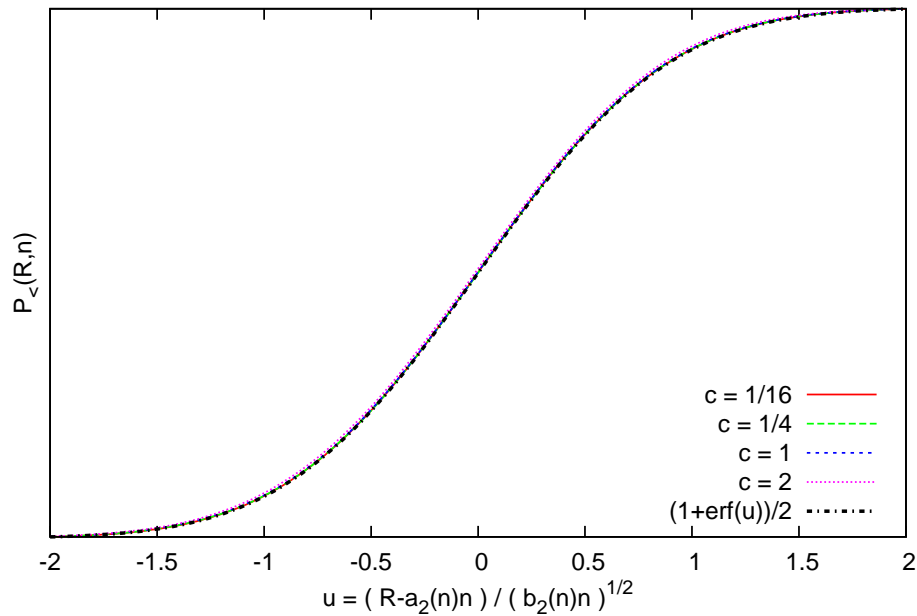}
\caption{Plot of the cumulative distribution of record numbers $P_<(R,n) ={\rm Proba}.\,[R_n \leq R]$ as a function of the shifted and scaled
variable $u =  {(R-a_2(c)\,n)}/{(\sqrt{b_2(c)\,n)}}$ for a random walk with Gaussian jump distribution (with $\sigma = 1$) of $n=10^4$ steps. The different curves correspond to different values of positive drift  $c=1/16,1/4,1$ and $2$. For each $c$ the data were obtained by averaging over $10^6$ samples. We compared the numerical results to the cumulative distribution of $V_2\left(\mu\right)$, which we obtained analytically (Eq. (\ref{rscaling4.reg4})). All curves collapse nicely, confirming that the asymptotic record number of a biased Gaussian random walk with a positive drift has the Gaussian distribution given by Eq. (\ref{rscaling2.reg4}). }
\label{IV_rescaled_cdf.fig}
\end{figure}

How are these typical fluctuations around the mean distributed? To answer this, we 
need to analyse $P(R,n)$ in Eq. (\ref{SA.dist}) in the scaling limit
where both $n$ and $R$ are large, but the ratio $(R-a_2(c)n)/\sqrt{n}$ is fixed.
To proceed, we set $s^*=0$ and substitute the small $s$ expansion of ${\tilde 
q}(s)$ in
Eq. (\ref{qs.taylor}) on the rhs of Eq. (\ref{SA.dist}), take $R$ large but fixed
to get
\begin{equation}
P(R,n) \approx {\tilde q}(0)\,\int_{-i\infty}^{+i\infty} \frac{ds}{2\pi i}
\, \exp\left[-s\,({\tilde q}(0)R-n)+ (1/2)\,b_2(c)\, {\tilde q}^3(0)\, R 
s^{2}\right]
\label{lptinv1.reg4}
\end{equation}
where $b_2(c)$ is given in Eq. (\ref{variance.reg4}). Next we set
\begin{equation}
R= a_2(c)\, n + \sqrt{b_2(c)}\, \sqrt{n}\, u \;,
\label{rscaling1.reg4}
\end{equation}
where $a_2(c)=1/{\tilde q}(0)$ is given in Eq. (\ref{a2c}) and take the scaling 
limit where $R\to \infty$, $n\to \infty$ but keeping the scaled 
variable $u$ above fixed. Substituting $R$ from Eq. (\ref{rscaling1.reg4}) into
Eq. (\ref{lptinv1.reg4}) and keeping only the two leading terms for large $n$
gives
\begin{equation}
P(R,n) \approx {\tilde q}(0)\,\int_{-i\infty}^{+i\infty} \frac{ds}{2\pi i}
\, \exp\left[- \sqrt{b_2(c)}\, {\tilde q}(0)\,\sqrt{n}\, s\, u +(1/2)\,b_2(c)\, 
{\tilde q}^2(0) n\, s^2\right]\,.
\label{lptinv2.reg4}
\end{equation}
Note that for fixed $u$, both terms inside the exponential are of the same
order.
Indeed, as in the section VB, the scaling in Eq. (\ref{rscaling1.reg4}) is chosen 
so as to make the
two leading terms precisely of the same order for large $n$.
Rescaling $ \sqrt{b_2(c)} {\tilde q}(0) \sqrt{n}\,s \to s$ simplifies to
\begin{equation}
P(R,n)\approx \frac{1}{\sqrt{b_2(c) n}}\, V_2(u) \quad {\rm where} \quad
u= \frac{R-a_2(c)\,n}{\sqrt{b_2(c)\,n}} \;,
\label{rscaling2.reg4}
\end{equation}
and the scaling function $V_{2}(u)$ is given by the Bromwich integral
\begin{equation}
V_2(u)= \int_{-i\infty}^{i\infty} \frac{ds}{2\pi i}\, e^{-u\, s+ s^2/2} \;,
\label{rscaling3.reg4}
\end{equation}
which can be exactly computed (since it is a Gaussian integral) to give
\begin{equation}
V_2(u)= \frac{1}{\sqrt{2\pi}}\, \exp[-u^2/2]\, .
\label{rscaling4.reg4}
\end{equation}
This then proves that $P(R,n)$ is asymptotically Gaussian as announced 
in Eq. (\ref{dist_IV}). Fig.~\ref{IV_rescaled_cdf.fig} confirms this result numerically. We 
plotted the cumulative distribution of record numbers $P_<(R,n) ={\rm Proba}.\,[R_n \leq R]$ as a function of the shifted and scaled
variable $u =  {(R-a_2(c)\,n)}/{(\sqrt{b_2(c)\,n)}}$ after $n=10^4$ steps for 
different values of positive drift $c$ and compared them to a Gaussian cdf (cumulative 
distribution function). All numerical results collapsed perfectly on the analytical curve.

\subsection{Regime V: $1<\mu\le 2$ and $c<0$}

\begin{figure}
\includegraphics[angle=-90,width=0.8\textwidth]{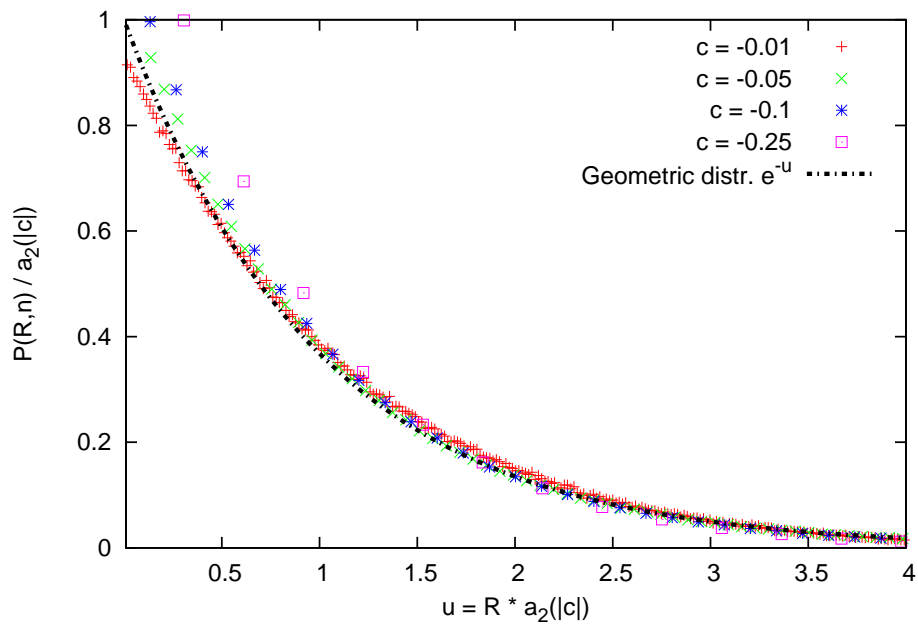}
\caption{ Rescaled distribution $a_2(|c|) P(R,n)$ of the record number $R_n$ after $n=10^4$ steps for a random walk with a Gaussian jump distribution, of variance $\sigma =1$, with different negative values of the drift $c=-0.01,c=-0.05,-0.1$ and $-0.25$. The data are plotted as a function of the rescaled variable $u=R \, a_2(|c|)$. For each value of $c$ the data were obtained by averaging over $10^4$ samples. We compared the numerical results with a simple geometric distribution. The good agreement confirms our analytical findings given by Eq. (\ref{prn2.reg5}). }
\label{V_pdf.fig}
\end{figure}

In this regime, we set $s^*=0$ in Eq. (\ref{SA.dist}) and substitute on its rhs 
the small $s$ 
expansion of ${\tilde q}(s)$
from Eq. (\ref{qs.5}). Keeping only leading order
behavior for small $s$ gives, for large~$n$,
\begin{equation}
P(R,n)\approx \alpha_\mu(c) [1-\alpha_\mu(c)]^{R-1}\, 
\int_{s^*-i\infty}^{s^*+i\infty} 
\frac{ds}{2\pi i}\, e^{s\,n}\,\frac{1}{s} \;,
\label{prn.reg5}
\end{equation}
where the constant $\alpha_\mu(c)=\exp[-W_{|c|,\mu}(0)]=\exp\left[-\sum_{n=1}^{\infty}
\frac{1}{n}\int_{|c|n}^{\infty}
P_n(x)\,dx\right]$ as 
given in Eq. (\ref{reg5.qn}).

Using the
fact that $LT^{-1}_{s\to n}[1/s]= 1$ gives the large $n$ (but $R$ fixed) behavior
of $P(R,n)$
\begin{equation}
P(R,n) \xrightarrow[n\to \infty]{} \alpha_\mu(c) \, [1-\alpha_\mu(c)]^{R-1}\,.
\label{prn2.reg5}
\end{equation}
Thus, the distribution becomes independent of $n$ for large $n$ and has a
simple geometric form with mean $\langle R_n\rangle \to 1/\alpha_\mu(c)$.
Comparing the expression of $\alpha_\mu(c)$ as given in Eq.~(\ref{reg5.qn}) and
those of $a_\mu(c)$ in Eq.~(\ref{amuc}) and $a_2(c)$ in Eq.~(\ref{a2c}) for $c>0$, 
one immediately finds 
that $\alpha_\mu(c)=a_\mu(|c|)$ for $1<\mu<2$ while $\alpha_2(c)=a_2(|c|)$,
the results mentioned respectively in Eqs. (\ref{meanconst1})
and (\ref{meanconst2}). 

In Fig.~\ref{V_pdf.fig} we compared Eq.~(\ref{prn2.reg5}) to numerical simulations of negatively biased Gaussian random walks with different values of $c$. For large $n$ the rescaled distribution of $u=R \, a_2(|c|)$ approaches the geometric (exponential) distribution $e^{-u}$.

\section{Extreme statistics of the age of a record}
\label{ages.section}

From the previous study of the mean number of records $\langle R_n \rangle$, one deduces that 
the typical age (see Fig. \ref{renewal.fig})) of a record is given by $l_{\rm typ} \sim 
n/\langle R_n\rangle$. However, following Ref. \cite{MZrecord} for the unbiased case,
it turns out that the extreme ages of records do not share the typical behavior. In this
section, we probe such atypical extremal statistics by considering the longest and shortest lasting records characterized by their
respective ages (durations)
$l_{{\rm max},n}$ and $l_{{\rm min},n}$. We 
focus on 
their mean values $\langle l_{{\rm max},n}\rangle$, $\langle l_{{\rm min},n} \rangle$ and find 
rather different asymptotic behaviors in the five regimes in the $(c, 0 <\mu\leq 2)$ strip
mentioned before (Fig. \ref{phd.fig}).

\subsection{Age of the longest lasting record $l_{{\rm max},n}$}
\label{longest_records.section}

We first consider the longest lasting record whose age $l_{{\rm max},n}$ is given by (see Fig.~\ref{renewal.fig})
\begin{eqnarray}
l_{\rm max} = \max(l_1, l_2, \cdots, l_R) \;.
\end{eqnarray} 
The cumulative distribution ${\cal F}_n(m) = {\rm Proba.}\,(l_{{\rm max},n} \leq m)$ was studied in Ref. \cite{MZrecord}, where an explicit formula for its generating function (GF) was obtained: 
\begin{eqnarray}
\sum_{n=0}^\infty {\cal F}_n(m) z^n = \frac{\sum_{l=1}^m Q(l) z^l}{1-\sum_{l=1}^m F(l) z^l} \;,
\end{eqnarray}
where $F(l) = Q(l-1)-Q(l)$, from which one deduces the generating function of the mean $\langle l_{{\rm max},n} \rangle = \sum_{m=1}^\infty [1-{\cal F}_n(m)]$
\begin{eqnarray}\label{lmax.1}
&& \sum_{n=0}^\infty z^n  \langle l_{{\rm max},n} \rangle = \sum_{m=1}^\infty \left[ \frac{1}{1-z} - \frac{\sum_{l=1}^m Q(l) z^l}{1-\sum_{l=1}^m F(l) z^l} \right]\\
&& = \frac{1}{1-z} \sum_{m=1}^\infty \frac{\sum_{l=m}^\infty F(l) z^l + (1-z) \sum_{l=m}^\infty Q(l) z^l}{(1-z) \tilde Q(z) + \sum_{l=m}^\infty F(l) z^l} \;,
\end{eqnarray}
where we have used that $\tilde F(z) = 1 - (1-z) \tilde Q(z)$ (\ref{qfrelation.2}). 

In the absence of drift, $c=0$, it was shown in Ref. \cite{MZrecord} that $\langle l_{{\rm max},n}\rangle$ behaves, for large $n$, linearly with $n$ with a non trivial coefficient, independently of the jump distribution $f(\eta)$
\begin{eqnarray}\label{pitman_yor}
\langle l_{{\rm max},n}\rangle \sim C_0 \, n \;, C_0 = \int_0^\infty dy \frac{1}{1 + y^{1/2} e^y \int_0^y dx \, x^{-1/2} e^{-x}} = 0.626508...
\end{eqnarray}
Interestingly, this constant $C_0$ appears also in the study of the longest excursion of 
Brownian motion \cite{pitman_yor,godreche_excursion}. Note that to obtain the large $n$ 
behavior of $\langle l_{{\rm max},n}\rangle$ from Eq. (\ref{lmax.1}) one has to analyse the 
above formula (\ref{lmax.1}) in the limit $z \to 1$. We will see that in this limit the above 
sum over $m$ is dominated by the large values of $m$, which thus depends crucially on the 
large $m$ behavior of the persistence probability $Q(m)$. Consequently $\langle l_{{\rm 
max},n} \rangle$ behaves quite differently in the five regimes in the $(c,0<\mu \leq 2)$ strip
in Fig.~\ref{phd.fig} and are
summarized as follows:

\begin{eqnarray}
\langle l_{{\rm max},n}\rangle &\sim & n\quad {\rm for}\quad 0<\mu<1\,\, {\rm and}\,\, c\,\,{\rm 
arbitrary}\quad ({\rm regime}\,\, {\rm I}) \;,
\nonumber \\
&\sim & n \quad {\rm for}\quad \mu=1 \,\, {\rm and}\,\,c \,\,{\rm.
arbitrary}  
\quad ({\rm regime}\,\, {\rm II}) \;,
\nonumber \\
&\sim & n^{\frac{1}{\mu}} \quad {\rm for}\quad 1<\mu<2\,\, {\rm and}\,\, c>0 
\quad ({\rm regime}\,\, {\rm III}) \;,
\nonumber \\
&\sim & \ln n \quad  {\rm for}\quad \mu=2 \,\, 
{\rm and}\,\, c>0 
\quad ({\rm regime}\,\, {\rm IV}) \;,
\nonumber \\
&\sim & n \quad {\rm for}\quad 1<\mu\le 2 \,\, {\rm and}\,\, c<0 
\quad ({\rm regime}\,\, {\rm V}) \;.
\nonumber \\
\label{l_minn_asymp}
\end{eqnarray}
In the following we will discuss the behavior of $\langle l_{{\rm max},n}\rangle$ separately for the five regimes.

\subsubsection{Regime I: $0 < \mu < 1$, $c$ arbitrary:}

In this regime, we remind that $Q(m)$ behaves, for large $m$, as
\begin{eqnarray}\label{asymptQF}
Q(m) \sim \frac{B_{\rm I}}{\sqrt{m}} \;, \; F(m) \sim \frac{B_{\rm I}}{2 m^{3/2}}\;,
\end{eqnarray}
where $B_{\rm I}$ is given in Eq. (\ref{BI}). 
\begin{figure}[ht]
\begin{center}
\includegraphics[width = 0.8\textwidth]{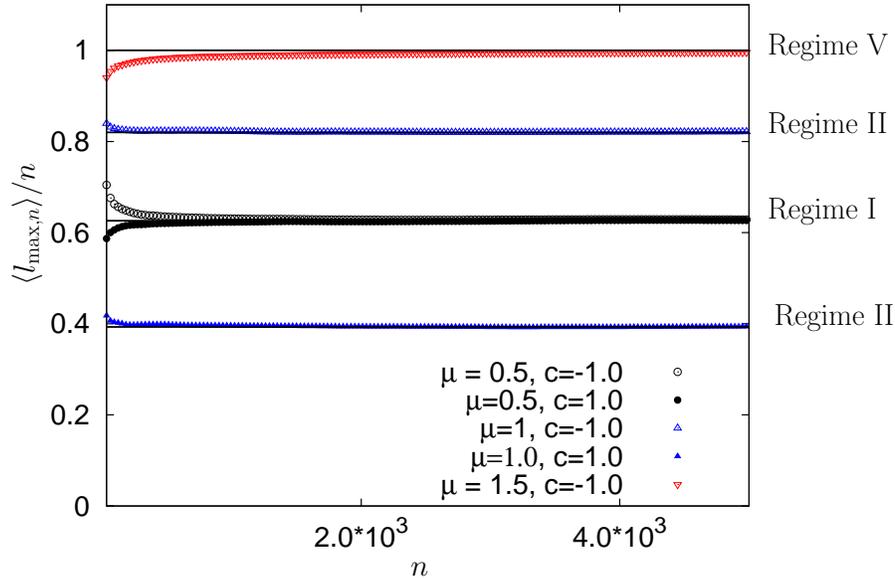}
\caption{Plot of $\langle l_{\max,n}\rangle/n$ in the different regimes I, II and V: 
the points are the results of our numerical simulations. For regime II ($\mu=1$), we present
two curves, one with a negative drift $(c=-1)$ (the second curve from top)
and one with a positive drift ($c=1$) (the bottom curve).
These data indicate that in all these cases $\langle l_{\max,n}\rangle \propto n$, for large $n$, with an amplitude which agree quite well with our analytical results, which are represented in solid line for each of these cases and corresponds to the formula given in Eq. (\ref{lmax.3}, \ref{lmax_regime2}, \ref{lmax_regime5}).}\label{fig_regime_c125}
\end{center}
\end{figure}
Setting $z = e^{-s}$ we are interested in the limit $s \to 0$ in the formula in Eq. (\ref{lmax.1}) where one can replace $F(m)$ and $Q(m)$ by their asymptotic behaviors
\begin{eqnarray}\label{lmax.2}
\sum_{n=0}^\infty \langle l_{{\rm max},n}\rangle e^{-s n} \sim \frac{1}{s} \sum_{m=1}^\infty \frac{\frac{1}{2} \sum_{l=m}^\infty l^{-3/2} e^{- s l} + s \sum_{l=m}^\infty l^{-1/2} e^{-sl}}{\sqrt{\pi}s^{1/2} + \frac{1}{2} \sum_{l=m}^\infty l^{-3/2} e^{-s l} } \;,
\end{eqnarray}
where we have used $\tilde q(s) \sim \sqrt{\pi} B_{\rm I}/ \sqrt{s}$ when $s \to 0$ (\ref{qs.1}, \ref{BI}). In the limit $s \to 0$, the discrete sums over $l$ and $m$ can be replaced by integrals and one finds that the right hand side in Eq. (\ref{lmax.2}) behaves like $1/s^2$ when $s \to 0$ with a prefactor which we can compute to obtain the large $n$ behavior of $\langle l_{{\rm max},n}\rangle$ as
\begin{eqnarray}\label{lmax.3}
\langle l_{{\rm max,n}}\rangle \sim C_{\rm I} \, n \;, \; C_{\rm I} = \int_0^\infty dy \frac{y^{-1/2} e^{-y}}{\sqrt{\pi} + \frac{1}{2} \int_y^\infty dx \, x^{-3/2} e^{-x}} = C_0 \;,
\end{eqnarray}
where $C_0$ is given above (\ref{pitman_yor}) and where the last equality is simply obtained 
by performing an integration by part in the integral over $x$ in the denominator. In Fig. \ref{fig_regime_c125}, we have plotted the results of our numerical estimate of $\langle l_{\max,n}\rangle$ (obtained by averaging over $10^4$ different realizations of random walks) for $\mu = 0.5$ and two different values of $c=\pm 1.0$. This plot shows that $\langle l_{\max,n}\rangle/n$ saturates rather quickly to the constant $C_0$, independently of $c$, in agreement with Eq. (\ref{lmax.3}).

Thus in this regime the large $n$ behavior of $\langle l_{{\rm max},n}\rangle$ is unaffected by the presence of the drift $c$. This result could have been anticipated as $l_{{\rm max}, n}$ can be considered as the longest excursion between two consecutive zeros of a renewal process with a persistence exponent $1/2$. This quantity was studied in Ref. \cite{godreche_excursion} and its average was computed, yielding the large $n$ behavior obtained in Eq. (\ref{lmax.3}).

\subsubsection{Regime II: $\mu = 1$ and $c$ arbitrary:}

In this regime, we recall that the persistence 
probability $Q(m)$ behaves algebraically for large $m$ with an exponent $\theta(c)$ which 
depends continuously on $c$
\begin{eqnarray}\label{persistence_age}
Q(m) \sim \frac{B_{\rm II}}{m^{\theta(c)}}\;, \quad \theta(c) = \frac{1}{2} + \frac{1}{\pi} 
\arctan(c) \;,
\end{eqnarray}
where the amplitude $B_{\rm II}$ is given in Eq. (\ref{cl.thetac}). 
\begin{figure}[ht]
\begin{center}
\includegraphics[angle=-90,width=0.8\textwidth]{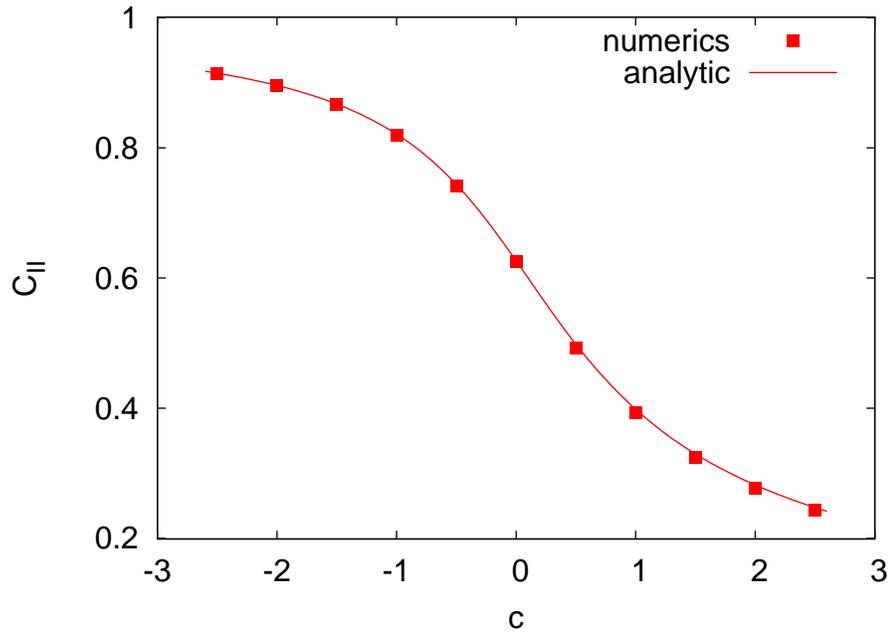}
\caption{Plot of $C_{\rm II}$ as a function of $c$. The red squares correspond to numerical data while the solid line corresponds to our analytical result in Eq. (\ref{lmax_regime2}) together with Eq. (\ref{persistence_age}).}\label{fig.plotc2}
\end{center}
\end{figure}
Here again we can use the result obtained in Ref. \cite{godreche_excursion} for the longest excursion between consecutive zeros of a renewal process with a persistence exponent $\theta(c)$ to obtain
\begin{eqnarray}\label{lmax_regime2}
\langle l_{{\rm max},n}\rangle \sim C_{\rm II} \, n \;, \quad C_{\rm II} = \int_0^\infty dy 
\frac{1}{1 + y^{\theta(c)} e^y \int_0^y dx \, x^{-\theta(c)} e^{-x}} \;,
\end{eqnarray} 
which depends continuously on $c$ and is independent of the non-universal amplitude $B_{\rm II}$~(\ref{persistence_age}). In Fig. \ref{fig.plotc2} we show a comparison of $C_{\rm II}$ obtained numerically (the squares symbols) and from our exact formula (solid line), which shows a very good agreement between both.

\subsubsection{Regime III: $1 < \mu < 2$ and $c>0$:}

In this regime the persistence probability $Q(m)$ behaves for large $m$ as
\begin{eqnarray}\label{persist_III}
Q(m) \sim \frac{B_{\rm III}}{m^\mu} \;,
\end{eqnarray}
where the amplitude $B_{\rm III}$ is given in Eq. (\ref{BIII}). Using again the results obtained in Ref.~\cite{godreche_excursion} one obtains that
\begin{eqnarray}\label{CIII}
\langle l_{{\rm max},n}\rangle \sim C_{\rm III} \; n^{1/\mu} \;,
\end{eqnarray}
where, however, the amplitude $C_{\rm III}$ was not given in Ref.~\cite{godreche_excursion}. 
\begin{figure}*[ht]
\begin{center}
\includegraphics[width = 0.8 \textwidth]{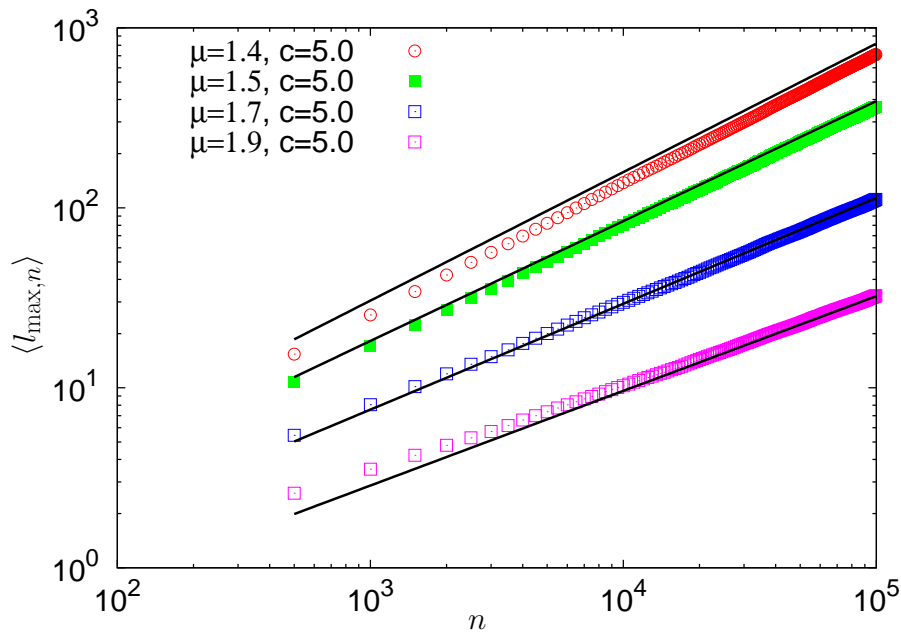}
\caption{Plot, in a log-log scale, of $\langle l_{\rm max,n}\rangle$ as a function of $n$ in regime III: the different curves correspond to different values of $\mu = 1.4, 1.5, 1.7, 1.9$ with a fixed value of $c=5.0$. The solid line are the exact results given in Eqs (\ref{CIII}, \ref{c3}), without any fitting parameter.}\label{fig.c3}
\end{center}
\end{figure}
A careful analysis of the above formula (\ref{lmax.1}) allows to obtain the amplitude $C_{\rm III}$ as
\begin{eqnarray}\label{c3}
C_{{\rm III}} = \frac{1}{c} \Gamma(1-1/\mu) \left[\frac{1}{\pi} \sin\left(\frac{\mu \pi}{2}\right) \Gamma(\mu) \right]^{1/\mu} \;,
\end{eqnarray}
which diverges as $C_{\rm III} \sim (\pi (\mu-1))^{-1}$ when $\mu \to 1$ and vanishes as $C_{\rm III} \sim \sqrt{\pi(2-\mu)/2}$ when $\mu \to 2$. In Fig. \ref{fig.c3} we show a plot of our numerical data for $\langle l_{\max,n}\rangle$ (averaged again over $10^4$ different realizations) for different values of $\mu = 1.4, 1.5, 1.7, 1.9$ and for a fixed value of the drift $c=5.0$. The solid lines indicate the corresponding exact asymptotic behaviors in Eq. (\ref{CIII}, \ref{c3}): the agreement between the two is quite good although the convergence to the asymptotic behavior gets slower as $\mu$ decreases to $1$.

\subsubsection{Regime IV: $\mu =2$ and $c>0$:}

In this case the persistence $Q(m)$ behaves quite differently as it vanishes exponentially for large $m$ as
\begin{eqnarray}\label{persist_IV}
Q(n) \sim \frac{B_{\rm IV}}{n^{3/2}}\, e^{-s_1 n}\, \quad {\rm where}\quad 
s_1=\frac{c^2}{2\sigma^2} \;,
\end{eqnarray}
where the amplitude $B_{\rm IV}$ is given in Eq. (\ref{BIV}). This case was not analyzed in Ref. \cite{godreche_excursion}. From Eq. (\ref{lmax.1}) one has in this case
\begin{eqnarray}\label{caseIV.1}
\sum_{n=0}^\infty \langle l_{{\rm max},n} \rangle e^{-s n} \sim \frac{1}{s} \sum_{m=1}^\infty \frac{\sum_{l=m}^\infty F(l)}{s \tilde q(0) + \sum_{l=m}^\infty F(l)} = \frac{1}{s} \sum_{m=1}^\infty \frac{Q(m)}{s \tilde q(0) + Q(m)} \;.
\end{eqnarray}
Therefore in the limit when $s \to 0$ one can estimate the leading behavior of the sum over $m$ as
\begin{eqnarray}\label{caseIV.2}
\sum_{n=0}^\infty \langle l_{{\rm max},n} \rangle e^{-s n} \sim \frac{m^*}{s} \;,
\end{eqnarray}
where $m^*$ is such that
\begin{eqnarray}
Q(m^*) \sim s \tilde q(0) \;.
\end{eqnarray}
\begin{figure}
\begin{center}
\includegraphics[width = 0.8\textwidth]{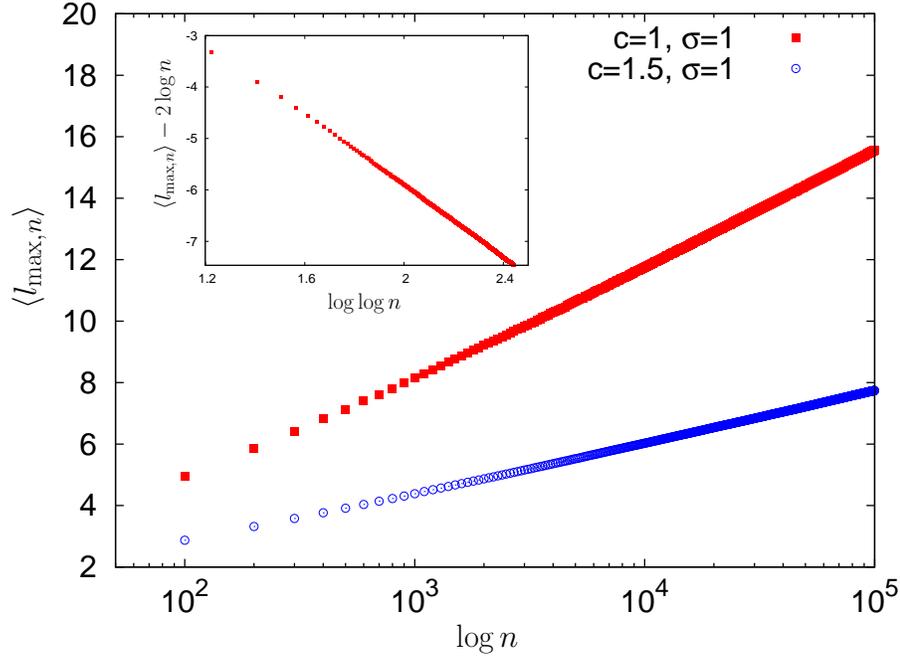}
\caption{Plot of $\langle l_{\max,n}\rangle$ as a function of $\ln{n}$ in the regime IV: here $\mu=2$ and the two curves correspond to 
$c=1$ and $c=1.5$ ($\sigma = 1$ in both cases). The two curves suggest a logarithmic growth, as expected from Eq. (\ref{log_growth}). {\bf Inset:} Plot of $\langle l_{\max,n}\rangle - 2 \ln{n}$ where $2 \ln n$ is the exact asymptotic result from Eq. (\ref{log_growth}) and $2\sigma^2/c^2 = 2$. This plot suggests rather strong corrections $\propto \ln{\ln{n}}$ to the leading logarithmic growth of $\langle l_{\max, n}\rangle$.}\label{fig.c4}
\end{center}
\end{figure}
From the asymptotic behavior above (\ref{persist_IV}) one finds that $m^* \sim - \frac{\sigma^2}{2c^2}\ln{s}$ so that finally
\begin{eqnarray}\label{log_growth}
\langle l_{{\rm max},n}\rangle \sim C_{\rm IV}\ln n \;, \; C_{\rm IV} = \frac{2\sigma^2}{c^2} \;,
\end{eqnarray}
which is in sharp contrast with the algebraic growth obtained above in Eq. (\ref{CIII}) for $1~<~\mu~<~2$ and $c > 0$. In Fig. \ref{fig.c4} we show a plot of $\langle l_{\max,n}\rangle$ as a function of $\ln n$: the straight line suggests indeed a logarithmic growth, in agreement with our analytic result (\ref{log_growth}). However, a more precise comparison with this exact asymptotic result, as shown in the inset of Fig. \ref{fig.c4}, suggests that the leading corrections are proportional to $\ln {\ln n}$, and hence quite strong.   

\subsubsection{Regime V: $1<\mu \leq 2$ and $c<0$:}

In this case the persistence probability $Q(m)$ tends asymptotically to a constant (\ref{reg5.qn}):
\begin{equation}
Q(m)\xrightarrow[m\to\infty]{} \alpha_\mu(c)= 
\exp[-W_{|c|,\mu}(0)]=\exp\left[-\sum_{n=1}^{\infty} 
\frac{1}{n}\int_{|c|n}^{\infty} 
P_n(x)\,dx\right]\,.
\end{equation} 
In addition from (\ref{reg3.singular}) one has that $Q(m) - \alpha_\mu(c) \propto n^{\mu-1}$ so that $F(m) \propto m^{-\mu}$ for large $m$. Therefore, the terms entering into the sum in Eq. (\ref{lmax.1}) are given, to leading order when $1-z = e^{-s} \to 0$ and large $m$ (which are terms which give the leading contribution to this sum over $m$)
\begin{eqnarray}
\frac{\sum_{l=m}^\infty F(l) z^l + (1-z) \sum_{l=m}^\infty Q(l) z^l}{(1-z) \tilde Q(z) + \sum_{l=m}^\infty F(l) z^l} \sim \frac{\alpha_\mu(c)}{\tilde q(0)} e^{-s m} = e^{-s m} \;.
\end{eqnarray}
Therefore this yields 
\begin{eqnarray}\label{lmax_regime5}
\langle l_{{\rm max},n}\rangle \sim C_{\rm V} \,n \;, \; C_{\rm V} = 1 \;.
\end{eqnarray}
This result, which is corroborated by our numerical simulations (see Fig. \ref{fig_regime_c125}), can be physically understood as in this regime where $c<0$ and $\mu > 1$ the number of records is finite and these records typically occur during the first steps of the random walks, where the walker might stay positive for a short while before it escapes to negative values when $n \to \infty$, and no record happens any more.

\subsection{Age of shortest lasting record $l_{{\rm min},n}$}
\label{shortest_records.section}

We now consider the shortest lasting record whose age $l_{{\rm min},n}$ is given by (see Fig.~\ref{renewal.fig})
\begin{eqnarray}
l_{{\rm min},n} = \min(l_1, l_2, \cdots, l_R) \;.
\end{eqnarray} 
Note that, given that the final incomplete interval $l_R$ is taken into consideration 
above, $l_{\min,n}$ can 
be zero: this happens when a record has been broken at the last step, such that $l_R = 0$. 

The cumulative distribution ${\cal G}_n(m) = {\rm Proba.}\,(l_{{\rm min},n} \geq m)$ was studied in Ref. \cite{MZrecord} and an explicit formula was obtained for its generating function: 
\begin{eqnarray}
\sum_{n=0}^\infty {\cal G}_n(m) z^n = \frac{\sum_{l=m}^\infty Q(l) z^l}{1-\sum_{l=m}^\infty F(l) z^l} \;,
\end{eqnarray}
from which one gets the generating function of the average value $\langle l_{\rm min,n}\rangle$ as
\begin{eqnarray}\label{lmin.1}
\sum_{n=0}^\infty z^n \langle l_{{\rm min},n}\rangle = \sum_{m=1}^\infty \frac{\sum_{l=m}^\infty Q(l) z^l}{1 - \sum_{l=m}^\infty F(l) z^l} \;. 
\end{eqnarray}
In the absence of drift, $c=0$, it was shown in Ref. \cite{MZrecord} that
\begin{eqnarray}
\langle l_{{\rm min},n}\rangle \sim D \sqrt{n} \;, \; D = \frac{1}{\sqrt{\pi}} \;.
\end{eqnarray}
As for $\langle l_{{\rm max},n}\rangle$ we will see that the behavior of $\langle l_{{\rm min},n}\rangle$, in the presence of non zero drift $c \neq 0$, is quite different in the five different regimes discussed above. Again we start by giving a brief summary of our results for $\langle l_{{\rm min},n}\rangle$:
\begin{eqnarray}
\langle l_{{\rm min},n}\rangle &\sim & \sqrt{n}\quad {\rm for}\quad 0<\mu<1\,\, {\rm and}\,\, c\,\,{\rm 
arbitrary}\quad ({\rm regime}\,\, {\rm I}) \;,
\nonumber \\
&\sim & n^{1-\theta\left(c\right)} \quad {\rm for}\quad \mu=1 \,\, {\rm and}\,\,c \,\,{\rm
arbitrary}  
\quad ({\rm regime}\,\, {\rm II}) \;,
\nonumber \\
&\sim & {\rm const.} \quad {\rm for}\quad 1<\mu<2\,\, {\rm and}\,\, c>0 
\quad ({\rm regime}\,\, {\rm III}) \;,
\nonumber \\
&\sim & {\rm const.} \quad  {\rm for}\quad \mu=2 \,\, 
{\rm and}\,\, c>0 
\quad ({\rm regime}\,\, {\rm IV}) \;,
\nonumber \\
&\sim & n \quad {\rm for}\quad 1<\mu\le 2 \,\, {\rm and}\,\, c<0 
\quad ({\rm regime}\,\, {\rm V}) \;,
\nonumber \\
\label{l_maxn_asymp}
\end{eqnarray}
again with $\theta\left(c\right)$ as defined in Eq. (\ref{cauchy_mean.1}). In the following we discuss the behavior of $\langle l_{{\rm min},n}\rangle$ in more detail for each of the five regimes.

\subsubsection{Regime I: $0< \mu < 1$ and $c$ arbitrary}

\begin{figure}
\begin{center}
\includegraphics[width = 0.8\textwidth]{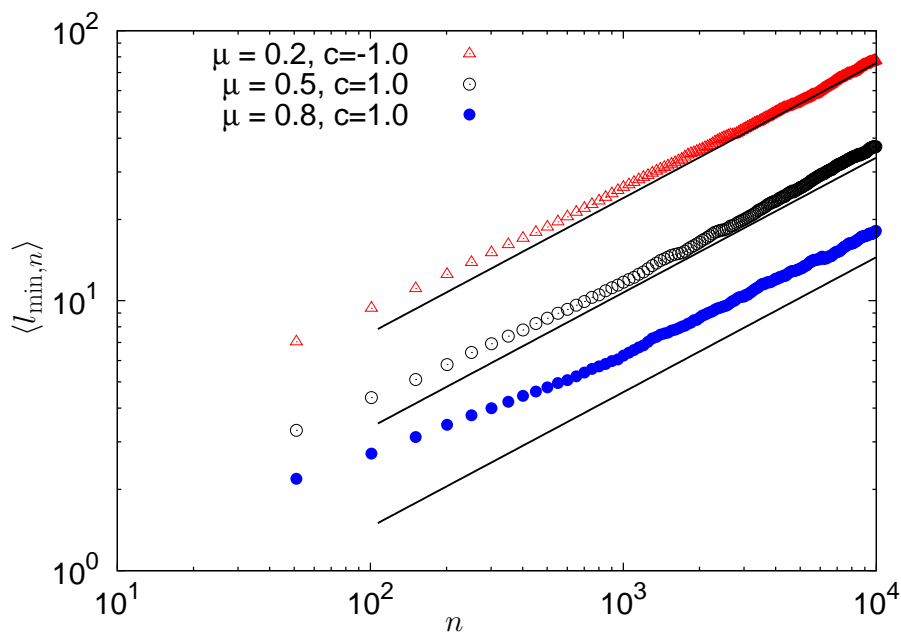}
\caption{Plot, on log-log scale, of $\langle l_{\min, n}\rangle$ as a function of $n$, for different values of $\mu < 1$ and $c$ (regime I). The points are the results of numerical simulations while solid lines correspond to our exact analytic result given in Eq. (\ref{expr_d1}). These data indicate that in this regime $\langle l_{\min, n}\rangle \propto \sqrt{n}$, although the corrections to the exact asymptotic behavior are clearly visible, in particular for $\mu = 0.8$, $c=1.0$.}\label{fig_d1}
\end{center}
\end{figure}

\begin{figure}[ht]
\begin{center}
\includegraphics[angle=0,width=0.8\textwidth]{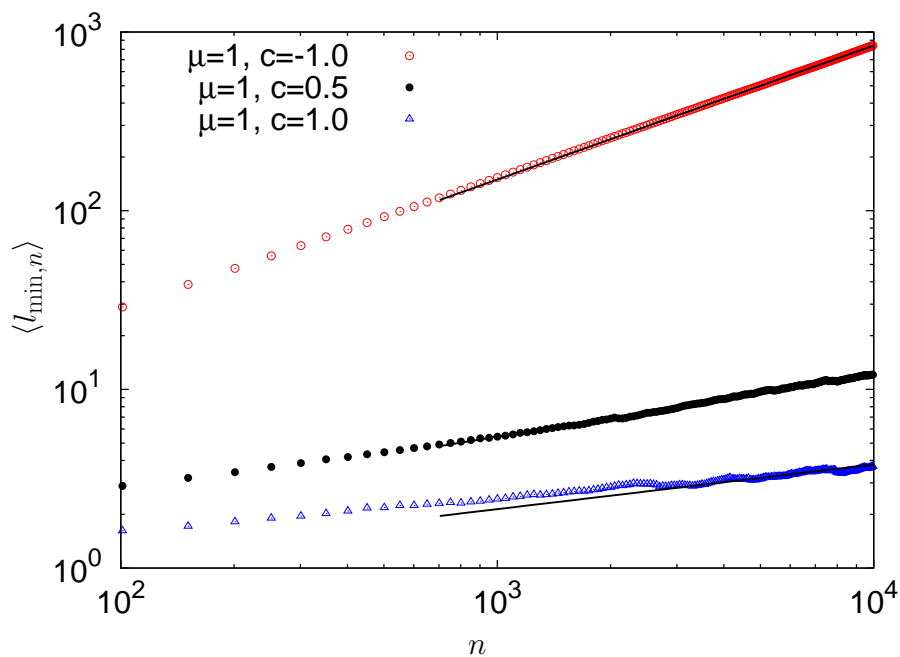}
\caption{Plot, on a log-log scale, of $\langle l_{\min,n}\rangle$ as a function of $n$ for $\mu = 1$ and different values of $c=-1,0.5$ and $c=1$. The solid line corresponds to the algebraic growth $n^{1- \theta(c)}$, from Eq.~(\ref{dII}).}\label{fig.d2}
\end{center}
\end{figure}

In this case the persistence probability decays algebraically as given in Eq. (\ref{asymptQF}) and the analysis of $\langle l_{{\rm min},n}\rangle$ can be obtained by noticing that, in the limit $z \to 1$, the denominator in Eq. (\ref{lmin.1}) can be simply replaced by $1$ while the remaining sums over $l$ (in the numerator) and over $m$ can be replaced by integrals. This yields straightforwardly
\begin{eqnarray}\label{expr_d1}
&&\langle l_{{\rm min},n}\rangle \sim D_{\rm I} \sqrt{n} \;, \; \\
&&D_{\rm I} = B_{\rm I} = \frac{1}{\sqrt{\pi}} \exp{\left[-\frac{1}{\pi} \int_0^\infty \frac{dk}{k} {\rm arctan} \left(\frac{\hat f(k) \sin{(kc)}}{1 - \hat f(k) \cos{(kc)}}  \right) \right]} \;,
\end{eqnarray} 
where the expression of $B_{\rm I}$ is given in Eq. (\ref{BI}). In Fig. \ref{fig_d1}, we show the results of our numerical simulations which are in a rather good agreement with Eq. (\ref{expr_d1}), although the corrections to this exact asymptotic behavior are clearly visible, in particular for $\mu = 0.8$, $c=1.0$. In Fig. \ref{fig.d2}, we show a plot of the numerical computation of $\langle l_{{\rm min},n}\rangle$ for $\mu=1$ and different values of $c=-1, 0.5$ and $c=1$: these data are in good agreement with the power law growth in Eq. (\ref{expr_d1}), although we have not attempted to estimate numerically the prefactor $D_{\rm I}$.

\subsubsection{Regime II: $\mu = 1$ and $c$ arbitrary}

In this regime where the persistence probability $Q(m)$ decays algebraically as in Eq. (\ref{cl.qn}), $\langle l_{\min, n}\rangle$ can be analyzed as in the regime I where 
 in the limit $z \to 1$, the denominator in Eq. (\ref{lmin.1}) can be simply replaced by~$1$ while the remaining sums over $l$ (in the numerator) and over $m$ can be replaced by integrals. This yields straightforwardly:
 \begin{equation}\label{dII}
 \sum_{n=1}^\infty e^{-s n} \langle l_{\min, n}\rangle \sim \frac{B_{\rm II}}{s^{2-\theta(c)}} \int_0^\infty dy \int_y^\infty dx \, x^{-\theta(c)} e^{-y} = \frac{B_{\rm II}}{s^{2-\theta(c)}} \Gamma[2 - \theta(c)] \;,
 \end{equation}
which yields
\begin{eqnarray}
\langle l_{\min, n}\rangle \sim D_{\rm II} \, n^{1-\theta(c)} \:, \; D_{\rm II} = B_{\rm II}  \;,
\end{eqnarray}
where $B_{\rm II}$ is given in Eq. (\ref{cl.thetac}) and 
$\theta(c)=1/2+\frac{1}{\pi}\,\arctan(c)$.

\subsubsection{Regime III: $1 \leq \mu < 2$ and $c>0$}

\begin{figure}[ht]
\begin{center}
\includegraphics[angle=0,width=0.8\textwidth]{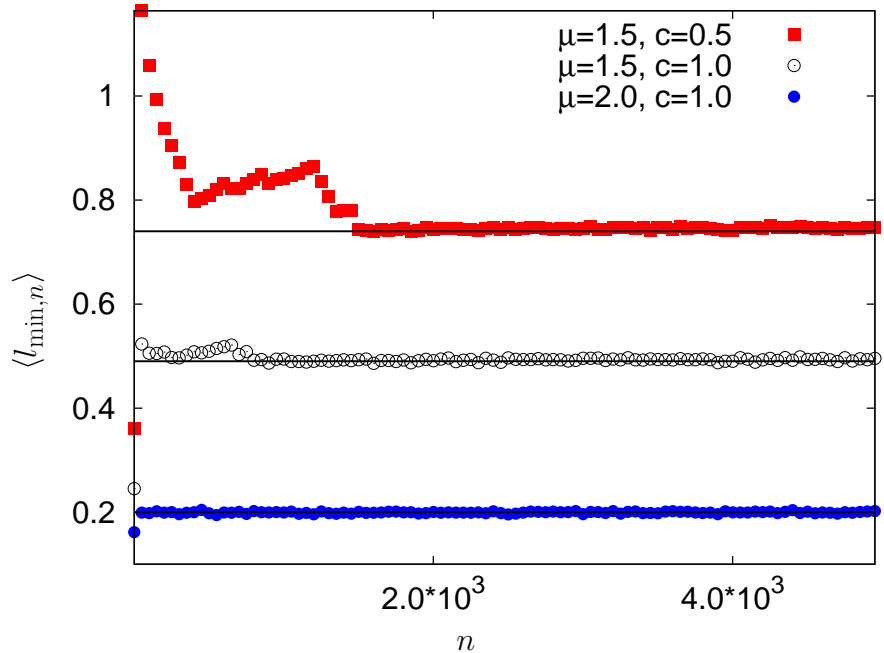}
\caption{Plot of $\langle l_{\min,n}\rangle$ as a function of $n$ for $\mu = 1.5$ and $\mu=2$ and different values of $c>0$, therefore corresponding to regime III and IV. The solid line corresponds to the exact result, from Eq.~(\ref{d3}, \ref{d4}).}\label{figd3d4}
\end{center}
\end{figure}
In this case we write the above formula (\ref{lmin.1}) as
\begin{eqnarray}\label{lmin.regIII}
&&\sum_{n=0}^\infty z^n \langle l_{{\rm min},n}\rangle = \frac{1}{1-z} \left(1 - \frac{1}{\tilde q(0)} \right)  + \sum_{m=2}^\infty \frac{\sum_{l=m}^\infty Q(l) z^l}{1 - \sum_{l=m}^\infty F(l) z^l} \;,
\end{eqnarray}
where we have simply isolated the term $m=1$ and used $1-\tilde F(0) = (1-z) \tilde Q(0)$ (\ref{qfrelation.2}). 
Now the above sum (\ref{lmin.regIII}), which starts with $m=2$, is dominated by the large values of $m$. Because of  
the algebraic decay of $Q(m) \sim m^{-\mu}$ in this case (\ref{persist_III}) and $\mu >1$ in this regime one gets that this second term behaves like $(1-z)^{\mu - 2}$, which is then subleading, compared to the first term which behaves like $(1-z)^{-1}$. Therefore one gets in this case
\begin{equation}\label{d3}
\langle l_{{\rm min},n}\rangle \sim D_{\rm III} \;, \;  D_{\rm III} = 1 - \frac{1}{\tilde q(0)} = 1 - \exp\left[-\sum_{n=1}^{\infty} 
\frac{1}{n}\,\int_{cn}^{\infty} P_n(x)\, dx\right] \;,
\end{equation}
where we have used the expression for $1/\tilde q(0)$ given in Eq. (\ref{amuc}). In Fig. \ref{figd3d4} we show a plot of the numerical computation $\langle l_{{\rm min},n}\rangle$ for $\mu = 1.5$ and different values of $c=0.5$ and $c=1$, which is in very good agreement with Eq. (\ref{d3}). Note that we have extracted the value of ${1}/{\tilde q(0)} $ which enters into the expression of $D_{\rm III}$ from the linear growth of the mean record number $\langle R_n \rangle$, according to (\ref{amuc}). 

\subsubsection{Regime IV: $\mu =2$ and $c>0$}

A similar analysis can be carried out in this case, starting from the same formula (\ref{lmin.regIII}). In this case, in the above sum (\ref{lmin.regIII}), which starts with $m=2$, one can safely put $z=1$, because of the behavior of the exponential decay of $Q(m)$ in this case (\ref{persist_IV}). Therefore one gets immediately
\begin{equation}\label{d4}
\langle l_{{\rm min},n}\rangle \sim D_{\rm IV} \;, \;  D_{\rm IV} = 1 - \frac{1}{\tilde q(0)} = 1 - \exp\left[-\sum_{n=1}^{\infty} 
\frac{1}{n}\,\int_{cn}^{\infty} P_n(x)\, dx\right] \;,
\end{equation}
where we have used the expression for $1/\tilde q(0)$ given in Eq. (\ref{a2c}). In Fig. \ref{figd3d4} we show a plot of the numerical computation $\langle l_{{\rm min},n}\rangle$ for $\mu = 2$ and $c=1$, which is good agreement with Eq. (\ref{d4}). Note that we have extracted the value of ${1}/{\tilde q(0)} $ which enters into the expression of $D_{\rm IV}$ from the linear growth of the mean record number $\langle R_n \rangle$, according to Eq.~(\ref{a2c}).

\subsubsection{Regime V: $1 < \mu \leq 2$ and $c<0$}

\begin{figure}
\begin{center}
\includegraphics[width=0.8\textwidth]{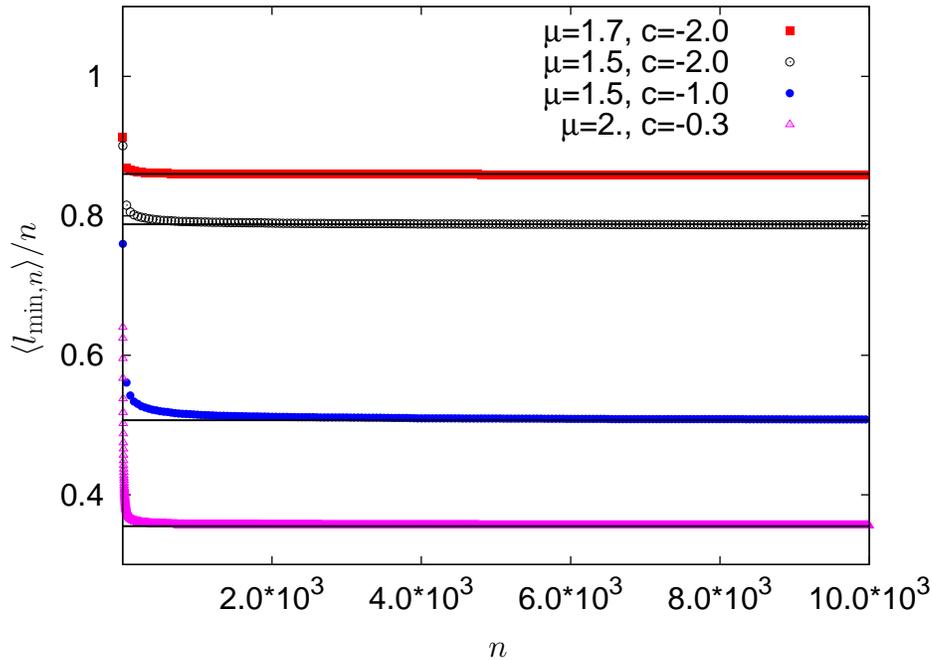}
\end{center}
\caption{Plot of $\langle l_{\min,n}\rangle/n$ as a function of $n$ for different values of $1< \mu \leq 2$ and different values of $c<0$, corresponding to regime V.}\label{fig.d5}
\end{figure}
In this regime where the persistence goes to a constant $Q(m) \to \alpha_\mu(c)$, for $m \gg 1$, one can simply replace $Q(l)$ by this constant value in the sum of the numerator in Eq. (\ref{lmin.1}) while the denominator can be simply approximated by $1$ in the limit $1-z = e^{-s} \to 0$. This yields straightforwardly
\begin{eqnarray}\label{lmin.regV}
\langle l_{\min,n}\rangle \sim \alpha_\mu(c) \, n \;.
\end{eqnarray}
In Fig. \ref{fig.d5} we show a plot of $\langle l_{\min,n}\rangle/n$ which we have computed numerically for different values of $\mu = 1.7, 1.5$ and $\mu=2$ and also for different values of the drift. These results are in very good agreement with our exact asymptotic result in Eq. (\ref{lmin.regV}), where the value of $\alpha_\mu(c)$ have been extracted from the mean record number $\langle R_n\rangle \sim 1/\alpha_\mu(c)$ (\ref{mean_IV}). This result (\ref{lmin.regV}) can be easily understood by realizing that $l_{\min,n} = n$ if the whole trajectory is on the negative side, which happens with probability $\alpha_\mu(c)$ while $l_{\min,n}$ is of order ${\cal O}(1)$ if the walker makes an excursion on the positive side. One also notices that, in this case, $l_{\rm typ} \sim \langle l_{\min, n} \rangle$.  
  
\section{Conclusion}
\label{conclusion.section}

\begin{figure}
\includegraphics[angle=-90,width=1\textwidth]{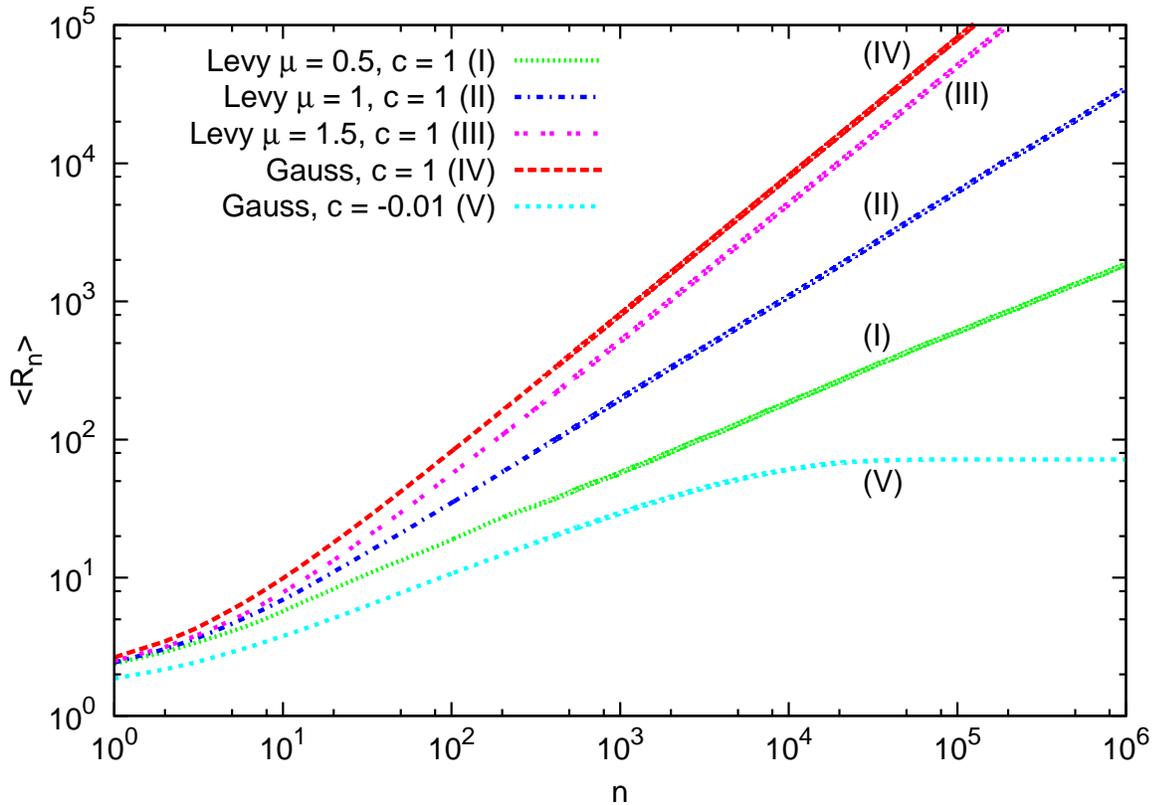}
\caption{The figure shows numerical results for the mean record 
number $\langle R_n\rangle$ for biased random walks from all five regimes. 
For regimes I to IV we used a positive bias of $c=1$, in regime V we simulated a 
Gaussian random walk (with $\sigma=1$) with a negative bias of $c=-0.01$. For each jump 
distribution we 
averaged over $10^4$ samples. In all these cases, as shown in detail in the previous figures, the asymptotic behavior
agree very well with our analytical predictions (which are not shown on this figure for clarity).}
\label{mean_R.fig}
\end{figure}

In this paper we considered a very simple 
model of a one dimensional discrete-time random walk in presence of a constant drift $c$.
At each time step the particle jumps by a random distance $c+\eta$ where
the noise $\eta$ is drawn from 
a continuous and symmetric jump distribution
$f(\eta)$, characterized by a L\'evy index $0<\mu \le 2$. The
jump has a finite second moment for $\mu=2$, while for $0<\mu<2$ the second moment diverges.
For this discrete-time series consisting of the successive positions
of the biased walker, we presented complete analytical studies of
the persistence and the record statistics. For the later, we studied
the mean and the full distribution of the number of records up to step $n$
and also the statistics of the duration of records, in particular those
for the longest and shortest lasting records. As a function of the
two parameters $c$ and $0<\mu\le 2$,  
we found that it is necessary to distinguish between five 
different universal regimes, as 
shown in the basic phase diagram in Fig.~\ref{phd.fig}.
In these 5 regimes, the persistence and the record statistics
exhibit very different asymptotic behaviors that are summarized
in Section 2 and we do not repeat them here. For instance, the growth of the
mean record number with $n$ in all five regimes is 
summarized in the simulation results in Fig.~\ref{mean_R.fig}, in
complete agreement with our analytical predictions.
The main conclusion is that even though this is a rather simple model, it 
exhibits
very rich and varied universal behaviors for record statistics and 
persistence depending 
on the two parameters $c$ and $0<\mu\le 2$.

Our results provide a simple yet nontrivial, but fully solvable model for
the record statistics, a subject which has gained considerable interest over the
last few years. Our results provide one generalization of the previous
results for record statistics for symmetric random walks~\cite{MZrecord}.
However, it is important to note that this extension does not yet cover 
all possible kinds of discrete-time random walks. 
In principle one could consider more 
complicated asymmetries of the jump distribution. It might be interesting to consider a jump 
distribution that has different tail-exponents in the left and in the right tail. Also a 
generalization of these results to an asymmetric lattice random walk is still missing. In 
\cite{MZrecord} a symmetric lattice random was also considered. It should be possible to 
compute the record statistics of a lattice random walk that has a higher probability to jump 
in one direction than in the other.

It might be interesting to see if our results can be applied to financial data, 
similar to the analysis in \cite{WBK,WMS}. Daily stock data however proved not to be useful 
for comparison because the asymptotic limit is hardly achieved in the
available  observational data. An application to stock data with a higher temporal resolution 
however 
should be possible and might provide new insights. Such an analysis is definitely an 
interesting subject for future research. Also the distribution of records in stock prices has 
not been analysed in detail before and it would be interesting to see if such
an analysis for available data can be fitted to our theoretical distributions.

\vskip 0.3cm \noindent\textbf{Acknowledgments:} SNM and GS acknowledge support by ANR grant 
2011-BS04-013-01 WALKMAT and in part by the Indo-French Centre for the Promotion of Advanced Research under Project 4604-3. GW is grateful for the kind hospitality of the Laboratoire de Physique Th\'eorique et Mod\`eles Statistiques during the completion of this work and for the 
financial support provided by DFG within the Bonn Cologne Graduate School of Physics and 
Astronomy.

\appendix

\section{The constant $A_{\rm I}$}

The constant $A_{\rm I}$ in Eq. (\ref{meanrec_2.reg1}) can be directly expressed
in terms of ${\hat f}(k)$ as announced in Eq. (\ref{ac.1}).
To derive this, we use the 
explicit expression of $P_n(x)$
from Eq. (\ref{pdf.x}) in the expression for $A_{\rm I}$ and integrate over $x$ to get
\begin{equation}
A_{\rm I}= \frac{2}{\sqrt{\pi}}\,\exp\left[\sum_{n=1}^{\infty} 
\frac{1}{n}\,\int_{-\infty}^{\infty} \frac{dk}{2\pi}\, [{\tilde f}(k)]^n 
\frac{1-e^{-i\,k\,c\,n}}{ik}\right].
\label{ampli1_reg1}
\end{equation}
Next we use the symmetry ${\hat f}(k)={\hat f}(-k)$ which leads to
\begin{equation}
A_{\rm I} = \frac{2}{\sqrt{\pi}}\,\exp\left[\frac{1}{\pi}\,\int_{0}^{\infty} 
\frac{dk}{k}\, \sum_{n=1}^{\infty} \frac{\sin(kcn)}{n}\, [{\tilde f}(k)]^n 
\right].
\label{ampli2_reg1}
\end{equation}
The sum on the rhs can be explicitly evaluated using the identity
\begin{equation}
\sum_{n=1}^{\infty} \frac{x^n}{n} \sin(an) = \arctan\left[\frac{x\sin(a)}{1-x 
\cos(a)}\right]
\label{iden.1}
\end{equation}
which then leads to the exact expression in Eq. (\ref{ac.1}).

We then analyze the behavior of $A_{\rm I}$ when $|c|$ is large and in the case where $\hat f (k) = \exp(-|k|^\mu)$, with $\mu < 1$. In that case one has $P_n(x) = n^{-1/\mu} {\cal L}_{\mu}(x/n^{1/\mu})$ for all $n$ and it is easier to start from the formula given in the text in Eq. (\ref{meanrec_2.reg1})
\begin{eqnarray}\label{redef_s0_app}
A_{\rm I} = \frac{2}{\sqrt{\pi}} e^{S_0} \;, \; S_0 \equiv S_0(c) = \sum_{n=1}^\infty \frac{1}{n} \int_0^{cn} {\cal L}_\mu(x/n^{1^{1/\mu}}) dx/n^{1/\mu}\;.
\end{eqnarray}
Note that, given that $P_n(x) = P_n(-x)$ one has $S_0(c) = S_0(-c)$ and we thus present the analysis for $c>0$. Performing the change of variable $y=x/n^{1/\mu}$ in the integral above (\ref{redef_s0_app}) we write
\begin{eqnarray}
S_0(c) = \sum_{n=1}^\infty \frac{1}{n} \int_0^{c n^{\frac{\mu-1}{\mu}}} {\cal L}_\mu(y) \, dy \;,
\end{eqnarray}
and take the derivative with respect to $c$
\begin{eqnarray}\label{derivative.s0.1}
S_0'(c) = \sum_{n=1}^\infty n^{-\frac{1}{\mu}} {\cal L}_\mu\left( \frac{c}{n^{\frac{1-\mu}{\mu}}}\right) \;.
\end{eqnarray}
In this expression, one notices that ${c}/{n^{\frac{1-\mu}{\mu}}} = (n/c^{\frac{\mu}{1-\mu}})^{\frac{\mu-1}{\mu}}$ so that when $c \to \infty$ the discrete sum over $n$ in Eq. (\ref{derivative.s0.1}) can be 
replaced by an integral (we recall that $\mu < 1$ here), which leads to
\begin{eqnarray}\label{derivative.s0.2}
S_0'(c) \sim \frac{1}{c} \int_0^\infty {\cal L}_{\mu} \left( y^{\frac{\mu -1}{\mu}} \right)\,
y^{-1/\mu}\, dy 
\;.
\end{eqnarray}
Finally, performing the change of variable $z = y^{\frac{\mu -1}{\mu}}$ in Eq. (\ref{derivative.s0.2}) yields
\begin{eqnarray}
S_0'(c) \sim \frac{1}{c} \frac{\mu}{1-\mu} \int_0^\infty {\cal L}_{\mu}(z) dz = \frac{1}{c} \frac{\mu}{2(1-\mu)} \;,
\end{eqnarray}
so that one gets
\begin{eqnarray}\label{AI.largec.1}
A_{\rm I} = \frac{2}{\sqrt{\pi}} e^{S_0} \propto c^{\frac{\mu}{2(1-\mu)}} \;, \; c \to \infty \;.
\end{eqnarray} 
This power law behavior (\ref{AI.largec.1}) can be understood from the following scaling argument. We are indeed interested in the records statistics of the variables $y_n$, with $y_n = x_n + c n$~(\ref{bias.1}) where  $x_n$ behaves for large $n$ as $x_n = {\cal O}(n^{1/\mu})$. Therefore for small $n$, $n < n^*$ when $c$ is large, $y_n$ is dominated by the drift term and $n^*$ is such that $c n^* \sim {n^*}^{1/\mu}$, which yields
\begin{eqnarray}
n^* \sim c^{\frac{\mu}{1-\mu}} \;.
\end{eqnarray} 
On the other hand, for small $n$, $n < n^*$, $y_n$ is dominated by the (positive) drift and hence is almost deterministic which yields $\langle R_n \rangle \sim n$, for $n < n^*$ while $\langle R_n \rangle \sim A_{\rm I}\sqrt{n}$ for $n > n^*$. By matching these two behaviors for $n = n^*$ one obtains
\begin{eqnarray}
A_{\rm I} \sim \sqrt{n^*} \propto c^{\frac{\mu}{2(1-\mu)}} \;,
\end{eqnarray} 
which yields the result obtained above (\ref{derivative.s0.2}). 

Note finally that, by using $S_0(c) = -S_0(-c)$ one obtains
\begin{eqnarray}
A_{\rm I} \sim (-c)^{\frac{-\mu}{2(1-\mu)}} \;, \; c \to -\infty \;.
\end{eqnarray}

\section{Computation of $\alpha_2(c) = a_2(|c|)$, $c<0$ for exponential jump 
distribution with $c<0$}\label{appendix_a2}

The expression for the amplitude $\alpha_2(c)$ in regime V
(with $c<0$) and for a general jump distribution is given in Eq. (\ref{reg5.qn}).
By comparing with Eq. (\ref{a2c}) we see that
$\alpha_2(c<0)= a_2(|c|)$ where $a_2(|c|)$ is the prefactor of the 
leading linear growth of mean record number in regime IV with drift positive $|c|$.  
For a general jump distribution $f(\eta)$, we then have
\begin{equation}
\alpha_2(c)= \exp\left[-\sum_{n=1}^{\infty}
\frac{1}{n}\int_{|c|n}^{\infty}
P_n(x)\,dx\right] \;,
\label{a2c_app.1}
\end{equation}  
where we recall that $P_n(x)= \int_{-\infty}^{\infty} \frac{dk}{2\pi}\, \left[{\hat 
f}(k)\right]^n \,
e^{-i\,k\, x}$ and ${\hat   
f}(k)=\int_{-\infty}^{\infty} f(\eta)\, e^{ik\eta}\, d\eta$ is the Fourier transform
of the jump distribution. Thus, in general, computing the prefactor $\alpha_2(c)=a_2(|c|)$ 
explicitly
is difficult for arbitrary $f(\eta)$. It can be done explicitly for Gaussian distribution
where $P_n(x)=(2\pi n\sigma^2)^{-1/2} \exp[-x^2/{2n\sigma^2}]$ itself is Gaussian
and $\alpha_2(c)=a_2(|c|)$ is then given by the formula in Eq. (\ref{a2cgaussian}).
In this appendix, we show that $\alpha_2(c)=a_2(|c|)$ can also be computed
explicitly for the symmetric exponential distribution
$f(\eta)= (2\,b)^{-1}\, \exp(-|x|/b)$.

For this exponential jump distribution, the Fourier transform has the Lorentzian form,
${\hat f}(k)= 1/[\pi (b^2\,k^2+1)]$. One can then substitute this in the
expression for $P_n(x)$ and eventually in Eq. (\ref{a2c_app.1}). After a
quite convoluted computation involving contour integration in the
complex plane, one can find $\alpha_2(c)$ explicitly. However, as we show below, 
for the exponential case,
there is an alternative simpler way to compute $\alpha_2(c)$ directly
(without going through the formula in Eq. (\ref{a2c_app.1}).   

The first observation is that $\alpha_2(c)$ is just the limiting value
of the persistence probability $Q(n)$ (the probability that
the walker stays {\em below} $0$ up to $n$ steps starting at $0$) when $n\to \infty$ in 
presence of 
a negative drift $c<0$. By symmetry, $Q(n)$ is then also the probability
that the walker, starting at the origin, stays {\em above} the origin up to
$n$ steps, but in presence of a positive drift $|c|>0$.  
So, the idea is to compute this probability $Q(n)$ directly for 
the exponential jump distribution and then take the limit $n\to \infty$
to compute $\alpha_2(c)=Q(n\to \infty)$. 

To compute $Q(n)$, we first define
\begin{equation}
q^+_n(y) = {\rm Proba.} \; {\text{that the random walker, starting at}} \; y\ge 0 \; 
{\text{stays 
positive up to step $n$}} \;.
\end{equation}
If we can compute $q_n^+(y)$, then $Q(n)$ is simply obtained by putting the
starting position to be $0$, i.e., 
$Q(n)= q_n^+(0)$. To compute $q_n^+(y)$, we can write a backward recurrence relation for
$q_n^+(y)$ by considering the jump that happens at the first step from $y$ to $y'\ge 0$
\begin{eqnarray}\label{backward.app}
&&q^+_n(y) = \int_0^\infty q^+_{n-1}(y') f(y+|c|-y') \, dy' \; \;, \\
&&q^+_0(y) = 1 \,\, {\rm for}\, y\ge 0\, .
\end{eqnarray}
In the limit of large $n$, we expect that $q^+_n(y)$ approaches
to an $n$ independent stationary value, $q^+_n(y)\to q^+(y)$,  
that just denotes the eventual probability with which the walker escapes to infinity
(starting from $y$) in presence of a positive drift $|c|$.
Taking $n\to \infty$ limit on both sides of Eq. (\ref{backward.app})
gives the integral equation for $y\ge 0$
\begin{equation}\label{backward1.app}
q^+(y) = \int_0^\infty q^+(y') f(y+|c|-y') \, dy'\, .
\end{equation}
Note that this equation is valid for arbitrary jump distribution $f(\eta)$. This half-space Wiener-Hopf type integral equation with asymmetric
kernel can not be solved in general. However, for the special
case of the exponential distribution, 
$f(\eta) = 1/(2b) \exp(-|\eta|/b)$, this integral equation (\ref{backward1.app}) 
can be transformed into a differential equation using
\begin{eqnarray}\label{identite.exp}
f''(\eta) = - \frac{1}{b^2} \delta(\eta) + \frac{1}{b^2} f(\eta) \;.
\end{eqnarray}
By differentiating twice Eq. (\ref{backward1.app}) with respect to $y$ one then obtains [using 
Eq. (\ref{identite.exp})]
\begin{eqnarray}\label{stationary.1}
\frac{d^2 q^+(y)}{dy^2} = - \frac{1}{b^2} q^+(y+|c|) + \frac{1}{b^2} q^+(y) \;.  
\end{eqnarray} 
Note that the solution $q^+(y)$ must approach to $1$ as $y\to \infty$:
$q^+(y\to \infty)=1$. This follows from the fact that if the particle starts
at the positive infinity, it escapes to positive infinity with probability $1$
in presence of any positive drift $|c|>0$.  

Note that the differential equation (\ref{stationary.1}), though linear, is actually {\em 
nonlocal} in $y$ due to the first term on the rhs and hence
is still not completely trivial to solve. Fortunately, it turns out
that it admits a solution of the form
\begin{eqnarray}\label{ansatz}
q^+(y) = 1 - F\, \exp{(-\lambda y/b)} \;,
\end{eqnarray}
where $F$ and $\lambda$ are two dimensionless constants (independent of $y$)
that are yet to be determined. Note that this ansatz manifestly
satisfies the boundary condition $q^+(y\to \infty)=1$.
Substituting this ansatz in Eq. (\ref{stationary.1}) we see that indeed
Eq. (\ref{ansatz}) is a solution provided $\lambda$ 
satisfies the equation
\begin{eqnarray}\label{def_lambda}
\exp{(-\lambda \, |c|/b)} = 1 - \lambda^2\, ;\,\, {\rm with}\,\, \lambda > 0 \;.
\end{eqnarray}
The transcendental equation has a unique positive solution which then
determines $\lambda$ uniquely. For example, for $b/c=1$, we get
using Mathematica the root $\lambda=0.714556\ldots$.
But we still need to determine the prefactor $F$ in the ansatz in Eq. (\ref{ansatz}).
The amplitude $F$ in Eq. (\ref{ansatz}) is obtained by 
injecting this solution back into the integral equation (\ref{stationary.1})
and performing the integral. Indeed, one finds that Eq. (\ref{ansatz}) 
is a solution of the integral equation provided  
\begin{eqnarray}\label{Fdeter}
F = 1 - \lambda \;.
\end{eqnarray}
This then uniquely determines the solution of the integral equation (\ref{stationary.1}) 
\begin{equation}
q^+(y)= 1- (1-\lambda)\, \exp{(-\lambda y/b)}
\label{sol_exp.1}
\end{equation}
where $\lambda$ is the unique positive solution of the transcendental equation
(\ref{def_lambda}). 

Noting finally that $\alpha_2(c)=Q(n\to \infty)= q^+(0)$ gives
\begin{eqnarray}\label{a2.exp}
\alpha_2(c) = a_2(|c|) = q^+(0) = \lambda \;,
\end{eqnarray}
where $\lambda > 0$ is the solution of Eq. (\ref{def_lambda}). 
We have checked that we indeed get exactly the same expression by evaluating the original 
general 
expression in Eq. (\ref{a2c_app.1}) for the exponential jump distribution, though
this was not completely trivial to check (we do not give details of this check here).

\end{document}